\documentclass[useAMS,usenatbib]{mn2e}

\usepackage{newtxtext,newtxmath}

\usepackage[T1]{fontenc}
\usepackage{ae,aecompl}

\usepackage{graphicx}	
\usepackage{amsmath}	
\usepackage{amssymb}	
\usepackage{morefloats}
\usepackage{rotating}
\usepackage{pdflscape}
\usepackage{mathtools}

\usepackage{amsfonts}
\usepackage{pifont}

\usepackage{color} 
\definecolor{purple}{RGB}{160,32,240}
\usepackage {comment}
\usepackage{url}
\usepackage[%
    font={small,sf},
    labelfont=bf,
    format=hang,    
    format=plain,
    margin=0pt,
    width=0.8\textwidth,
]{caption}
\usepackage{subcaption}

\def\mnras{\mbox{MNRAS}}
\def\apj{\mbox{ApJ}}
\def\aap{\mbox{A\&A}}
\def\apjl{\mbox{ApJ}}
\def\apjs{\mbox{ApJS}}
\def\aj{\mbox{AJ}}
\def\nat{\mbox{Natur}}
\def\pasp{\mbox{PASP}}

\def\araa{\mbox{ARAA}}

\graphicspath{{./final_plots/}}

\def\msun{\mbox{M$_{\odot}$}}

\def\mstar{\mbox{$M_{\rm C}$}}

\defcitealias{RDA12}{RDA12}

\def \lt sima{$\; \buildrel \lt \over \sim \;$}    
\def\lesssim{\lower.5ex\hbox{\lt sima}}           
\def\gtsima{$\; \buildrel \gt \over \sim \;$}    
\def\grtsim{\lower.5ex\hbox{\gtsima}}           

\title[Clumps with ML]{Stellar Masses of Giant Clumps in CANDELS and Simulated Galaxies Using Machine Learning}

\author[]{Marc Huertas-Company$^{{1},{2},{3},{4}}$, Yicheng Guo$^{5}$, Omri Ginzburg$^{6}$, Christoph T. Lee$^{7}$
\newauthor Nir Mandelker$^{8,9}$, Maxwell Metter$^{5}$, Joel R. Primack$^{7}$, Avishai Dekel$^{6,7}$
\newauthor  Daniel Ceverino$^{10}$, Sandra M. Faber$^{11}$, David C. Koo$^{11}$, Anton Koekemoer$^{12}$
\newauthor Gregory Snyder$^{12}$, Mauro Giavalisco$^{13}$, Haowen Zhang$^{14}$\\
$^{1}$LERMA, Observatoire de Paris, PSL Research University, CNRS, Sorbonne Universit\'es, UPMC Univ. Paris 06,F-75014 Paris, France\\
$^{2}$Univerist\'e de Paris, 5 Rue Thomas Mann - 75013, Paris, France\\
$^{3}$Departamento de Astrof\'isica, Universidad de La Laguna, E-38206 La Laguna, Tenerife, Spain\\
$^{4}$Instituto de Astrof\'isica de Canarias, E-38200 La Laguna, Tenerife, Spain\\
$^{5}$Department of Physics and Astronomy, University of Missouri, Columbia, MO 65211, USA \\
$^{6}$Racah Institute of Physics, The Hebrew University, Jerusalem 91904 Israel\\
$^{7}$Physics Department, University of California, Santa Cruz, CA 95064, USA\\
$^{8}$Department of Astronomy, Yale University, PO Box 208101, New Haven, CT, USA\\
$^{9}$Heidelberger Institut für Theoretische Studien, Schloss-Wolfsbrunnenweg 35, 69118 Heidelberg, Germany\\
$^{10}$Departamento de Física Teórica, CIAFF, Facultad de Ciencias, Universidad Aut\'onoma de Madrid, E-28049 Madrid, Spain\\
$^{11}$UCO/Lick Observatory, Department of Astronomy and Astrophysics, University of California, Santa Cruz, CA 95064, USA\\
$^{12}$Space Telescope Science Institute, Baltimore, MD, United States\\
$^{13}$Department of Astronomy, University of Massachusetts Amherst, 710 N. Pleasant St., Amherst, MA 01003, USA\\
$^{14}$Department of Astronomy and Steward Observatory, University of Arizona, Tucson, AZ 85721, USA}

\date{Accepted XXX. Received YYY; in original form ZZZ}

\pubyear{2019}

\begin{document}

\def \aj {AJ}
\def \mnras {MNRAS}
\def \pasp {PASP}
\def \apj {ApJ}
\def \apjs {ApJS}
\def \apjl {ApJL}
\def \aap {A\&A}
\def \nat {Nature}
\def \araa {ARAA}
\def \iaucirc {IAUC}
\def \aaps {A\&A Suppl.}
\def \qjras {QJRAS}
\def \na {New Astronomy}
\def \aapr {A\&ARv}

\label{firstpage}
\pagerange{\pageref{firstpage}--\pageref{lastpage}}

\maketitle

\begin{abstract}
A significant fraction of high redshift star-forming disc galaxies are known to host giant clumps, whose nature and role in galaxy evolution are yet to be understood. In this work we first present a new method based on neural networks to detect clumps in galaxy images. We use this method to detect clumps in the rest-frame optical and UV images of a complete sample of $\sim1500$ star forming galaxies at $1<z<3$ in the CANDELS survey as well as in images from the VELA zoom-in cosmological simulations. We show that observational effects have a dramatic impact on the derived clump properties leading to an overestimation of the clump mass up to a factor of 10, which highlights the importance of fair comparisons between observations and simulations and the limitations of current HST data to study the resolved structure of distant galaxies. After correcting for these effects with a mixture density network, we estimate that the clump stellar mass function follows a power-law down to the completeness limit ($10^{7}$ solar masses) with the majority of the clumps being less massive than $10^9$ solar masses. This is in better agreement with recent gravitational lensing based measurements. 
The simulations explored in this work overall reproduce the shape of the observed clump stellar mass function and clumpy fractions when confronted under the same conditions, although they tend to lie in the lower limit of the confidence intervals of the observations. This agreement suggests that most of the observed clumps are formed in-situ. 

\end{abstract}




\section{Introduction}

A prominent feature of distant star-forming galaxies is the frequent presence of high-surface brightness concentrations, or "clumps" embedded in a more uniform light distribution.
Their origin and evolution are important to understand many aspects of
galaxy formation, e.g., gas accretion, feedback, bulge formation, and
supermassive black hole formation. Clumps are mostly identified in rest-frame UV
and emission-line (CO or H$\alpha$) images of galaxies over a wide redshift
range
\citep[e.g.,][]{giavalisco96,conselice04,elmegreen05,elmegreen07,elmegreen09a,ravindranath06,fs11b,ycguo12clump,ycguo15fclumpy,ycguo18clump,wuyts12,murata14,shibuya16, soto17,2019MNRAS.489.2792Z,larson20,zick20}.

Clumps are a few orders of magnitude more massive than star-forming regions in
nearby galaxies
\citep[e.g.,][]{elmegreen07,genzel08,genzel11,fs11b,ycguo12clump,ycguo18clump,dz17,soto17,2019MNRAS.489.2792Z}.
Their specific star formation rates are higher than their surrounding areas by
a factor of several
\citep[e.g.,][]{genzel08,genzel11,ycguo12clump,ycguo18clump,wuyts12,wuyts13,hemmati14,mieda16,fisher17b}.
Their size is still uncertain and under intense debate. Some studies with
unlensed galaxies found the size to be $\sim$1 kpc
\citep[e.g.,][]{elmegreen07,fs11b}, but other studies with gravitationally 
lensed galaxies found 
smaller sizes of a few hundred or even tens of pc
\citep[e.g.,][]{livermore12,zick20}. \citet{fisher17b} argue that observation
resolution and sensitivity as well as clustering of clumps result in an
overestimate of clump sizes. However \citet{cosens18} conclude that the size of clumps
scales with their H$\alpha$ luminosity and there is no difference
between lensed and unlensed data.

The formation of clumps is related to the mass assembly processes of galaxies
at high redshifts. Clumps are thought to form through gravitational instability
in gas-rich turbulent disks, sometimes triggered by external perturbations
\citep[e.g.,][]{noguchi99,immeli04a,immeli04b,bournaud07,bournaud09a,elmegreen08,2009ApJ...703..785D,ceverino10,ceverino12,dekel14,inoue16}.
This scenario of violent disk instability is supported by some observations,
especially for massive clumpy galaxies
\citep[e.g.,][]{elmegreen07,bournaud08,genzel08,genzel11,ycguo12clump,ycguo15fclumpy,hg16,mieda16,fisher17b}.
Clumps, however, can also have an ex-situ origin, that is, as minor mergers
\citep[e.g.,][]{hopkins13}. Evidence of this scenario is also present in the
literature 
\citep[e.g.,][]{puech09,puech10,wuyts14,straughn15,ribeiro16}.
\citet{ycguo15fclumpy} suggest that the formation mechanisms depend on the mass
of clumpy galaxies: clumps in massive galaxies are formed through violent disk
instability, while those in lower mass galaxies are formed through minor
mergers. If that is the case, one would expect that the clump stellar mass function of clumps changes as a function of galaxy mass. 

The evolution of clumps determines their importance on the broad picture of
galaxy formation and evolution: whether clumps are building blocks of galactic
bulges or just a transient phenomenon. Basically, two scenarios have been proposed.
Some models and simulations suggest that clumps can live long enough to migrate
towards the centers of their host galaxies, eventually merging into the
progenitors of today's bulges
\citep[e.g.,][]{bournaud07,bournaud14a,elmegreen08,ceverino10,ceverino12,mandelker14,mandelker17}.
This scenario is supported by observed radial color gradients of clumps
\citep[e.g.,][]{fs11b,ycguo12clump,ycguo18clump,shibuya16,soto17}. On the other
hand, some other models and simulations suggest that clumps are self-disrupted
by their powerful starburst-induced outflows on a timescale of a few tens of
Myr
\citep[e.g.,][]{murray10,genel12,hopkins12clump,hopkins14fire,buck16,oklopcic17}. 
The two scenarios distinguish between different feedback models and strength,
i.e., whether the feedback of star formation is strong enough to destroy clumps
in a short timescale \citep[e.g.,][]{moody14}. 

An important parameter to understand clump formation and evolution is their
stellar mass. For example, as discussed in \citet{dz18}, if clumps were formed
in-situ through gravitational instability, the clump mass function would have a
slope of $-2$ (in logarithmic space) at its massive end. The contribution of
clump mass to the total mass of their galaxies is also important. Clumps are
found to contribute $\sim 10\%$ of the total UV luminosity or SFR of their
galaxies
\citep[e.g.,][]{fs11b,ycguo12clump,ycguo18clump,wuyts12,mieda16,rujopakarn19}, but
their contribution to the total stellar mass is still unknown, although it is
believed to be small. Clump mass can also be used as a diagnostic of clump
evolution. For example, whether the mass of clumps shows any radial variation
can be used to test the inward migration scenario \citep{bournaud14a}.
Currently, however, robust measurements of stellar mass for a large sample of
clumps are still insufficient in the literature. A challenge of obtaining a
complete census of clump mass is to detect clumps in a wavelength that traces
stellar mass more than star formation. Rest-frame optical provides a good
choice and is accessible by current observing facilities for clumps over a wide
range of redshifts.

The capabilities of current observations pose another significant challenge to
properly measure physical parameters of clumps.
Even with the spatial resolution and sensitivity of {\it HST}, clumps at high
redshifts can only be marginally resolved or may even be unresolved. 
Therefore, a clump observed by {\it HST}, let alone by ground-based
telescopes without Adaptive Optics correction, could be either a single object or the blend of a few nearby smaller
clumps. Some authors argued that
with a limited spatial resolution of $\sim$1 kpc (which is equivalent to {\it
HST}'s resolution for galaxies at $z\sim1$), many of the giant clumps (with clump stellar mass
\mstar$\gtrsim10^8$\msun) identified in observations are actually the result of
blending of smaller structures or clustering of clumps
\citep[e.g.,][]{tamburello15,dz17,benincasa19,2020MNRAS.494.1263M}. \citet{dz17} also discuss
that the sensitivity threshold used for the clump selection strongly biases
against clumps at the low-mass end. 
Similarly, \citet{fisher17b} even argue that due to the effects of clump
clustering, with a 1 kpc resolution, the SFR surface density of clumps would be
overestimated by up to a factor of 20. Some authors
\citep[e.g.,][]{buck16,oklopcic17} argue that disk stars contaminate clump age
measurement, resulting in an artificially old clump age. 

In order understand the nature of clumps, a {\it direct} comparison between
observations and models or simulations is needed. Observational effects
related to limited image resolution and sensitivity, such as PSF effects and
realistic (and correlated) noises, should be applied to simulations for {\it
each} specific observation. These forward-modeled simulations have been used in
previous studies for integrated galaxies( e.g. ~\citealp{2018ApJ...858..114H}).
For small, faint sub-structures of galaxies, such as clumps, using these
forward-modeled simulations is particularly important, because these
sub-structures are severely affected by PSF and noise in observations.

In this work we perform several steps towards better
quantifying the stellar mass distribution of clumps in distant galaxies. First, we develop a novel method for the detection of clumps based on deep
neural networks. The main advantage of this approach is that it is
significantly faster and more sensitive than previous methods and therefore can
be easily applied to large samples of galaxies in different wavelengths such
as the ones that will be soon available (e.g. from JWST, Euclid, and WFIRST). We then apply
our method to the five fields of the CANDELS survey (Grogin, et al. 2011;
Koekemoer, et al. 2011) in up to seven different detection bands, increasing the
sample of clumps by a factor of 3 compared to previous works on the same survey
(Guo et al. 2015).  More importantly, we statistically quantify for the first
time the contribution of clumps to the galaxy stellar mass in a complete sample.  We will use  state-of-the art cosmological simulations
forward modeled in the CANDELS observational plane, namely the VELA hydrodynamic zoom-in simulations
\citep{2014MNRAS.442.1545C}, to carry out a
direct comparison between observations and simulations of clumps (see also Ginzburg et al. 2020 in prep.). The
forward-modeled VELA simulations also help to improve our clump study in two
other ways: (1) evaluating the completeness of our sample selection and (2)
understanding and correcting systematic and random errors of physical parameter
measurements.

The analysis has two main parts: in sections~\ref{sec:Method} and~\ref{sec:Comparison} we present the neural network based clump detector and quantify its accuracy, and in section~\ref{sec:CANDELS} we apply the new method to a CANDELS subsample of star-forming galaxies and a sample of simulated galaxies, quantify the stellar mases of clumps and discuss clump abundances, clumpy fractions and contribution of clumps to the total galaxy mass in both simulations and observations. Section~\ref{sec:summary} is a summary and discussion.

Throughout the paper, we adopt a flat ${\rm \Lambda CDM}$ cosmology with
$\Omega_m=0.3$, $\Omega_{\Lambda}=0.7$ and use the Hubble constant in terms of
$h\equiv H_0/100 {\rm km~s^{-1}~Mpc^{-1}} = 0.70$. All magnitudes in the paper
are in AB scale \citep{oke74} unless otherwise noted. 

\section{Method for clump detections}
\label{sec:Method}

This section first describes the main method developed in this work to detect clumps.

\subsection{Model Architecture}
\label{sec:Architecture}

The main purpose of this work is to identify the positions of all off-center clumps belonging to a galaxy in order to characterize their properties. In terms of image processing, the problem can be described as an instance segmentation problem, if one assumes that all clumps have the same geometrical properties. This assumption might not entirely be true. \cite{2019MNRAS.489.2792Z} identified for example two populations of clumps with different size distributions. However at first order, clumps are small unresolved sources when compared to the host galaxy.

Therefore we decided to use state-of-the-art segmentation networks based on encoder-decoder Convolutional Neural Networks (CNNs). We assume that the detection of the galaxy is done in a previous step. The input for the network is hence a $128\times128$ pixels stamp with a galaxy at its center. The desired output is another image with all pixels set to zero except for the pixels belonging to clumps which are set to one.  The required configuration is therefore a fully convolutional network whose inputs and outputs are images.  We choose here a U-net type architecture~\citep{2015arXiv150504597R} that has been proven to be very efficient for image segmentation, especially in the bio-medical field but also in astronomy (e.g.~\citealp{2020MNRAS.491.2481B,2019arXiv190611248H})

\begin{figure}
	\centering
	\includegraphics[width=0.45\textwidth]{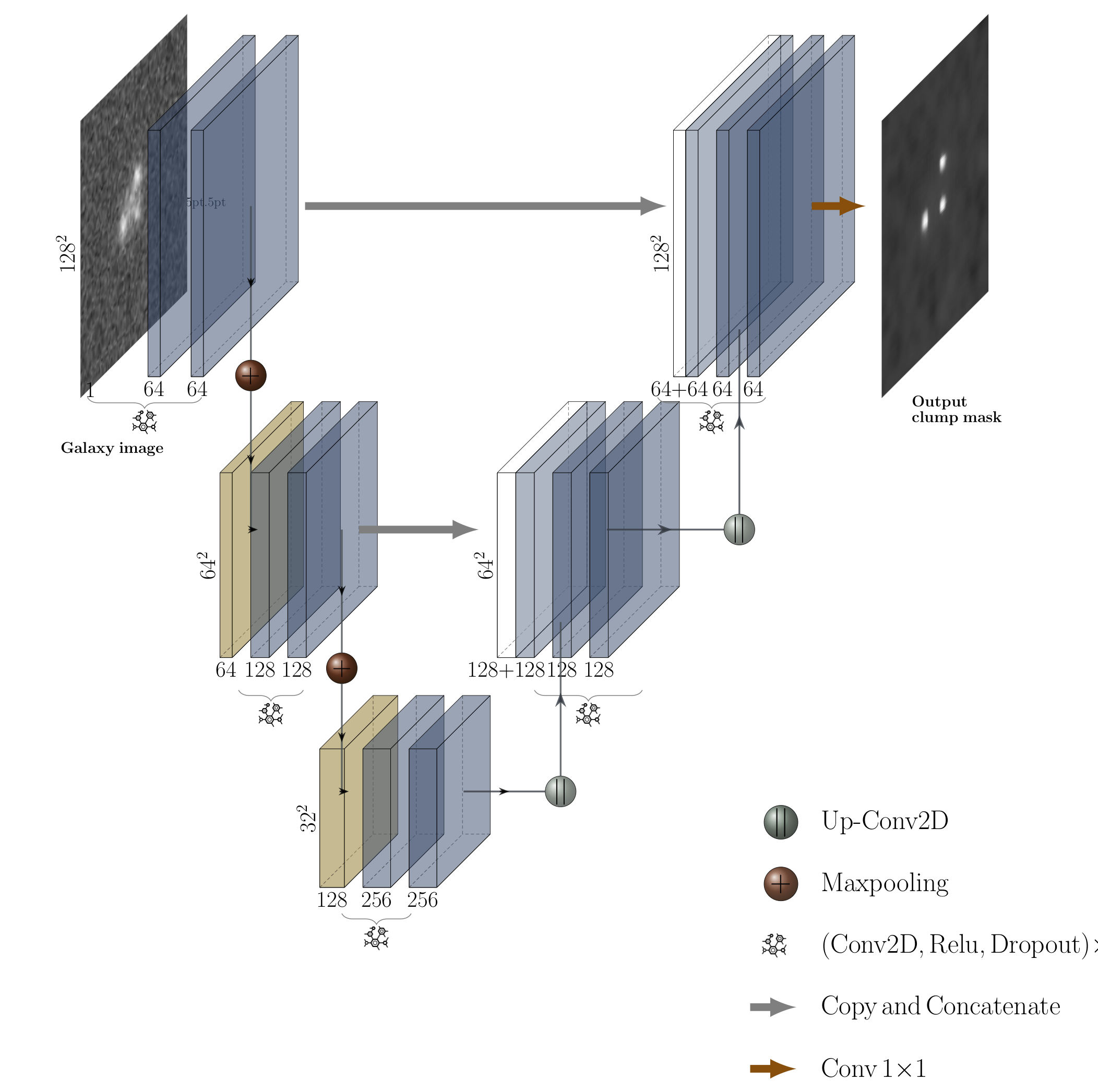}
    \caption{Schematic representation of the network model used to detect clumps in this work. We use a supervised U-net type of architecture~\citep{2015arXiv150504597R}. The input stamp is contracted through successive convolutions and then a new image containing the clump segmentation mask is created through up sampling operations.}
    \label{fig:unet}
\end{figure}

 The encoder part of the network is a standard CNN which takes the image and compresses it to a lower dimension latent space through consecutive convolutions and pooling operations. The decoder part reconstructs an image from the latent space through up-sampling operations. The particularity of the U-net is that there are skipped connections linking the encoder and decoder branches. This has been shown to help in the reconstruction phase and therefore improves the segmentation quality. We use a ReLu activation function in all layers except the last one which is left without activation. We decided to leave the last layer without activation (instead of a Softmax) because it improves the post processing of the detection image to build a clump catalog (see next section). A detailed representation of the specific architecture used in this work is shown in figure~\ref{fig:unet}.  

\subsection{Simulated training set}
\label{sec:Training Set}

The neural network needs to be trained to detect clumps following a supervised approach. To this purpose, a sample of galaxies where the positions of clumps are known is needed.  Although there exist catalogs of clumps in observed galaxies from the CANDELS survey, for instance (e.g.~\citealp{ycguo15fclumpy}, hereafter Guo15), the samples are typically too small for proper training. Furthermore training in real data would propagate any biases existing in the original technique to our clump detector. 

We decided then to train the network with simulated galaxies only. We first generate galaxies using a single Sersic analytic profile~\citep{1968adga.book.....S} using the code GalSim~\citep{2015A&C....10..121R}. We allow parameters (e.g total flux, effective radius $R_e$, Sersic index $n$, axis-ratios $b/a$) to vary following uniform random distributions to cover all the observed range from star-forming galaxies in the CANDELS survey (e.g.~\citealp{2012ApJS..203...24V, 2018MNRAS.478.5410D}). The exact selection of galaxies is described in section~\ref{sec:CANDELS}.

More precisely, we generate galaxies randomly within the following limits:
22 < $m_{AB}$ < 26 (optical bands) and 20 < $m_{AB}$ < 25 (NIR bands); 
$ 0.3 < n  < 2.0$;
$0.6 < \log_{10} R_e < 1.1$;
$0.6 < b/a < 1.0$
where n is the galaxy Sersic index, $R_e$ is the semi-major effective radius measured in kpc, and b/a is the axis ratio.  To model clumps, we add one or more small sources to each galaxy, generated using GalSim as $n = 1$ galaxies with effective radii between 1 and 2 pixels. The number, fluxes, sizes, and positions of clumps are selected randomly, but according to a variety of rules.  Generally, we add between 1 and 4 clumps to each galaxy, except for small galaxies ($\log_{10} R_e < 0.8$) to which we add only 1 or 2 clumps to avoid crowding and to reduce obscuring of the clumps by the galaxy center.  For a given galaxy, we limit the combined flux of all clumps to be at most 45\% of the total flux (galaxy flux plus clump fluxes).  We allow the flux of individual clumps to range from a minimum of 6\% to a maximum of 30\% of total flux.  Finally, we choose clump positions randomly within the annulus 0.5 $R_e$ to 2.0 $R_e$.  For every added clump, we keep in a binary mask the position in the image where it was added. Finally, the image is convolved with a real PSF from the CANDELS survey and noise is added. To model the noise, real empty regions from the CANDELS fields are used in order to include existing spatial correlations.

The procedure is the result of an iterative manual search to come up with a configuration that produces the best results on real data. In particular, we realized that simulating too faint clumps or galaxies without clumps reduces the performance, even if fainter clumps and clump-free galaxies do exist in the observations. In any case, the above procedure should not be considered as unique, and other simulated samples can also produce accurate results. 

 The above procedure is repeated for seven different filters, four in HST/ACS optical and three HST/WFC3 infrared: $F435W$ (b band), $F606W$ (v band), $F775W/F814W$ (i band), $F850LP$ (z band), $F105W$ (Y band), $F125W$ (J band) and $F160W$ (H band) using the corresponding noises and PSFs. 
 
\begin{figure}
\centering
\subcaptionbox{Simulated $F606W$ image}{\includegraphics[width=0.20\textwidth]{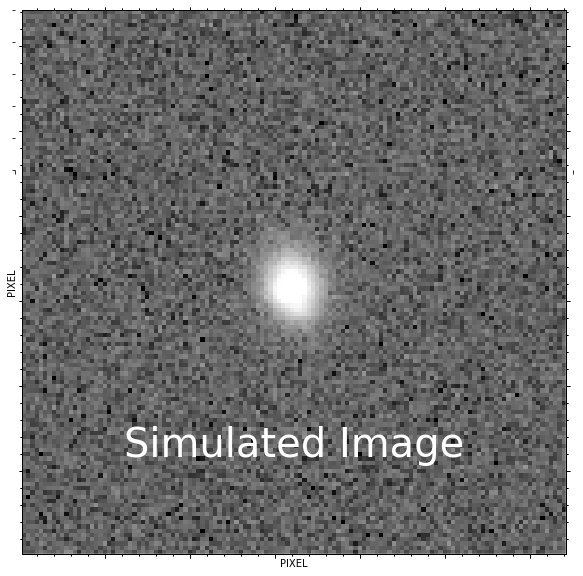}}%
\hfill
\subcaptionbox{Ground truth mask of clumps}{\includegraphics[width=0.20\textwidth]{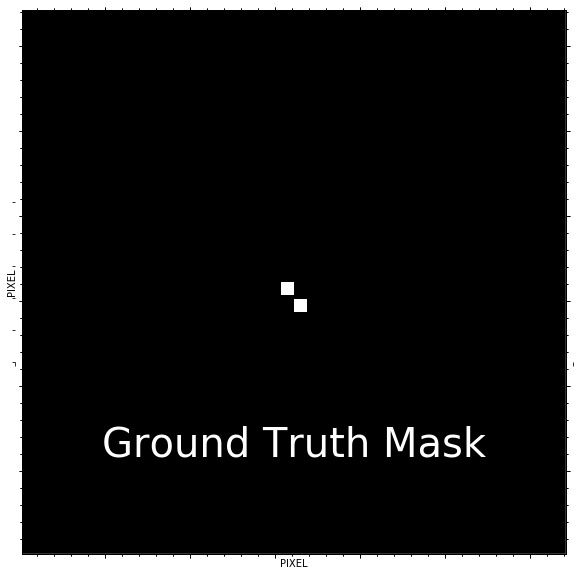}}%
\hfill
\subcaptionbox{Unet output}{\includegraphics[width=0.20\textwidth]{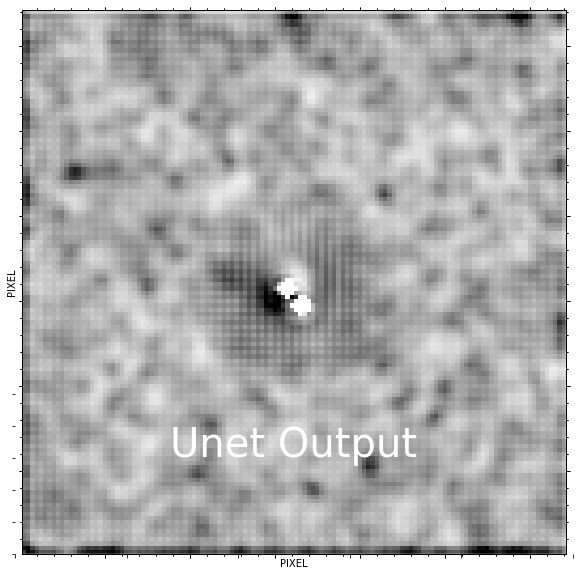}}%
\hfill
\subcaptionbox{SExtractor detection on Unet output}{\includegraphics[width=0.20\textwidth]{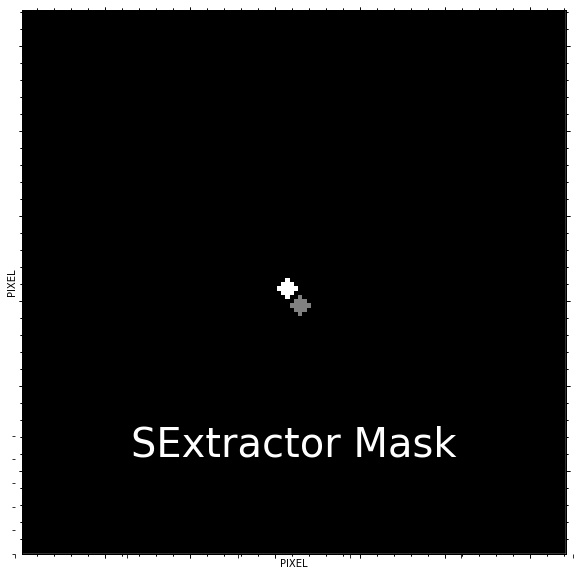}}%
\caption{Example of clump detection performance on simulated GalSim test dataset. The top left and right panels show the raw GalSim test image and true clump mask, respectively.  The bottom left and right panels show the model output (predicted clumps) and SExtractor detection, respectively.  This example galaxy contains two true clumps, both of which are identified with high confidence and precision by the model even if they are barely visible in the original image.}
\label{fig:galsim_test_example}
\end{figure}

The top row of figure~\ref{fig:galsim_test_example} shows an example of a simulated galaxy in the v band with two clumps together with the final mask. Our goal is thus to train the network to predict the mask (top right panel in the figure) given the image of the galaxy (top left panel). Notice that the clumps are barely visible in the image.

\subsection{Training and detection procedure}
With the training set, we proceed to train the U-net. We train seven different models for the seven different detection bands independently ($F435W$ (b), $F606W$ (v), $F775W$ (i) ($F814W$ for COSMOS, EGS and UDS where $F775W$ is not available), $F850LP$ (z, for GOODS-N and GOODS-S), $F105W$ (Y), $F125W$ (J) and $F160W$ (H)). In all seven cases, we use a sample of 98,000 galaxies for training while 2,000 galaxies are kept for testing and evaluating the accuracy (see section~\ref{sec:Performance}). The network is trained following a standard procedure. We use a weighted binary cross-entropy loss to account for imbalance in the training set (the number of pixels belonging to clumps is much smaller than the number of pixels outside clumps) and an Adam optimizer. The initial learning rate is set to $10^{-5}$ and decreased two times by a factor of ten to ease convergence.

As described in the previous section, the output of the network is simply an image with values roughly ranging from 0 to 1. Zero means that no clump is detected at that position, and one means that a clump is detected with high confidence. The output of the U-net needs to be post-processed to decide whether there is a clump or not and also to determine the position of the clump. 

To do so, we use SExtractor~\citep{1996AAS..117..393B} on the detection image as a tool to only identify the positions of the detected clumps in each stamp. We emphasize that SExtractor is only used here to identify the centroids of the detected clumps by the neural network. Any other labeling method could have been used at this stage. This allows us to build a catalog containing the positions of all clumps in the stamp. The bottom panels of figure~\ref{fig:galsim_test_example} illustrate an example of the prediction steps for one galaxy with two simulated clumps.

\subsection{Accuracy of clump detections and flux measurements}
\label{sec:Performance}

We then quantify the accuracy of our clump detector with a set of 2000 galaxies not used for training. 

\begin{figure}
    \centering
    \vspace{-10 pt}
    \subcaptionbox{}{\includegraphics[width=0.45\textwidth]{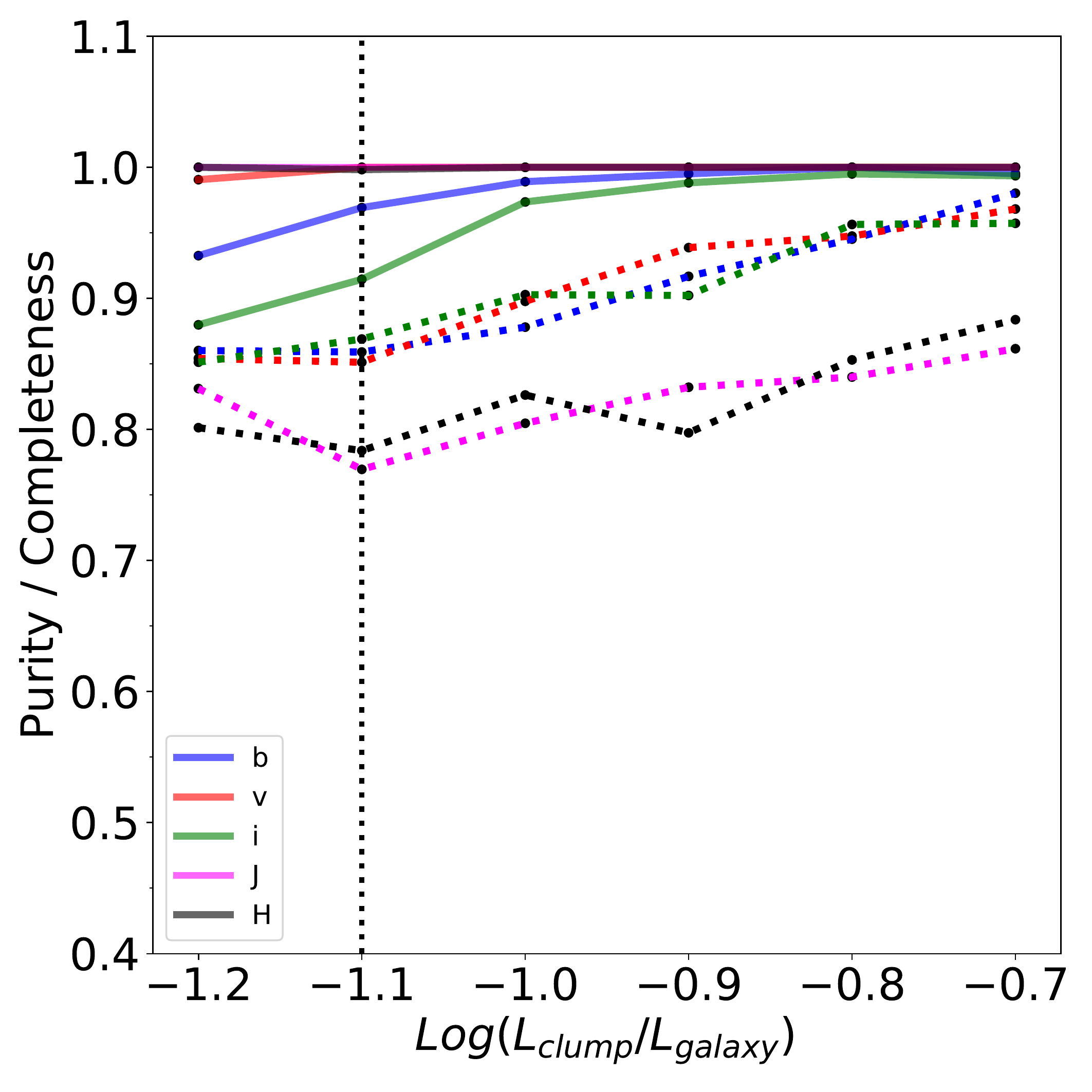}}%
     \qquad
    \subcaptionbox{}{\includegraphics[width=0.45\textwidth]{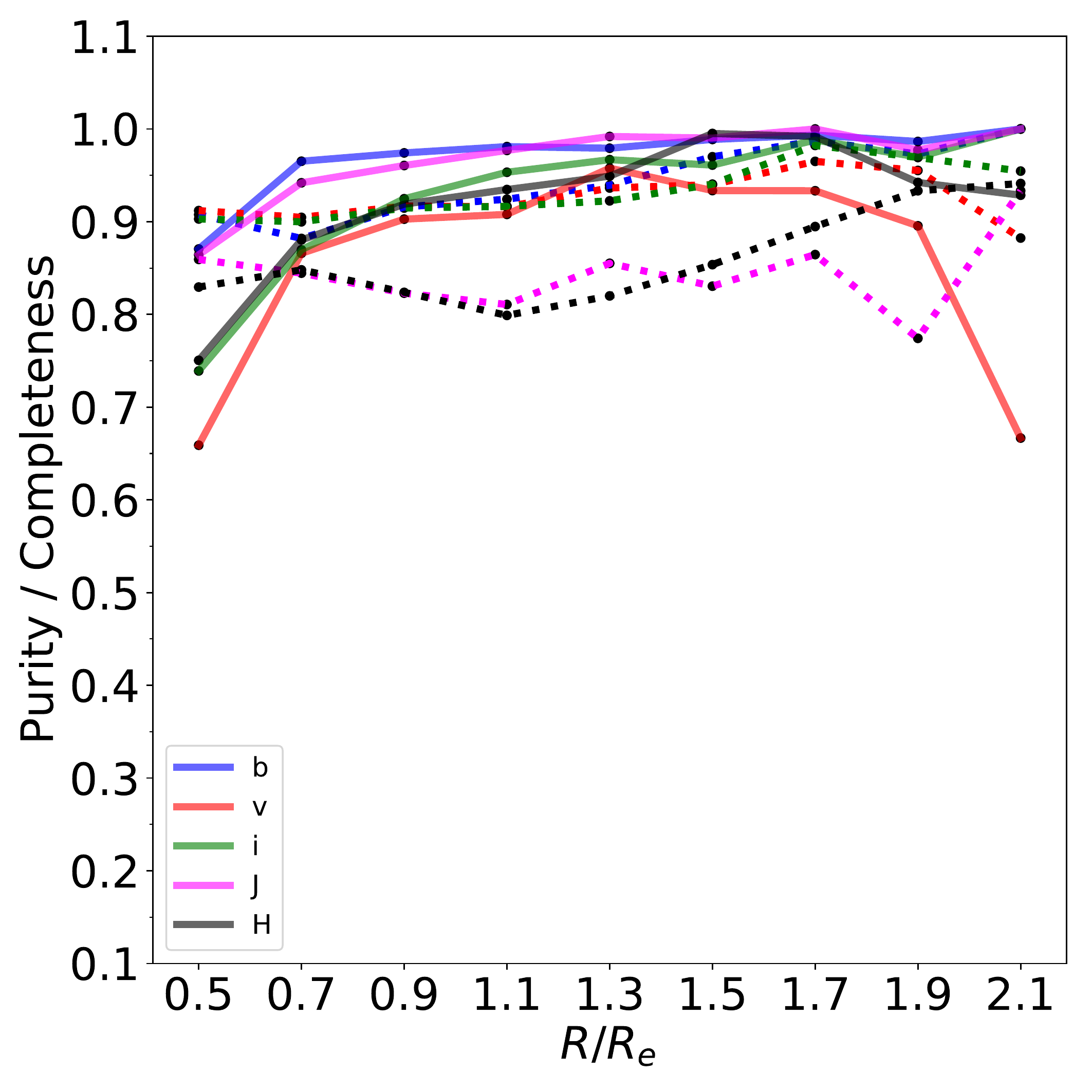}}%
     \qquad
      \vspace{-10 pt}
    \caption{Completeness (dashed lines) and purity (solid lines) of clump detections measured on the test simulated dataset. The different panels show the results as a function of (a) relative clump luminosity and (b) clump radial position within the galaxy.  The different colors correspond to different detection bands as labeled. Completeness describes the fraction of true clumps the model also finds, while Purity represents the fraction of predicted clumps that are also true simulated clumps. The dashed vertical line in the top panel indicates a value of relative luminosity of 8\%. Overall purity and completeness remain above $80-90\%$ depending on the detection band.  }%
    \label{fig:galsim_sex_PC_comparison}%
\end{figure}

Figure~\ref{fig:galsim_sex_PC_comparison} shows first the completeness and purity of the detection as a function of the relative clump luminosity (top) and the clump distance to the galaxy center (bottom) for five photometric bands. Completeness is defined as the fraction of simulated clumps that are detected. Purity measures the fraction of true clumps among all the detected clumps by the U-net. Generally speaking, the figure shows that completeness is overall above $90\%$ in the observed optical bands and slightly lower ($\sim80\%$) in the infrared bands. This is probably a consequence of the difference in spatial resolution between the two cameras (ACS and WFC3) . Purity is generally above 90-95\%. This value should be considered as a best case, since the host galaxies are pure analytic profiles so it is very unlikely that there are artifacts detected as clumps except at the very faint end in which noise fluctuations can be sometimes detected as clumps. 

The top panel of figure~\ref{fig:galsim_sex_PC_comparison} shows the completeness and purity as a function of relative luminosity which has been used in previous works (e.g. Guo15) to select giant clumps. We define the relative clump luminosity as the ratio between the measured clump luminosity and the galaxy luminosity, i.e. $L_{clump}/L_{galaxy}$). Overall, completeness increases for brighter clumps. However, even for the faintest clumps ($L_{clump}/L_{galaxy}\sim6\%$), completeness remains above $90\%$ ($80\%$) for the ACS (WFC3) clumps. 

The bottom panel of figure~\ref{fig:galsim_sex_PC_comparison} shows that the accuracy of detections does not significantly depend on the position of the clump within the galaxy except perhaps at the very central parts where the central bulges might be sometimes misidentified as clumps. We also observe a decrease of purity at large galactocentric distances for the v band detections. Since this is only appreciated in one filter it is likely to be a statistical fluctuation.

Besides detection, we require an accurate measurement of fluxes given that one of the main purposes of this work is to derive clump stellar masses via SED fitting. Clump fluxes are determined using a simple aperture photometry on each position with a 4 pixel aperture radius as done in Guo15. The measured flux is background subtracted by removing the best fit Sersic model from the host galaxy. In the case of simulations, since the models are pure analytic profiles with known parameters, the correction is straightforward and the flux coming from the galaxy disc is decently removed. This is not the case for real observations, which will be discussed in section~\ref{sec:CANDELS}. We find that, overall, the clump fluxes are recovered accurately with a scatter of $\sim0.1$ dex, with a small increase in scatter for faint clumps. When several clumps are present, the flux of each clump tends to be overestimated because flux from the neighboring clumps is included. It is worth noticing that the Galsim simulations do not include a realistic distribution of clumps. If anything, the number of bright clumps is overestimated as compared to reality (see section~\ref{sec:Training Set}). It is therefore reasonable to expect that the effect of blending of two bright clumps in the observations is small.

\section{Performance of clump detector on real observations}
\label{sec:Comparison}

The main goal of this work is to apply the clump detector trained on simplistic analytic simulations to real observed galaxies. Training on simulations is particularly dangerous in machine learning where a perfect match between the training and application datasets is assumed. It is therefore critical to properly assess the performance of our clump detector on real observations before moving to any scientific analysis. 

We adopt two different approaches to quantify the reliability of the detections. We first quantify how fake clumps inserted on real observed galaxies are detected. We then compare our results with Guo15 detections on the same observed galaxies.

\subsection{Detection of fake clumps on real galaxy images}
\label{sec:Detection Probability}

Following the procedure presented in Guo15, we add fake clumps into a sample of $\sim1500$ real CANDELS galaxies and measure how well we recover them using the developed method. The procedure to add clumps is fully described in Guo15 and we refer the reader to the aforementioned work for more details. This test is intended to quantify how our U-net based detector behaves when confronted with realistic galaxy morphologies while keeping information on the ground truth. For the sake of clarity, we only show the results of this exercise in the $F606W$ filter (i.e., rest-frame UV for $z>1$ CANDELS galaxies) but similar results are observed in the other bands. We emphasize that such a dataset could not have been used for training since it contains a mix of true clumps and fake clumps, making it difficult for the network to understand why some clumps need to be ignored. Since the same galaxies were analyzed also by Guo15, we can also perform a direct comparison between the two methods on the same sample.

Table~\ref{tbl:fake_clumps} shows the statistics of the detections for the Guo15 method and our approach. We show the statistics for all clumps first and then for only the brightest clumps. As it can be seen, our method recovers almost all clumps detected by Guo15 on the same galaxies (only $\sim4\%$ of clumps are detected by Guo15 and not by the deep learning approach). This is an indication that even if trained on simplistic simulations, the neural network model behaves as expected. It detects indeed the same clumps as a more traditional and well tested approach would do. Although this might appear surprising, a possible explanation is that this specific task does not depend on the actual galaxy profile. The network essentially learned to detect unresolved off-center sources independently of the galaxy shape.  In addition, the neural network is able to detect a significant number of clumps that were undetected by the Guo15 algorithm. This is especially true when all clumps are considered, irrespective of the relative luminosity. The fraction of clumps detected by the U-net but not detected by Guo15 reaches $\sim20\%$. It suggests that the clump detector presented in this work is more sensitive than the previous algorithms. But when only bright clumps are considered (rightmost column of table~\ref{tbl:fake_clumps}), the fraction of clumps undetected by Guo15 drops to $\sim10\%$. 

\begin{figure}
	\centering
	\includegraphics[width=0.45\textwidth]{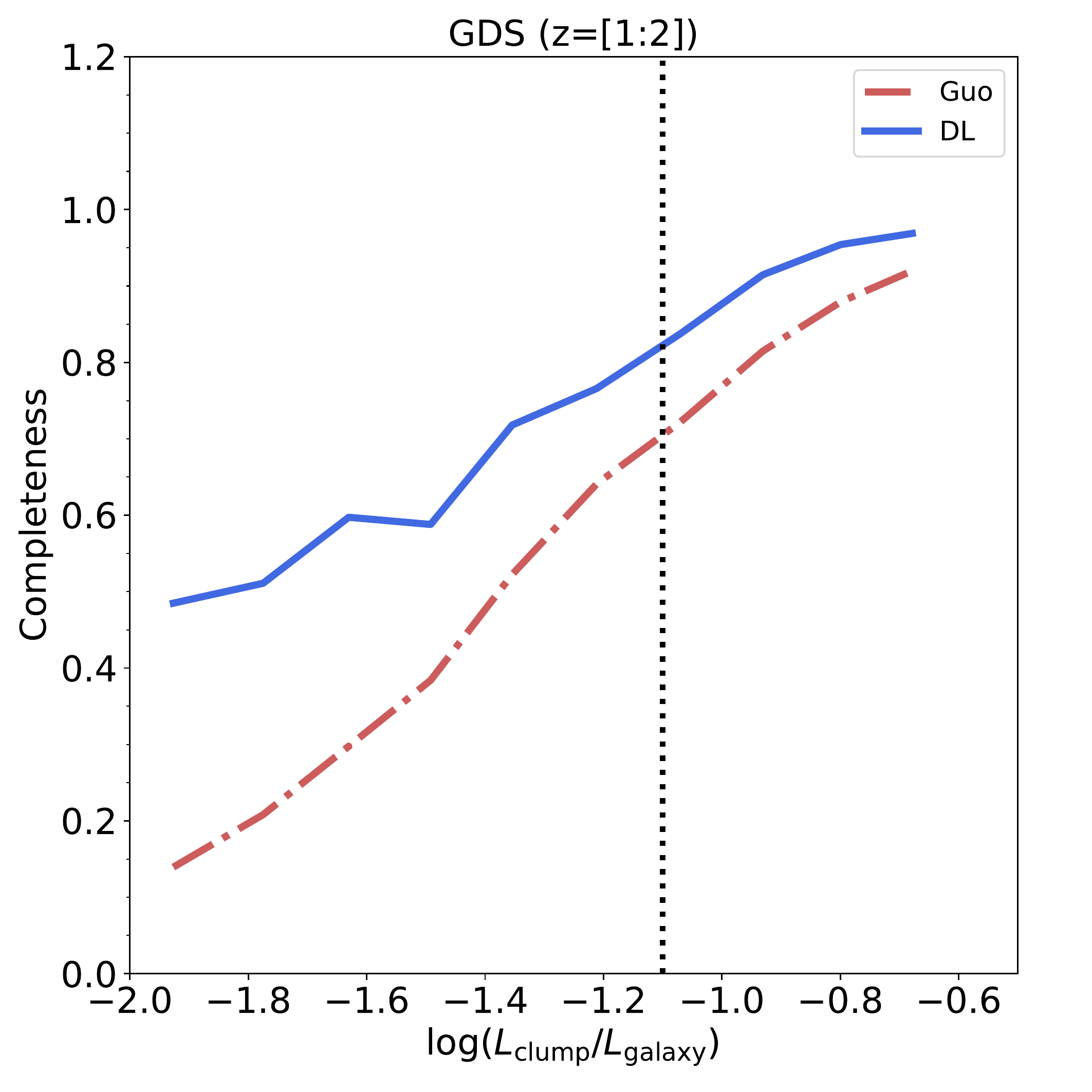}
    \caption{Detection completeness of fake clumps inserted on real CANDELS galaxies for the deep learning based detector (blue solid line) and for the Guo15 detector (red dotted-dashed line) in the v band ($F606W$) as a function of the relative clump luminosity. The U-net clump detector trained on simple simulations achieves an overall higher completeness than the Guo15 detector on the same galaxies. The dotted vertical line indicates the clump relative luminosity of $8\%$ used in Guo15 to select giant UV clumps.}
    \label{fig:guo_sim_clump_prob}
\end{figure}

We further quantify the difference in completeness between the two methods in figure~\ref{fig:guo_sim_clump_prob} which shows the completeness of the detections as a function of the relative clump luminosity. We clearly observe that the DL clump detector has a higher completeness over all the range of clump relative luminosities explored. The difference is particularly pronounced at low relative luminosities which confirms the higher sensitivity of our clump detector.  It is particularly interesting that faint clumps are recovered with fairly high completeness even if they were not included in the original simulated training set. In this particular application, using a cleaner training set with only bright clumps, helped the network to generalize. We note also that the values of completeness reported in figure~\ref{fig:guo_sim_clump_prob} differ slightly from the values reported in figure~\ref{fig:galsim_sex_PC_comparison} at similar relative luminosities. This is because the samples are different. One is made only of analytic profiles while the other is composed of real galaxies with fake clumps.

\begin{table}
\begin{center}
\begin{tabular}{ccc} 
\hline
 & All & $\frac{L_{clump}}{L_{gal}}>0.08$ \\
 \hline
detected in both Guo15 and DL & 51\% &82\%  \\ 
detected in Guo15, not detected in  DL &4\% & 2\%  \\ 
 not detected in Guo15, detected in DL&22\% & 10\%  \\ 
not detected in Guo15 nor in DL &23\% & 5\%  \\ 
\end{tabular}
\end{center}
\caption{Comparison of the detection statistics of fake clumps inserted in real CANDELS galaxies (see text for details) using the Guo15 detector and our deep learning based method. The first column shows the results for all clumps and the second column only for bright clumps ($L_{clump}/L_{galaxy}>8\%$). The first row shows the fraction of clumps detected by both algorithms. The second row indicates the fraction detected by Guo15 and not by the U-net, the third row clumps detected by the network and not by Guo15, and finally the last row shows clumps which are not detected by either of the two methods. Our DL method is overall more sensitive than the Guo15 approach. }
\label{tbl:fake_clumps}
\end{table}

\subsection{Comparison with Guo15}
\label{sec:P&C}
We now perform a direct comparison with the existing real clump detections on galaxies in common with the Guo15 sample. This comparison is more complicated since the ground truth is not known. Therefore, this section is more focussed on exploring the abundances and properties of the clumps detected with both methods to quantify potential differences and biases. We recall that Guo15 only detected clumps in the UV rest-frame bands while with DL we run all bands for all galaxies. The comparison is performed here only for UV rest-frame detections.

\begin{figure}
	\centering
	\includegraphics[width=0.53\textwidth]{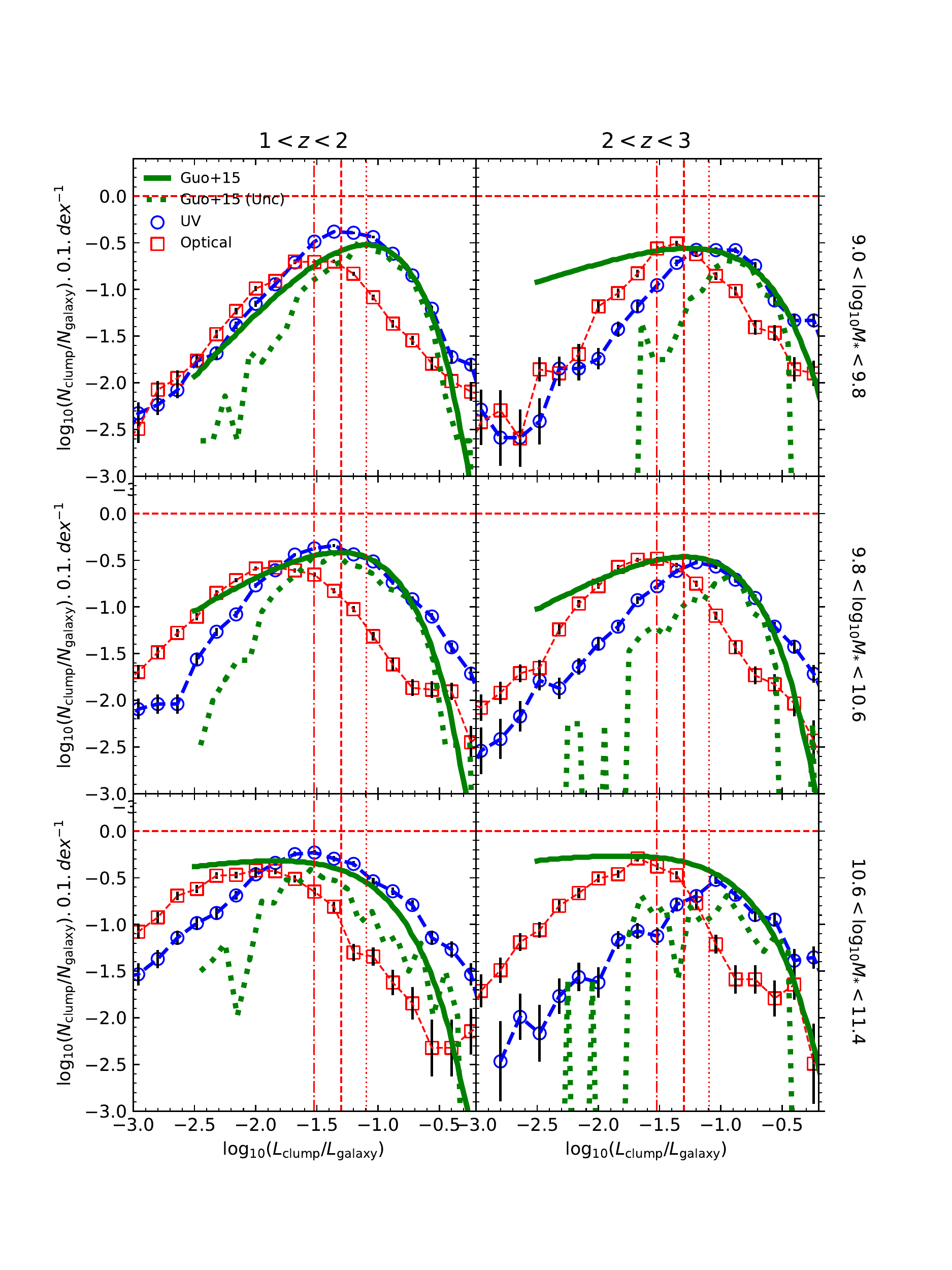}
    \caption{Relative optical and UV rest-frame Clump Luminosity Functions in CANDELS. As detailed in the text, for the optical (UV) rest-frame detections we use filters F125W (F606W) for $z<2$ and F160W (F775W) for $z>2$. The number of clumps in each panel is normalized by the total amount of galaxies in that bin so that it gives an indication of the average number of clumps per galaxy. Each column shows a different redshift bin and each row corresponds to a different galaxy stellar mass bin. The red squares are optical-rest-frame selected clumps while blue circles show UV rest-frame selected clumps. Error bars indicate Poissonian uncertainties. The green dotted lines indicate the measurements on GOODS-S performed by Guo15 in the UV rest-frame without incompleteness correction. The solid green line show the results when a correction for incompleteness is applied. The dashed-dotted, dashed and dotted vertical lines indicate a relative clump luminosity of $3\%$, $5\%$ and $8\%$ respectively. The dashed horizontal line is shown for reference and indicates a value of one clump / galaxy. Our measurements in the UV rest-frame agree reasonably well with the incompleteness corrected lines from Guo15 down to a relative luminosity of $3\%$, except for some cases at $z>2$ in which our detections lie slightly below Guo15 corrected detections. This confirms the better sensitivity of the deep learning clump detector. }
    \label{fig:clump_LF}
\end{figure}

We first compare in figure~\ref{fig:clump_LF} the obtained relative clump Luminosity Functions (cLFs)  for different galaxy stellar mass and redshift bins to the one of Guo15 for UV selected clumps ($F606W$ for $1<z<2$ and $F775W$ for $2<z<3$). Following Guo15, cLFs are derived by counting the number of clumps in a given relative luminosity bin and dividing by the total number of galaxies in the bin. Guo15 uses only GOODS-S and UDS data. We use here all five CANDELS fields. We see a relatively good agreement between the Guo15 results and ours. Interestingly, the figure also confirms that the deep learning based detector presented here is more complete at low clump relative luminosities. The incompleteness corrected lines from Guo15 indeed better track our raw measurements down to a clump relative luminosity of $\sim3\%$. However, the uncorrected measurements begin to deviate from the DL sample at a higher relative luminosity threshold. The figure also shows for comparison the cLF of optically detected clumps ($F125W$ for $1<z<2$ and $F160$ for $2<z<3$) which to our knowledge is the first time it is shown. The figure clearly shows that UV rest-frame clumps tend to concentrate a larger fraction of the total galaxy luminosity than optical rest-frame clumps at all redshifts and stellar masses. The abundance of very bright clumps ($L_{clump}/L_{galaxy}\sim8\%$) is around a factor of 10 larger in the UV. This suggests a mild contribution of clumps to the total stellar mass budget that motivates the scientific analysis of the following sections.

\begin{figure*}
    \centering
    \vspace{-10 pt}
    \subcaptionbox{}{\includegraphics[width=0.30\textwidth]{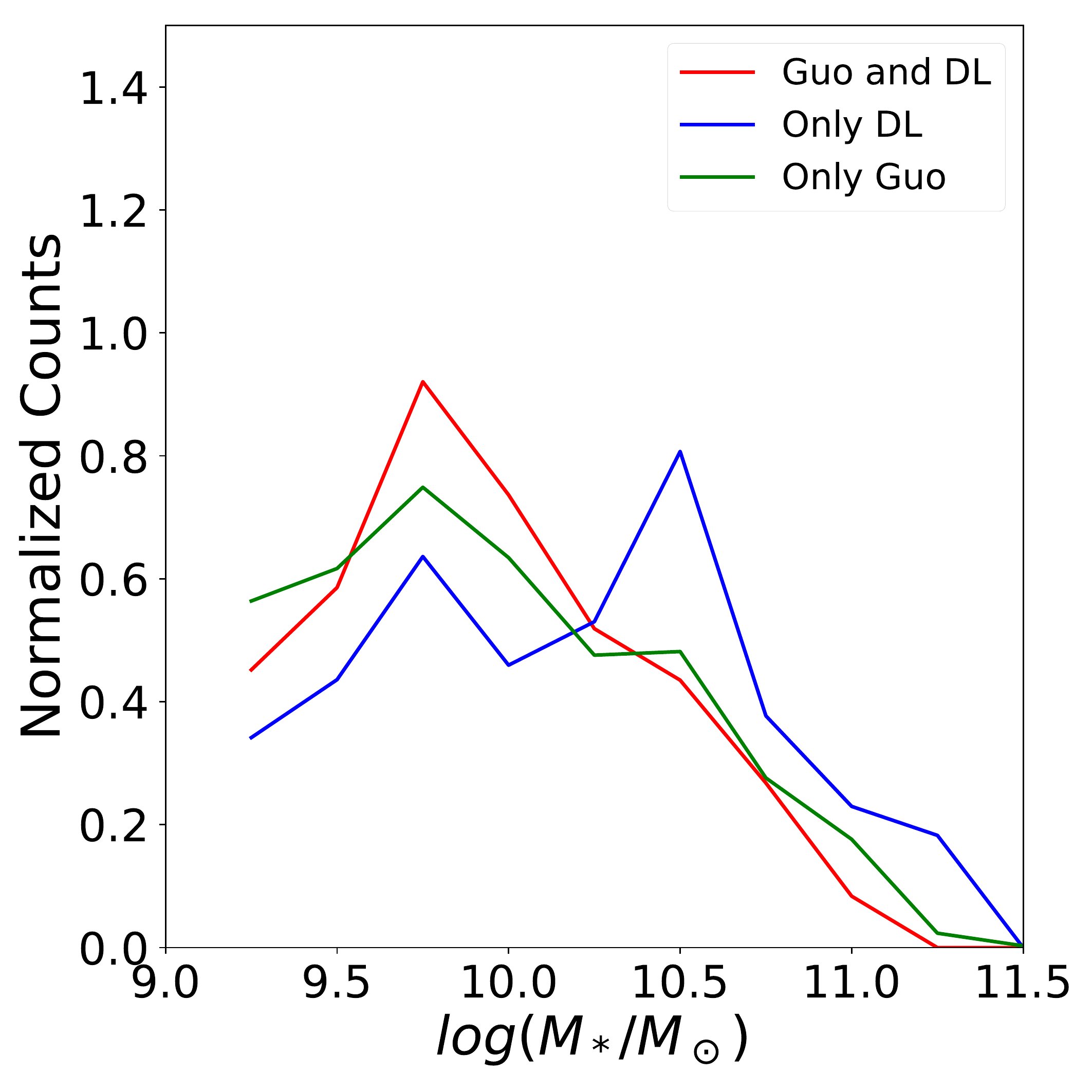}}%
     \qquad
    \subcaptionbox{}{\includegraphics[width=0.3\textwidth]{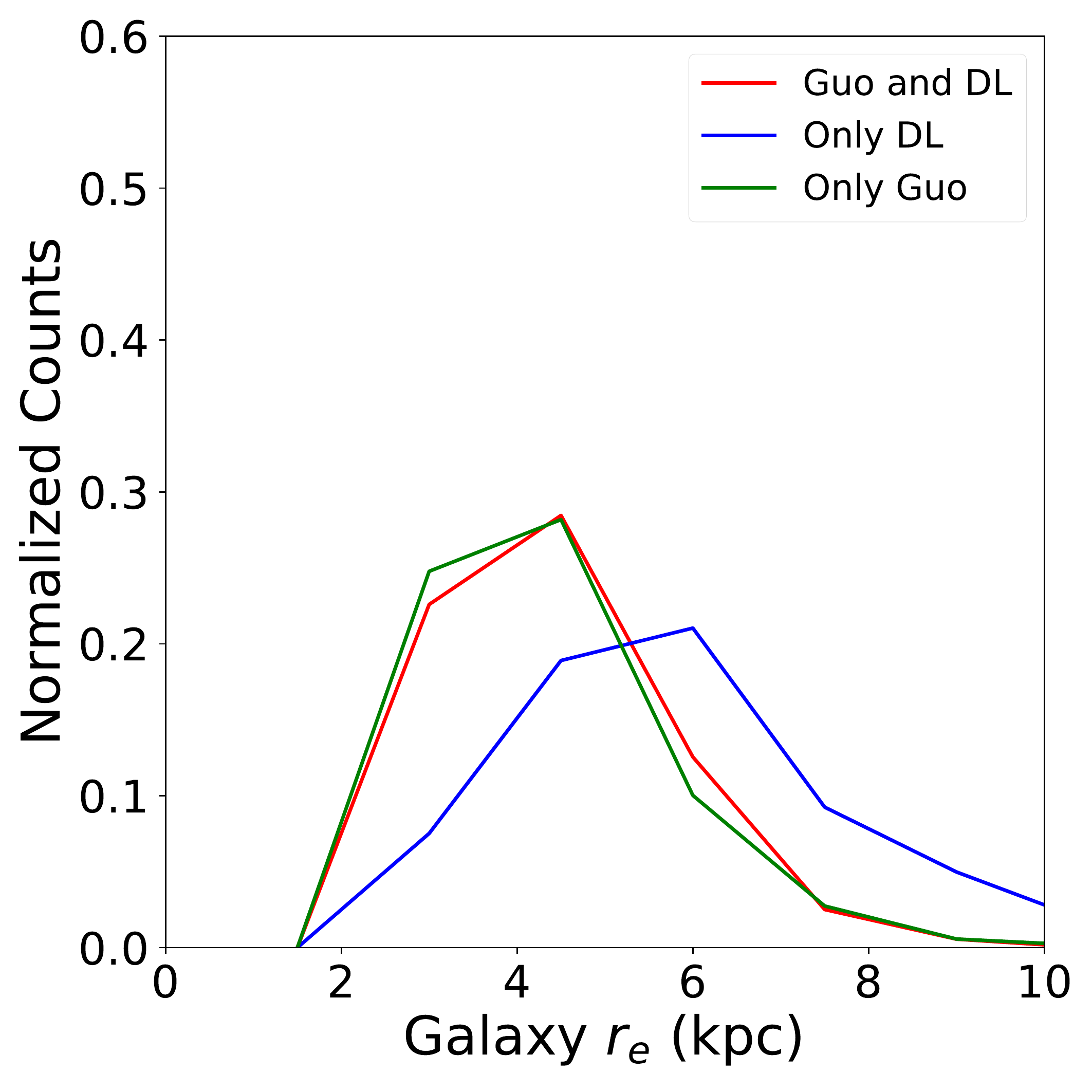}}%
     \qquad
      \subcaptionbox{}{\includegraphics[width=0.3\textwidth]{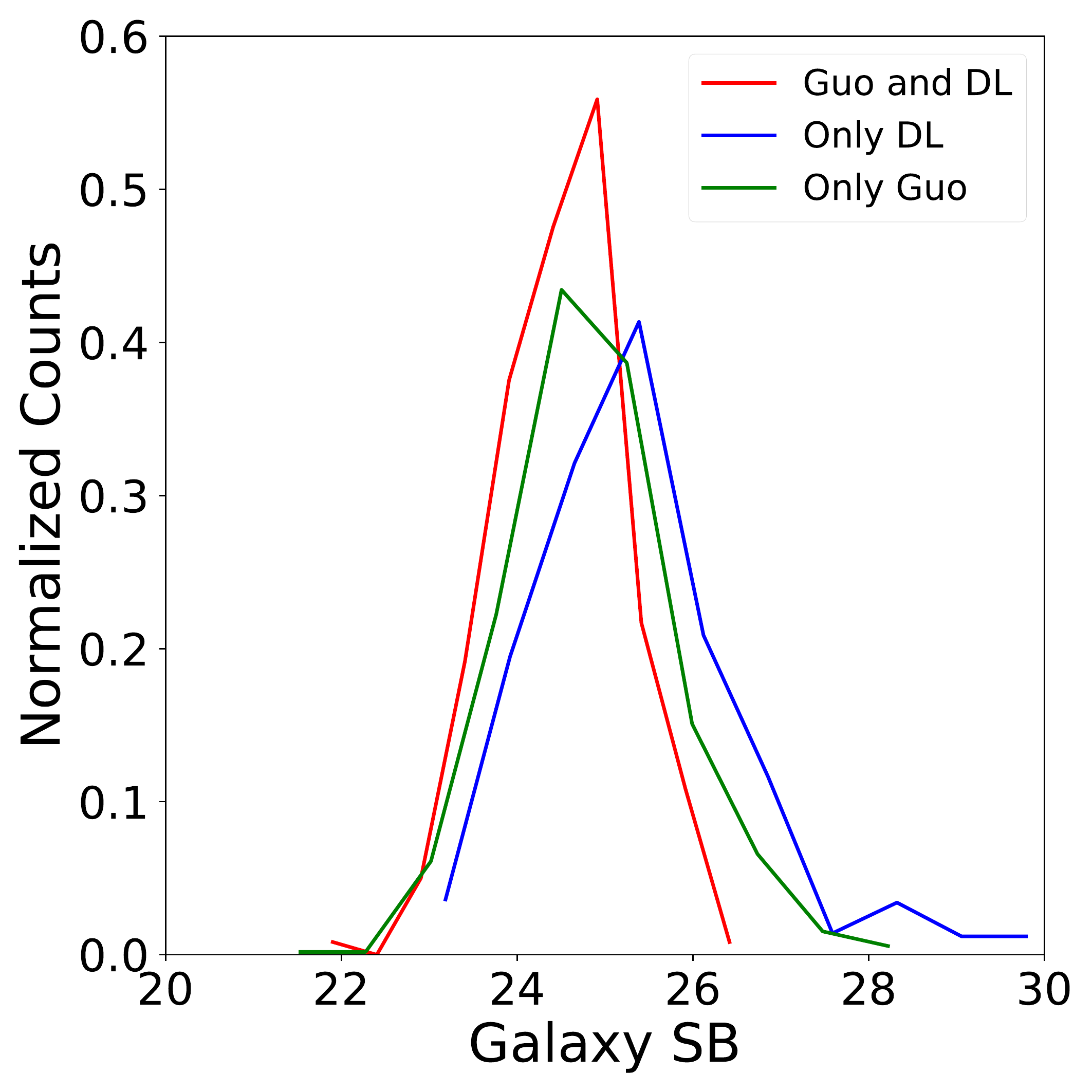}}%
     \qquad
     
      \subcaptionbox{}{\includegraphics[width=0.30\textwidth]{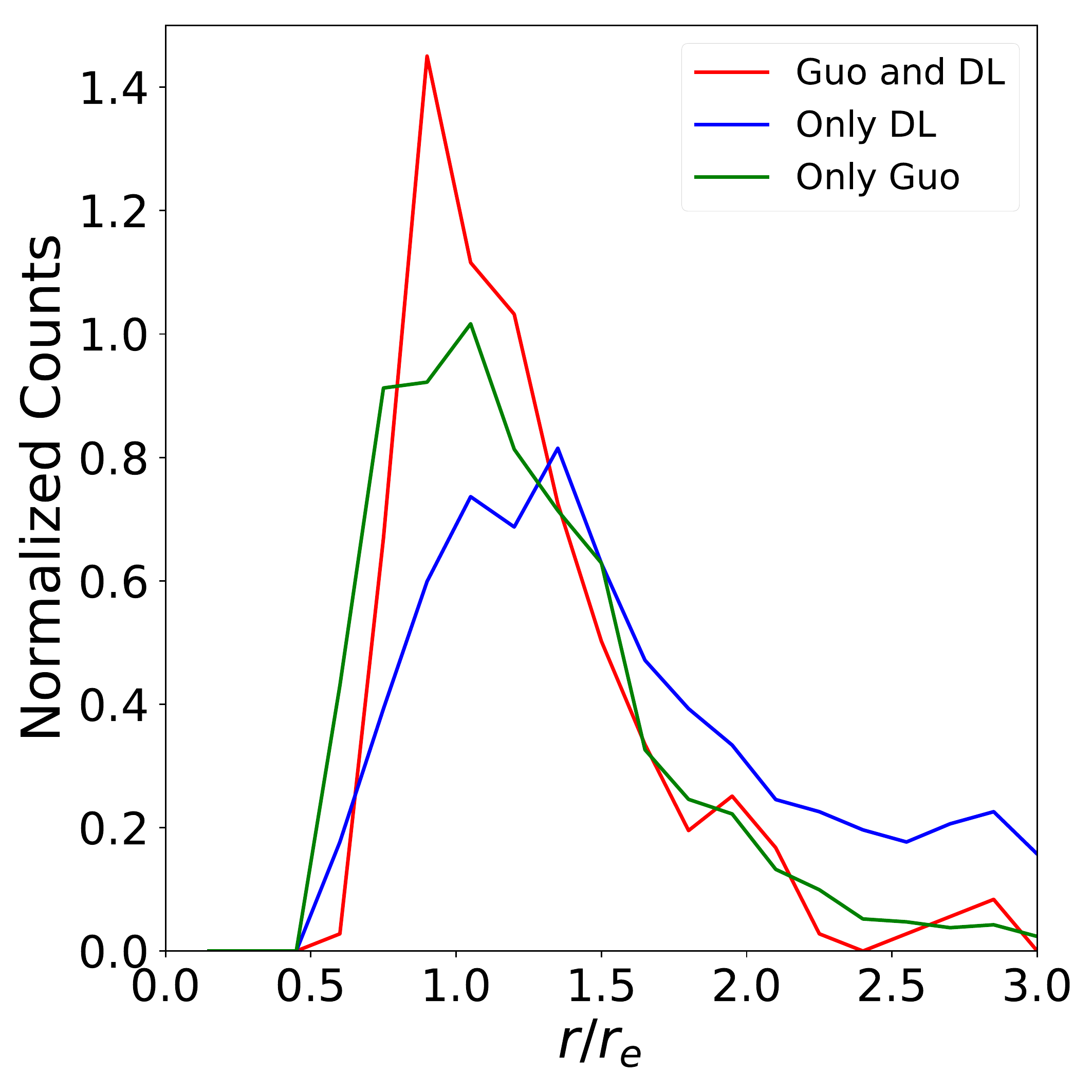}}%
     \qquad
    \subcaptionbox{}{\includegraphics[width=0.3\textwidth]{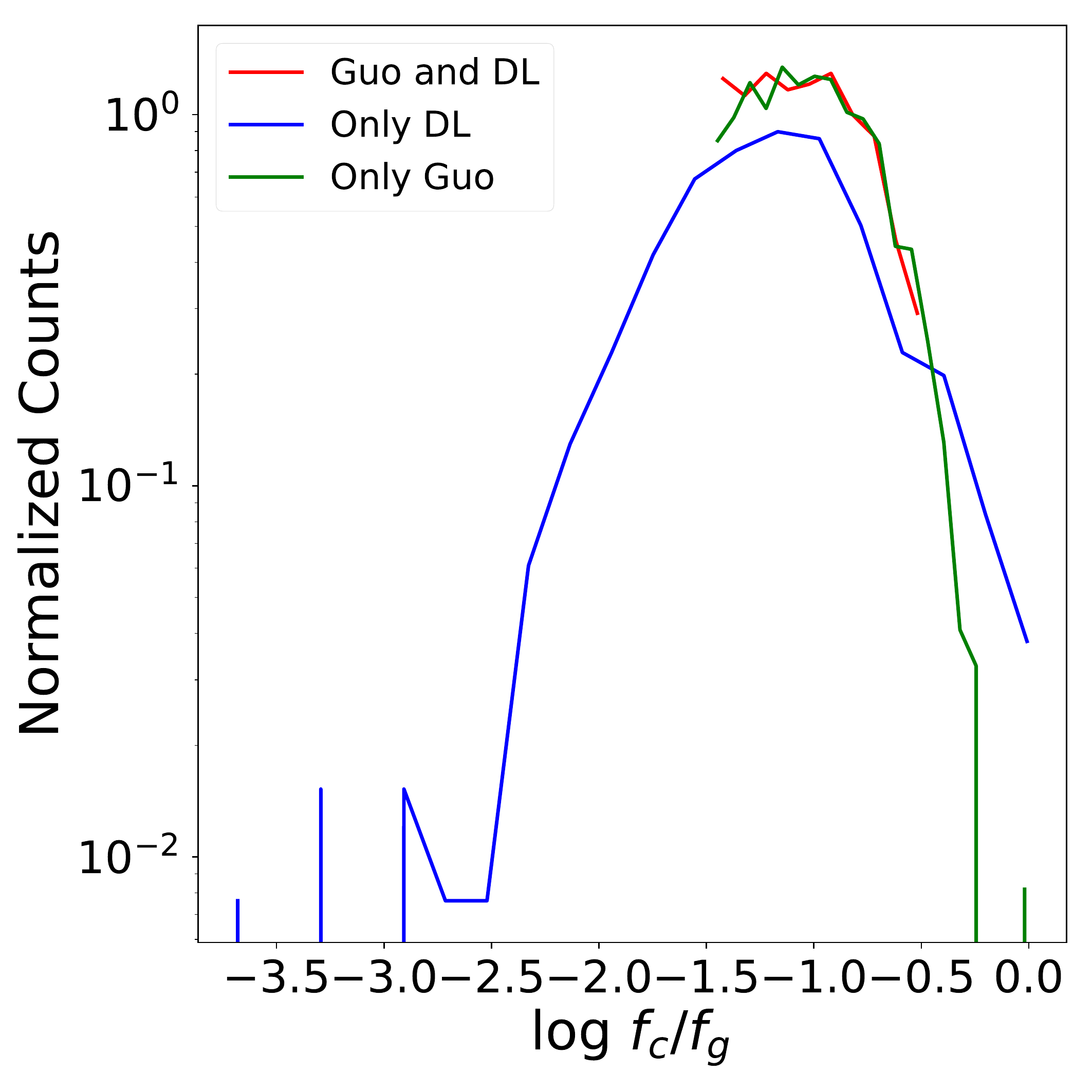}}%
     \qquad
      \subcaptionbox{}{\includegraphics[width=0.3\textwidth]{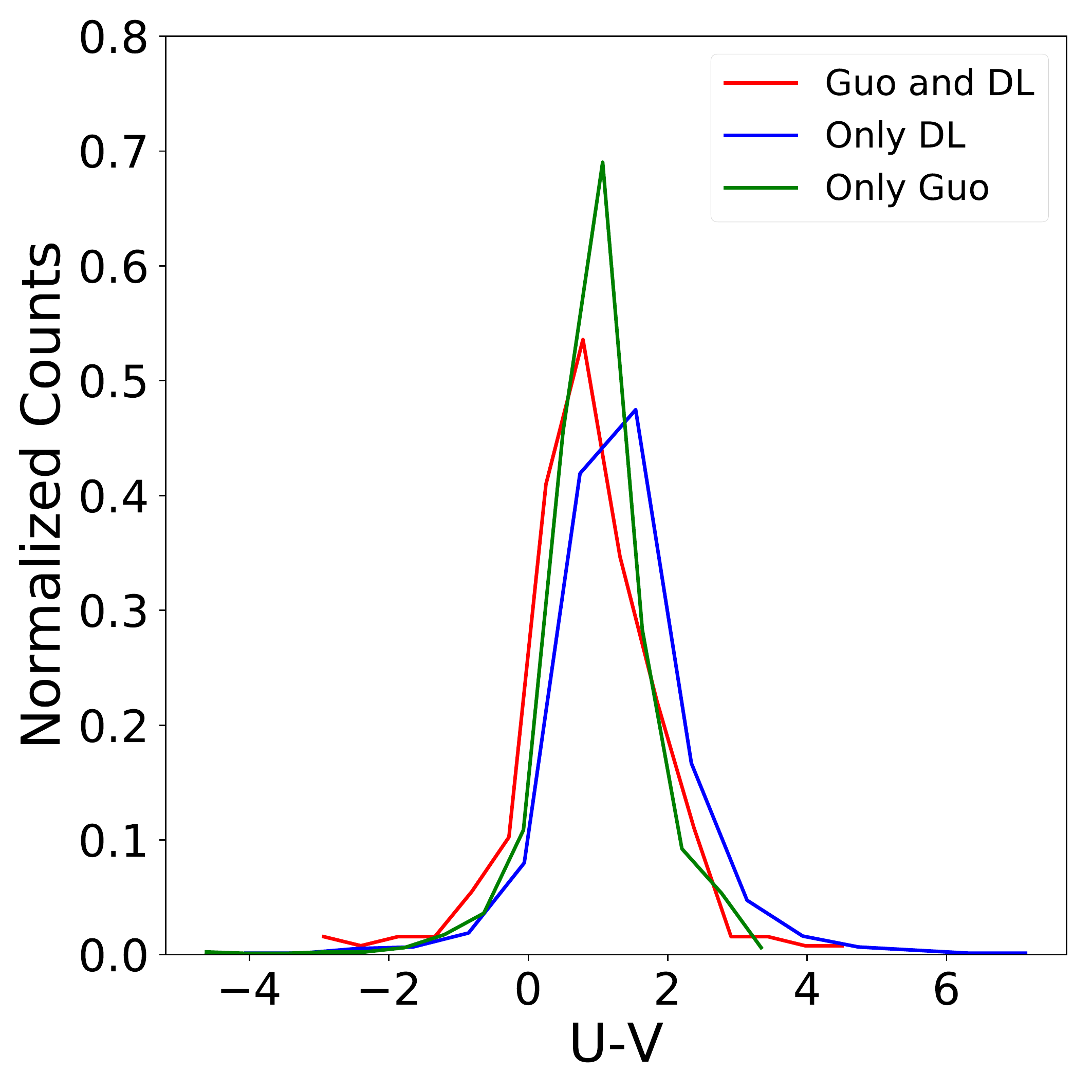}}%
     \qquad

    \caption{Comparison of the properties of clumps and host galaxies detected with deep learning and with the Guo15 method. The red line shows clumps detected by both methods. The blue line indicates clumps only detected by deep learning. The green line shows only the clumps detected by Guo15. Host galaxy properties are shown in the top row. From left to right: galaxy stellar mass, galaxy effective radii and galaxy surface brightness. Clump properties are shown in the bottom row. From left to right: distance to galaxy center, relative UV flux and U-V color. }%
    \label{fig:guo_DL_referee}%
\end{figure*}

The comparison of figure~\ref{fig:clump_LF} suggests that the deep learning based detections are more sensitive but recover globally the same population of clumps as the Guo15 algorithm in a given range of clump luminosities. Since clumps are detected in this work with this new method, it is important to properly understand the differences between the populations of clumps detected with both methods before moving to a scientific analysis. To that purpose, we plot in figure~\ref{fig:guo_DL_referee} some clump and host galaxy properties. We first observe that the DL detections cover essentially the same parameter space as the Guo15 detections. However, they also probe regions in which the Guo15 algorithm detects very few clumps. The clumps not detected by the Guo15 algorithm tend to be at larger galactocentric distances ($r/r_e>1.5$) and in more extended galaxies as revealed by the top panels of figure~\ref{fig:guo_DL_referee}. This again suggests that the DL algorithm is more sensitive and can detect clumps at lower surface brightnesses. As also hinted by the luminosity functions, clumps detected by the DL algorithm in rest frame optical extend to lower relative clump luminosities and such clumps therefore tend to be slightly redder. We note however that in the official Guo15 catalog very faint clumps ($L_c/L_g<3\%$) have been removed because they are considered unreliable. The DL algorithm also tends to detect a population of bright clumps missed by the Guo15 catalog. A visual inspection reveals that those are bright and large clumps at large distances from the galaxy center. They are most likely minor mergers that the Guo15 algorithm did not select.

Overall, these tests confirm that our clump detector, even if trained only on simplistic simulations, extrapolates surprisingly well to real data and validates its application for a scientific analysis. We emphasize that the goal of this comparison is not to reach a perfect agreement with Guo15 since the approaches are different and both contain misidentifications. The main idea is to verify that there is a reasonable level of agreement between the two approaches which supports the fact that the U-net is behaving well when confronted with real data.

\section{Stellar masses of clumps in CANDELS and simulated galaxies}
\label{sec:CANDELS}

Based on the validation tests presented in the previous sections, we now move to analyze the properties of clumps in the CANDELS survey \citep{2011ApJS..197...35G, 2011ApJS..197...36K} and in the VELA zoom-in hydrodynamic cosmological simulations~\citep{2014MNRAS.442.1545C, 2015MNRAS.450.2327Z} using our deep learning based detector. 

\subsection{Observational sample}
\label{sec:obs_sample}
We perform a similar selection to the one done by Guo15 but covering the five CANDELS fields.  More precisely, objects are H-band selected ($15<F160W<24.5$) between redshift 0.5 and 3. We then apply additional stellar mass cuts ($9<\log_{10}(M_*/M_\odot)<12$) for completeness reasons. Additionally we include only galaxies with axis ratios larger than 0.5 to avoid highly inclined objects although this also selects agains prolate galaxies~\citep{2015MNRAS.453..408C, 2016MNRAS.458.4477T, 2019MNRAS.484.5170Z}. We exclude very small galaxies ($R_e<0.2$ arscec in the H-band) to avoid spatial resolution related biases. We finally restrict our sample to star-forming galaxies only, defined as objects with Specific Star Formation Rate (sSFR) larger than $10^{-10}$ ${\rm yr}^{-1}$. All physical quantities are taken from the official CANDELS catalogs~\citep{2013ApJS..206...10G,2013ApJS..207...24G,2017ApJS..228....7N,2017ApJS..229...32S, 2019ApJS..243...22B, 2015ApJ...801...97S}. We refer the reader to the mentioned works for details on how these parameters are derived.

We run our seven trained models to detect clumps in seven different bands when available.  One of the main purposes of this work is to  quantify the distribution of clump masses and their contribution to total galaxy masses. We therefore build a sample of clumps detected in rest-frame optical, which traces older stellar population more effectively than UV. To that purpose, we use detections in the NIR bands: $F125W$ (J) for $1<z<2$ and  $F160W$ (H) for $2<z<3$. This selection corresponds roughly to a rest-frame band of $500$ nm which should allow to better probe the contribution of clumps to the galaxy stellar mass budget than previous UV based selections. We only consider clumps in the distance range $0.5R_e<r_c<3R_e$ from the galaxy center. The lower limit avoids contamination by the galaxy bulge and the upper limit reduces contaminations from neighboring galaxies. 

Although the neural network model is run in all five CANDELS fields and the catalog is released with this work, the following analysis uses data only from the 2 fields (GOODS-S and GOODS-N) 
that include 7 bands / clump in order to have a better photometric coverage (the other three fields do not have $F435W$ nor $F850LP$ coverage). The final sample used for scientific analysis consists of $1,575$ galaxies and $3,733$ detected clumps. 

\subsection{Simulated sample}
\label{sec:sim_sample}
We use here the VELA  zoom-in hydrodynamic cosmological simulation suite presented and analyzed in a variety of previous works~\citep{2014MNRAS.442.1545C, 2015MNRAS.453..408C, 2015MNRAS.450.2327Z, 2016MNRAS.457.2790T, 2016MNRAS.458..242T, 2016MNRAS.458.4477T, 2018ApJ...858..114H}. We refer the reader to the aforementioned works for a detailed description of the simulations. Very briefly, the simulation is made of 35 galaxies simulated with best spatial resolution of 17-35 physical pc and was run with the Adaptive Refinement Tree (ART) code \citep{1997ApJS..111...73K, 2003ApJ...590L...1K, 2009ApJ...695..292C}.  One important feature for this work is that the high spatial resolution allows tracing the cosmological streams that feed galaxies at high redshift, including mergers and smooth flows, and they resolve the 
physical phenomena that govern the formation of clumps in the simulation. We stress that full box simulations such as Illustris TNG100 have a typical resolution of $\sim1$ kpc~\citep{2018MNRAS.475..648P} which prevents properly resolving clump formation. Even the recently completed TNG50 simulation~\citep{2019MNRAS.490.3196P} does not have enough spatial resolution to resolve the formation of clumps, which requires resolution better than $100$ pc. In addition to the high spatial resolution, another advantage of using the VELA simulations is that the properties of clumps are well studied and understood. In \citet{mandelker14,mandelker17} clumps were detected in the VELA galaxies at many timesteps by using both the gas and stellar components in 3D. More details can be found in the aforementioned works.  

Following the approach presented in \cite{2019ApJ...874...59S} and \cite{2018ApJ...858..114H}, we forward model the 35 galaxies with the radiative transfer code Sunrise\footnote{Sunrise is freely available at https://bitbucket.org/lutorm/sunrise. Sunrise images of the VELA simulated galaxies are available online at MAST -- see \url{https://archive.stsci.edu/prepds/vela/}}\citep{2006MNRAS.372....2J,2010NewA...15..509J, 2010MNRAS.403...17J}  and generate HST-like images of galaxies in time steps of $\sim 100$ Myrs in the redshift range $1<z<3$ and with the same filters used for the observational sample. We add real noise from the CANDELS survey to the different generated stamps using the procedure outlined in \cite{2018ApJ...858..114H}. We call the images generated that way \emph{VELA Candelized} images. The images have been produced using 19 different projections (camera orientations - see \cite{2018ApJ...858..114H, 2019ApJ...874...59S} for details). We use here only cameras 12 to18 which are fully randomly oriented between time steps and are thus independent of the box coordinates and angular momentum. This should indeed be closer to real observations. Additionally, structural parameters are derived for simulated galaxies by fitting Sersic models using GALFIT as for the observations that we use to apply the same selections ($R_e>0.2$ arcsec, $b/a>0.5$).

\subsection{Stellar masses of clumps through SED fitting} 
\label{sec:seds}
Quantifying the contribution of clumps to the galaxy mass requires a proper estimation of the stellar masses of the detected clumps. We use standard SED fitting to estimate the stellar population properties of the clumps in both observed and simulated galaxies. 

We first match the optical selected clumps in the six other detection bands. We consider that a clump has been detected in another filter if there is a detection within four pixels of the optical rest-frame detections. Whenever a clump is not detected, we set its flux to 0 in that filter and it is thus not used for the fit.  Following this procedure, $\sim70\%$ of the optical detected clumps are detected in at least 3 bands. 

We follow the procedure described in section~\ref{sec:Performance} for the estimation of clump fluxes. In summary, the clump flux is obtained by performing a 4 pixel aperture photometry at the clump position after removing the flux coming from the disk at the same position and correcting for the PSF aperture. To account for the PSF, we apply a factor of $1.28$ ($1.55$) to the aperture fluxes measured with the ACS (WFC3) cameras. These factors are calibrated by computing the ratio of 4 pixel aperture fluxes over total fluxes in ACS and WFC3 PSFs. The factor is slightly larger for WFC3 because the PSFs are wider. The disk flux is computed by using the best Sersic model for each galaxy in the H-band (\citealp{2012ApJS..203...24V} for the observations). Since structural parameters can change with wavelength, we use the multi-wavelength fits to the 2D surface brightness profiles of H-band selected galaxies brighter than $H=23.5$ published in~\cite{2018MNRAS.478.5410D} to check that using the values derived in the H-band does not significantly change the final clump flux for the brighter galaxies for which we have the structural parameters measured in all wavelengths. Therefore, we assumed that the same will be true for the fainter galaxies and decided to keep the H-band measurements for all galaxies to preserve homogeneity.

\begin{figure}[H]
	\centering
	 \subcaptionbox{}{\includegraphics[width=0.45\textwidth]{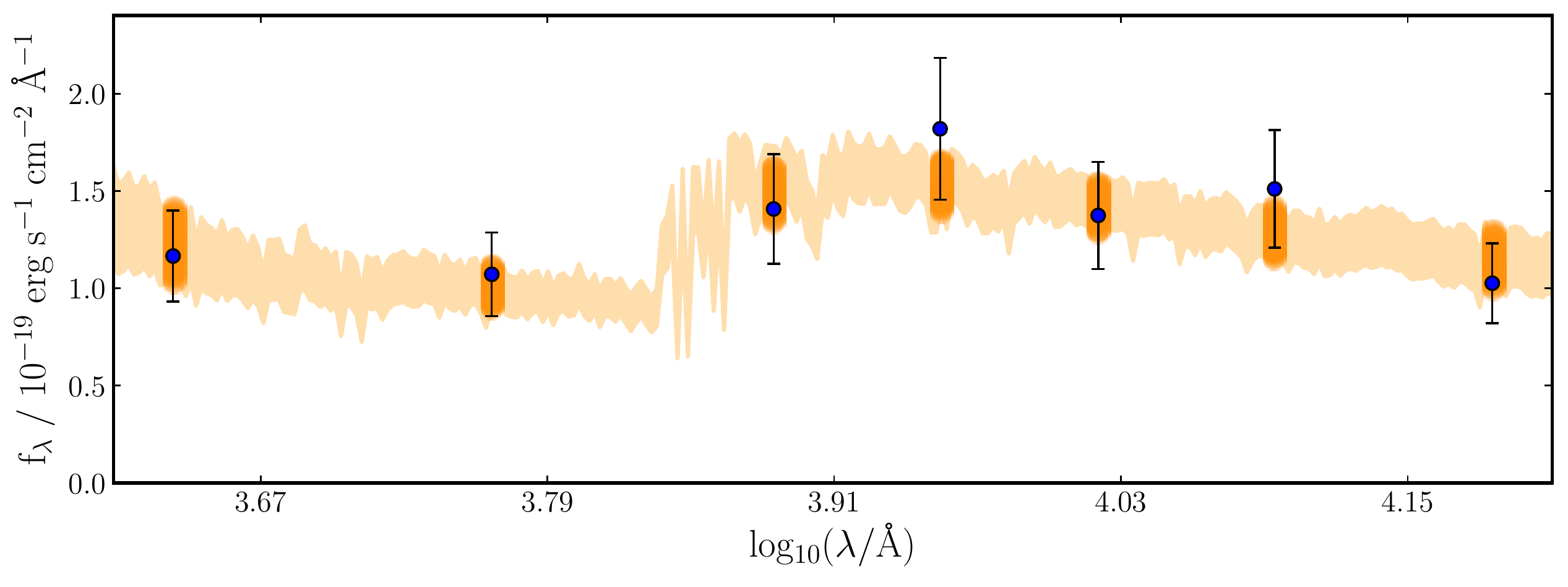}}
         \qquad
          \subcaptionbox{}{\includegraphics[width=0.45\textwidth]{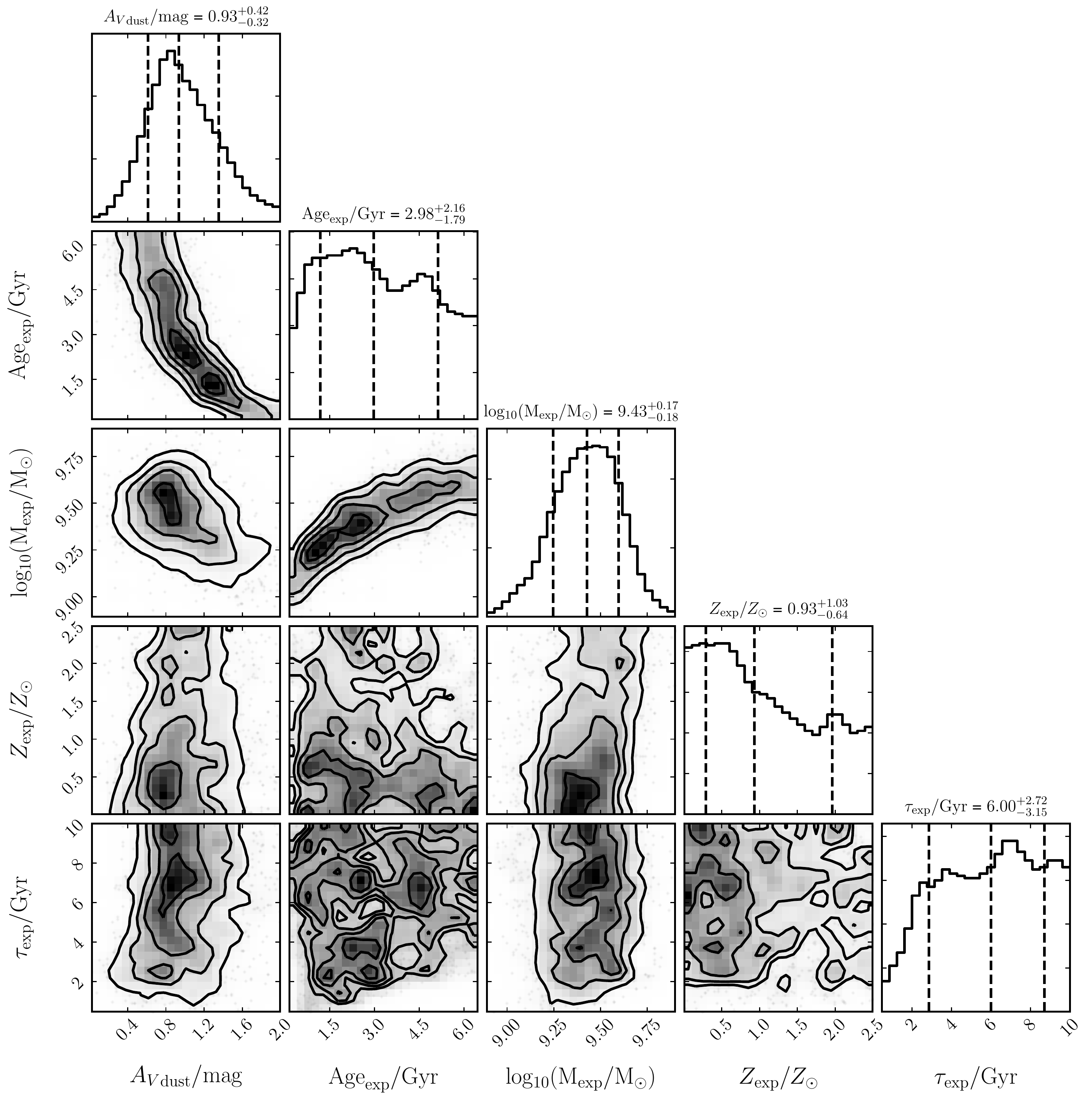}}%
    \caption{Example of fitting of a clump SED. The (a) measured fluxes of clumps in seven different bands (blue filled circles) along with the best fit spectrum. The shaded yellow region indicates the $1\sigma$ confidence interval from the posterior distribution. (b) Corner plot including several parameters estimated from the best fit model. Stellar mass is the best constrained parameter with a typical uncertainty of $0.3$ dex.}
    \label{fig:spectrum}
\end{figure}

We then perform a fit of the 7 band clump SEDs built as explained in the previous section using the bayesian code Bagpipes~\citep{2018MNRAS.480.4379C}. Given the poor photometric coverage of our data, we use a simple tau model for all the clumps with a ~\cite{2000ApJ...533..682C} dust attenuation law. We use BC03 models~\citep{2003MNRAS.344.1000B} and a Kroupa IMF~\citep{2001MNRAS.322..231K}.  The redshift is fixed at the galaxy spectroscopic redshift when available; if not, the best photometric redshift is used. For the simulations we use the known redshift. We are fully aware  that a simplistic tau model is probably not the best SFH model for a clump. We tried an alternative model with constant SFH and the derived stellar masses are essentially the same. As we will discuss in the following sections, a possible route for improvement is the use of non-parametric approaches (e.g.~\citealp{2020arXiv200603599L}) but this is for now out of the scope of the present work. The top panel of figure~\ref{fig:spectrum} shows an example of the best fit model of an observed clump which has been detected in the 7 photometric bands. 

\begin{figure*}
	\centering
	\includegraphics[width=0.90\textwidth]{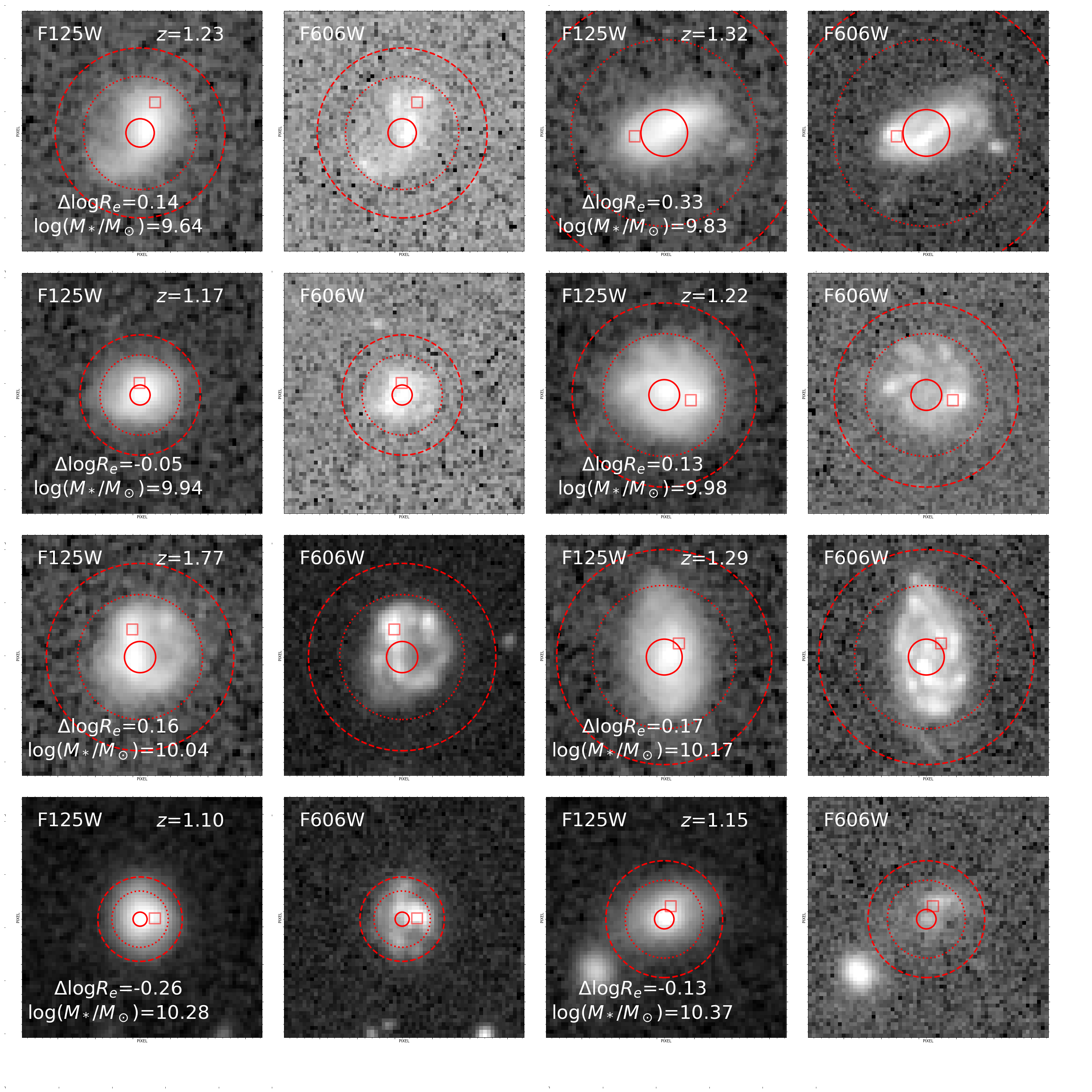}%
    \caption{Example of clumpy galaxies with $1<z<2$ in the $F125W$ and $F606W$ filters with detected massive clumps ($\log_{10}M_c/M_\odot>7$) sorted by increasing galaxy stellar mass. The red squares indicate the positions of the clumps. Only massive clumps are shown. The solid, dotted and dashed red circles show distances of $0.5R_e$, $2R_e$ and $3R_e$ respectively. Each stamp is $64\times64$ pixels ($\sim30\times30$ kpc).}
    \label{fig:ex_stamps}
\end{figure*}

The analysis in the forthcoming sections uses exclusively the stellar mass of the clump. Although the spectral resolution of our data is poor, we find that stellar masses are relatively well constrained by the SED fitting procedure with a typical $1\sigma$ confidence interval from the posterior distribution of $\sim0.3$ dex. For the cases where fewer than three photometric values are used, the stellar mass uncertainties are around $\sim0.6-0.7$ dex. This is certainly high and must be kept in mind when analyzing the results presented in the following sections. Having access to the full posterior enables us to propagate the errors onto the different measurements, as we will discuss as well in the following sections. Figure~\ref{fig:ex_stamps} shows some examples of galaxies with detected massive clumps observed both in the optical and UV rest-frames.

Other quantities such as metallicities and ages are significantly more degenerate, as can be appreciated in the corner plot of figure~\ref{fig:spectrum}. This is naturally expected from a model constrained with only a few data points. It therefore becomes difficult to use such data for any scientific analysis. Hence we have decided not to discuss clump ages in this work. This will be done in forthcoming dedicated work using an alternative approach (Ginzburg et al. in prep.). 

\subsection{Impact of observational effects on clump stellar masses and completeness}
\label{sec:compl}

The comparison with simulations 
allows us to quantify the impact of observational effects on the derived clump properties from the CANDELS images. Given that the \emph{true} stellar mass of the clumps is known in the simulation, we can use this as an estimator of the completeness of our deep learning based detector and to assess the accuracy of the clump mass measurements. We use the clumps identified in 3D by \cite{mandelker17} as a reference and then match with the 2D detections. Given the much lower resolution of the Candelized images, it is difficult to associate a 2D detection with one unique 3D clump because of blending (see e.g.~\citealp{moody14, 2020MNRAS.494.1263M}). Therefore, in order to perform the matching between 2D and 3D we follow an alternative approach. We divide the Candelized image in small boxes of $10\times10$ pixels ($\sim 4.8$~kpc) and compute the total stellar mass in clumps in the region from the original 3D simulations that falls within each box. We essentially use the clump stellar mass in the catalogs from~\cite{mandelker17} and add all the clump stellar masses in the $10\times10$ pixels region. We then consider that there has been a detection if there is a 2D clump inside the box and associate to the 2D detection the total clump mass computed. We find that $\sim10\%$ of the 2D clumps cannot be associated with a 3D clumps which is consistent with our purity estimations of $\sim90\%$. 

\subsubsection{Completeness}
\begin{figure*}
	\centering
	 \subcaptionbox{}{\includegraphics[width=0.31\textwidth]{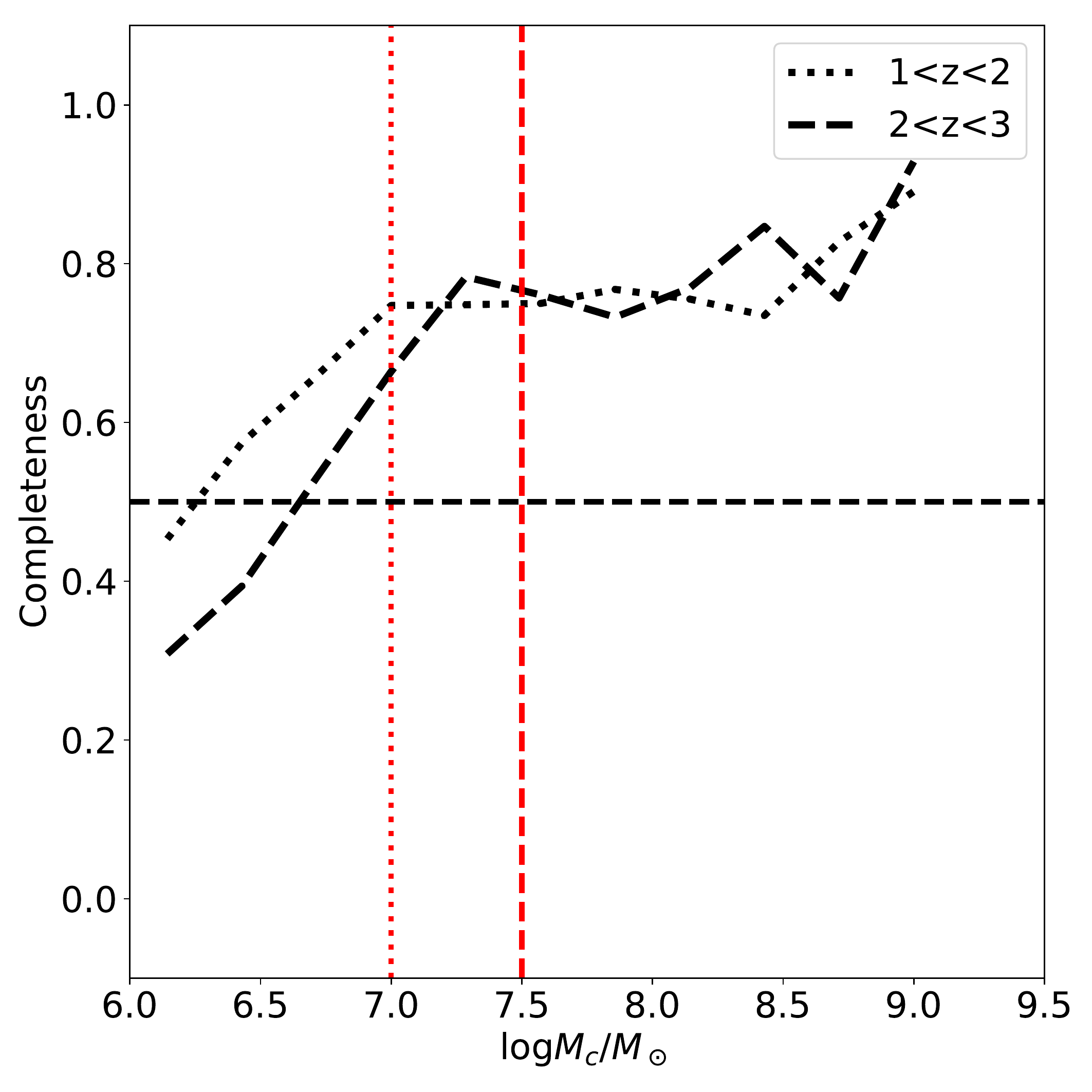}}%
	 \qquad
	  \subcaptionbox{}{\includegraphics[width=0.31\textwidth]{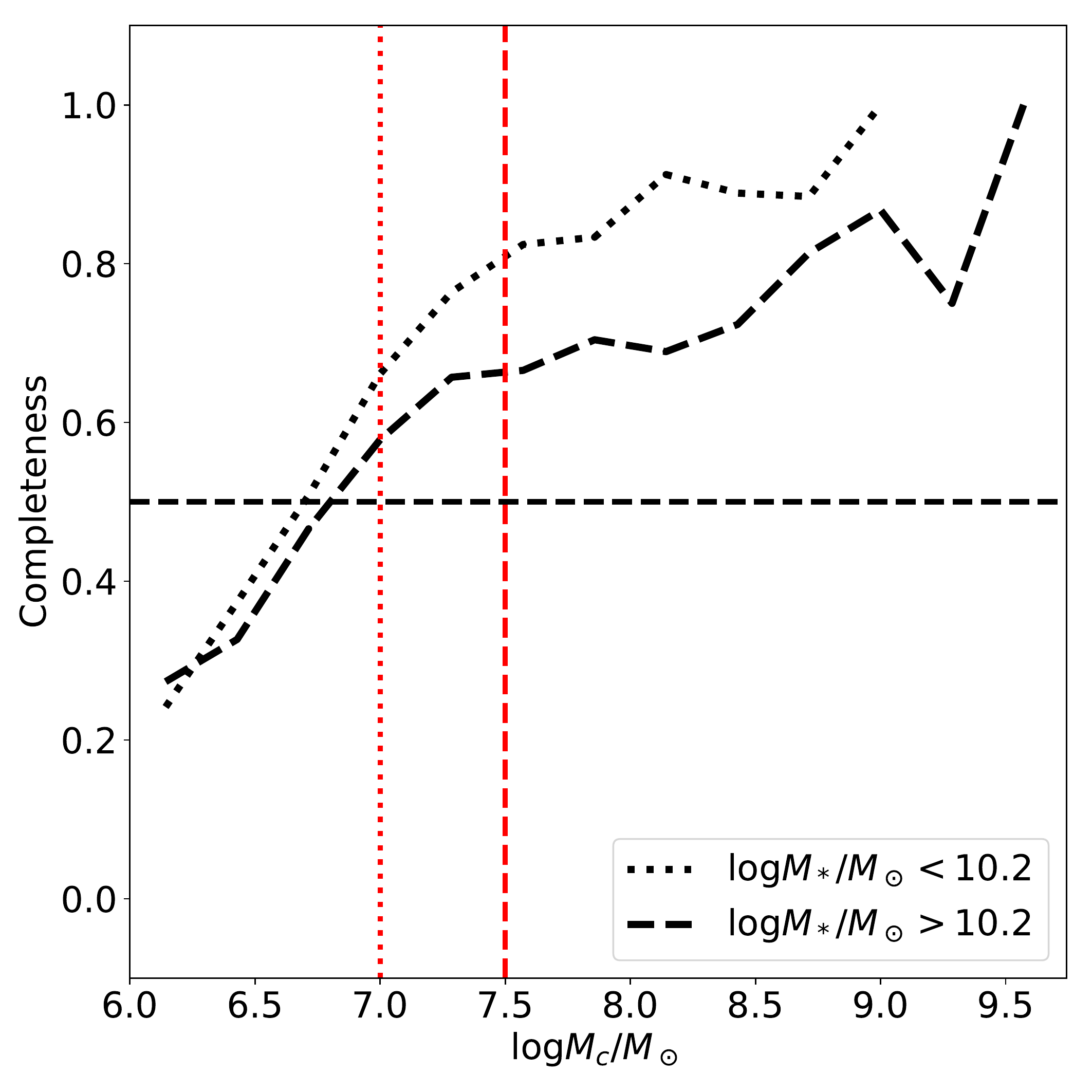}}%
	  \qquad
	   \subcaptionbox{}{\includegraphics[width=0.31\textwidth]{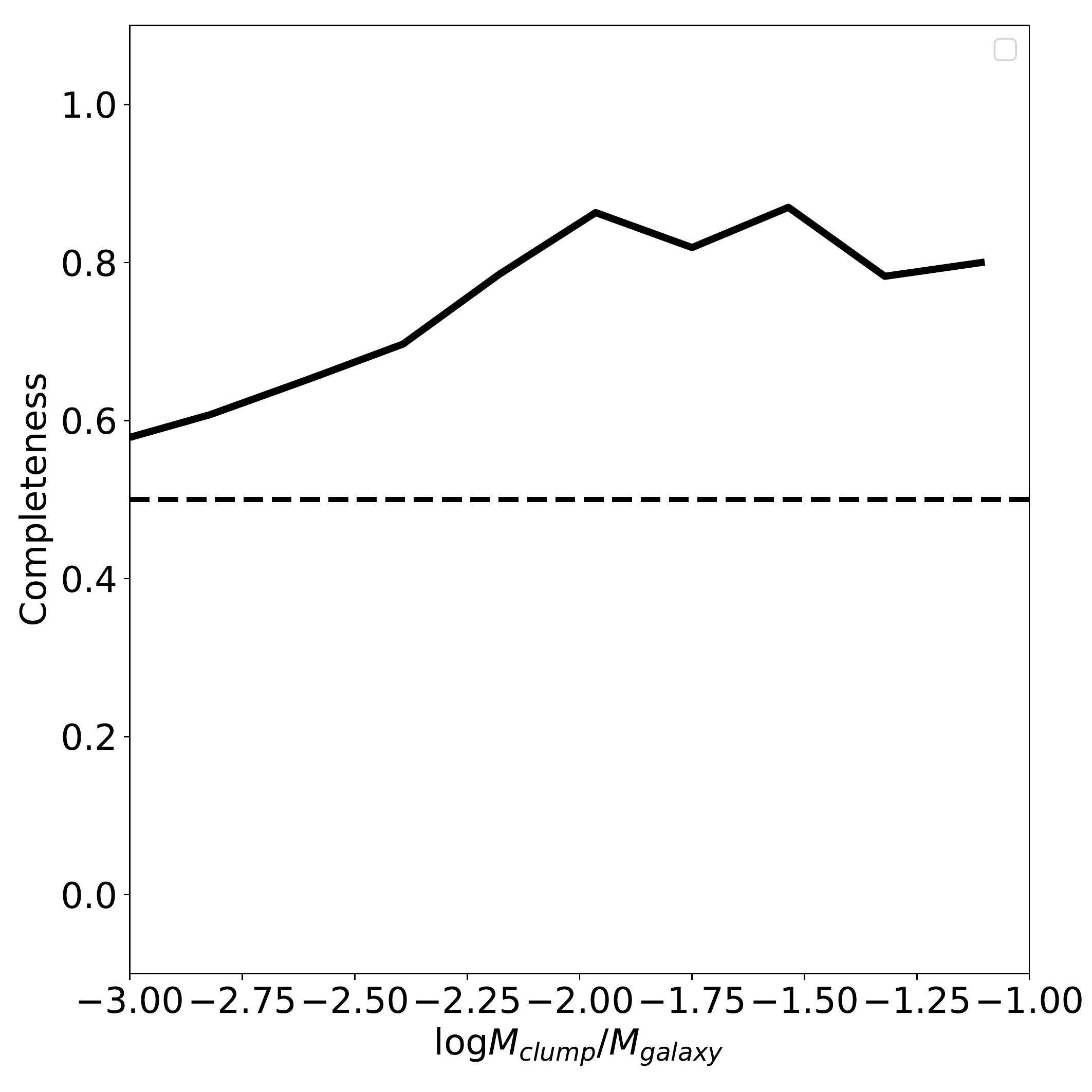}}%
  \caption{Calibration of the deep learning based clump detector completeness with numerical simulations. The figure shows the fraction of 3D clumps detected in the 2D mock Candelized images from the VELA numerical simulation as a function of the \emph{true} clump stellar mass from the 3D simulations (left and middle panels) and as a function of the ratio between clump mass and galaxy mass (right panel). The true mass is computed by adding all clump masses within a $10\times10$ pixel region (see text for details). In the left panel each line shows a different redshift bin as labeled. In the middle panel the lines indicated different galaxy stellar mass bins. In the right panel, the completeness is shown as a function of ratio between the clump mass and the galaxy mass. The dotted and dashed vertical lines show clump masses of $10^7$ and $10^{7.5}$ solar masses respectively. The clump detector detects $\sim80\%$ of all clumps more massive than $10^{7}$ solar masses and relative masses larger than $\sim 1\%$. Below these thresholds, the completeness starts decreasing. Since the threshold of $\sim1\%$ corresponds to different clump masses depending on the galaxy mass, we measure a difference in completeness for low mass and massive galaxies in the middle panel. }
    \label{fig:VELA}
\end{figure*}

 We first quantify the completeness of the detections. Figure~\ref{fig:VELA} plots the fraction of 3D clumps detected in 2D as a function of the the total 3D clump stellar mass in the $10\times10$ boxes described previously for two different redshift and galaxy stellar mass bins. The figure also shows the completeness as a function of the ratio between the mass of the clump and the mass of the galaxy. We see that our detector detects $\sim80\%$ of 3D clumps above a stellar mass of $\sim10^{7}$ solar masses at $z<2$ and that the success rate drops below this mass. At higher redshifts ($z>2$) the completeness starts decreasing at slightly larger masses because galaxies are fainter, but the impact is mild. The dependence with the galaxy stellar mass is more pronounced. Clump detections in massive galaxies are less complete. The reason for this is that the contrast of a clump at fixed mass is weaker in a massive system. This is indeed reflected in the right panel of figure~\ref{fig:VELA}. The detections are $\sim80\%$ complete down to a relative clump mass of $\sim 1\%$, which corresponds to different clump stellar masses for massive and low mass galaxies. For simplicity, in the following, we will restrict our analysis to clumps more massive than $10^{7}$ solar masses. This selection avoids incompleteness related biases on the global population. We shall keep in mind however when interpreting the results that the massive end of the galaxy population might be more affected by incompleteness.

\subsubsection{Clump stellar mass accuracy}
We now look at how well clump stellar masses are recovered from the images using our procedure. We show in the 
left panel of figure~\ref{fig:mass_2d_3d} the comparison between the stellar mass for a given 2D detection and the true 3D mass in the associated region (computed by adding up the masses of all 3D clumps projected into the region). Although there is a clear correlation between the true clump masses and the estimated ones, the figure very clearly shows that our SED based method severely overestimates the clump stellar mass by about an order of magnitude. This is true even after adding up all the masses of all clumps in the box. Although surprising, these results are in agreement with previous works which also estimate that the different observational effects can lead to a factor of 10 overestimation of the mass~\citep{2020MNRAS.494.1263M, 2018NatAs...2...76C}. 

There are several reasons that can explain this big difference. One is obviously spatial resolution, which causes clumps to be blended in the 2D images. We have estimated that each 2D detection corresponds on average to three 3D clumps (see also~\citealp{moody14}). Even if this is partially taken into account by adding up the stellar mass of all 3D clumps in a region, blending causes the 2D clump region to also be contaminated by emission from the galaxy which in turn overestimates the mass. The color code of figure~\ref{fig:mass_2d_3d} shows the distance of each clump to the galaxy center. Clumps for which the stellar mass is most overestimated indeed tend to be in the inner regions where the emission from the galaxy is stronger. This confirms that contamination from the galaxy flux certainly contributes to the overestimation. Another possibility is that the Star Formation Histories (SFHs) adopted to fit the SEDs are not adapted for clumps which are expected to have bursty star formation histories. We recall that we also tried constant SFHs without major changes in the  resulting masses. However, the effect of the adopted SFH is something to be investigated in future work by using for example non-parametric SFHs (e.g.~\citealp{2020arXiv200603599L}). Overall, these large errors suggest that with current data it is very difficult to establish constraints on individual clump properties. However, it is reasonable to assume that the same biases will be present in the real CANDELS observations. Therefore we can still learn about clump formation by comparing simulations and observations provided they are confronted under comparable conditions, as we will show in the following sections. 

\subsubsection{Clump stellar mass correction}

\begin{figure*}
	\centering
	 \subcaptionbox{Uncorrected}{\includegraphics[width=0.31\textwidth]{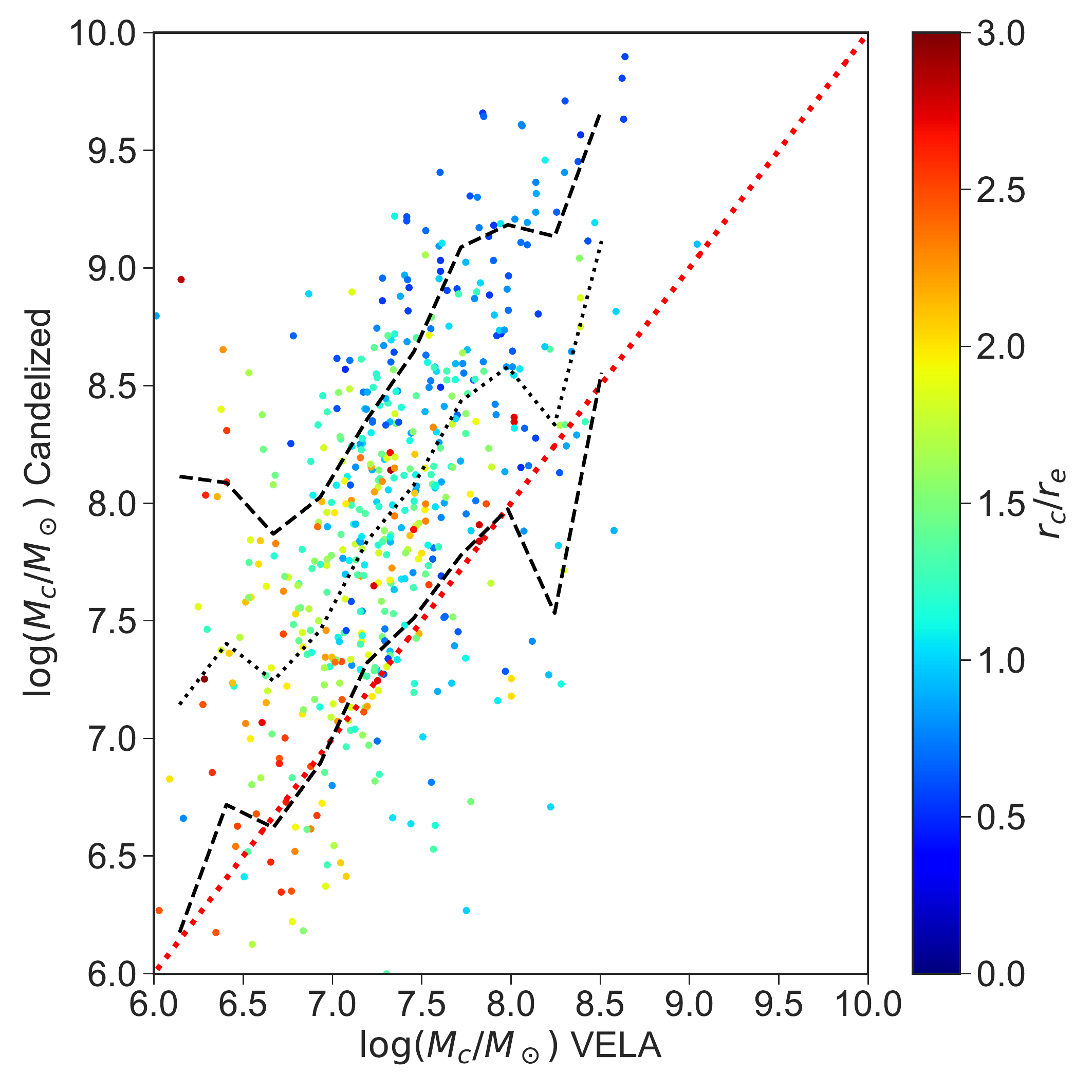}}%
         \qquad
          \subcaptionbox{Corrected with VELA prior}{\includegraphics[width=0.31\textwidth]{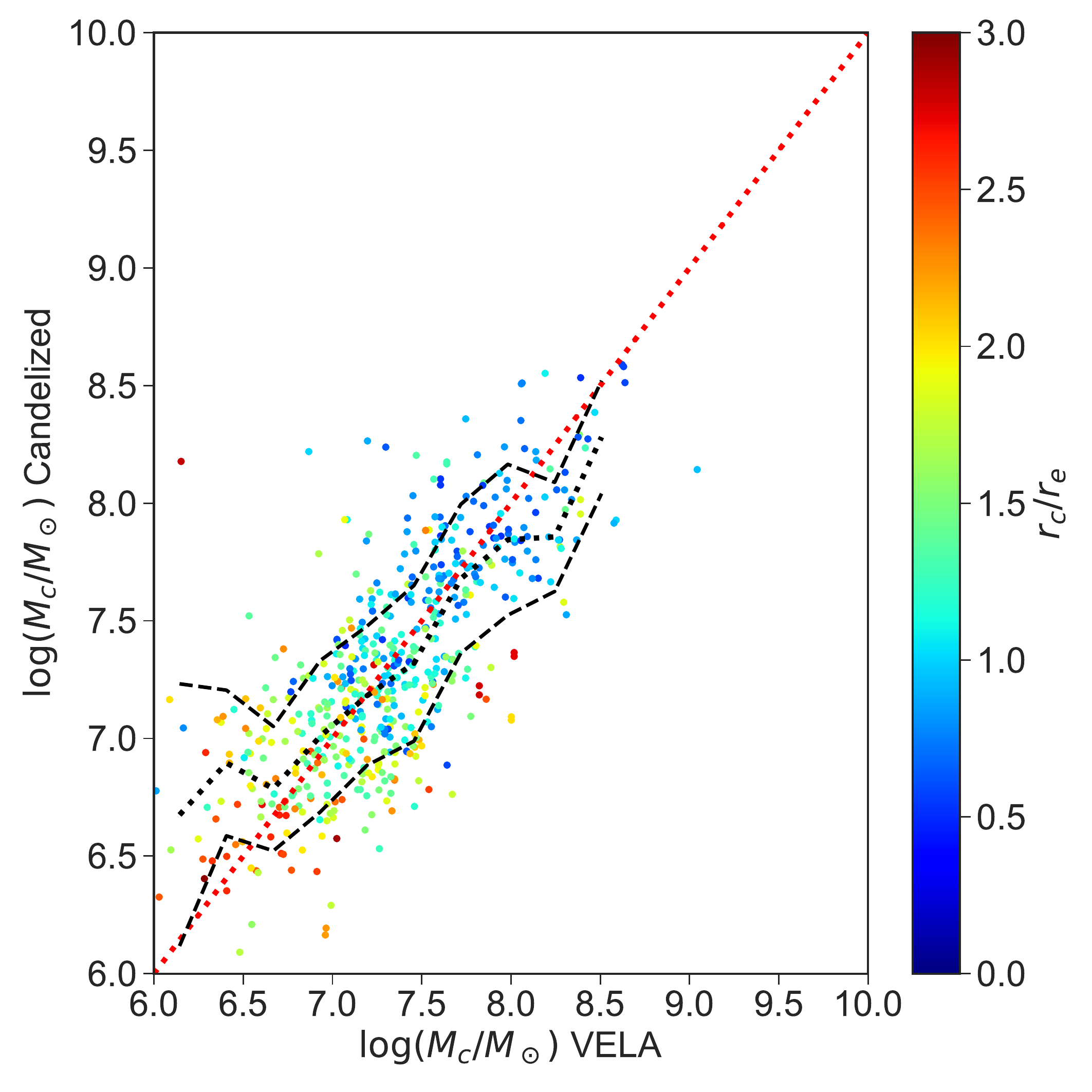}}%
          \qquad
          \subcaptionbox{Corrected with flat prior}{\includegraphics[width=0.31\textwidth]{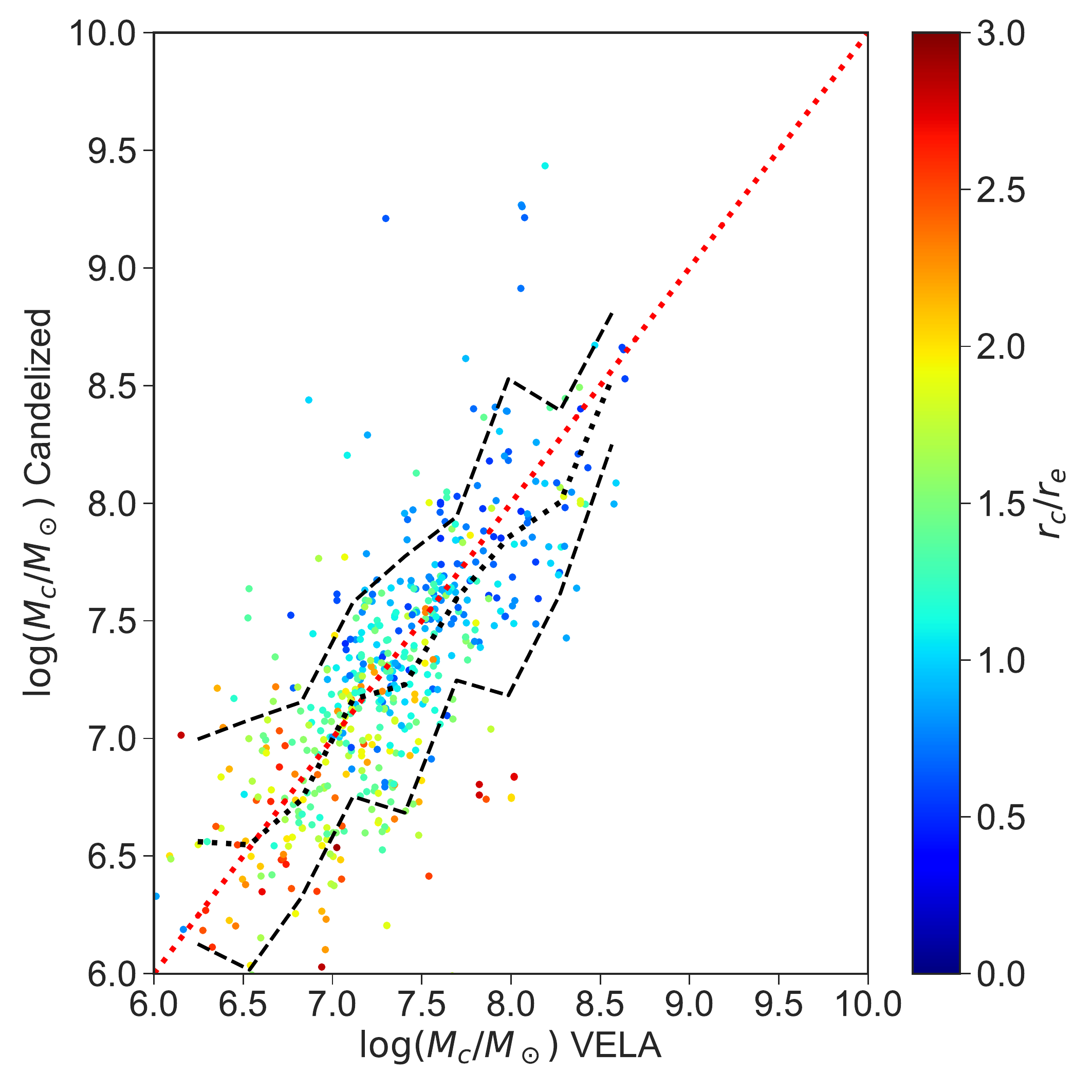}}%
  \caption{Relation between true clump stellar masses derived from the 3D simulation and stellar masses estimated through SED fitting on the mock Candelized images (see text for details on how both 3D and 2D clumps are matched). Each point is a clump. The color code indicates the distance of the clump to the galaxy center normalized by the effective radius. The dotted line indicates the median values and the dashed black lines the standard deviation. (a) The y-axis shows the values directly obtained from the SED fitting procedure. Although there is a correlation between the true and estimated clump masses, the masses derived in 2D severely overestimate the true clump stellar masses. (b) The y-axis shows the corrected values with a mixture density network (see text for details). Once the correction is applied, the mass measurements from the 2D images better agree with the 3D mass measurements and also the scatter is reduced. A bias remains, especially at the high mass end. (c) The y-axis shows the corrected values with the same mixture density network but modified to account for the prior of the VELA distribution. The bias at the high mass end is reduced.  }
    \label{fig:mass_2d_3d}
\end{figure*}

In addition, we can go a step further and use the relation shown in figure~\ref{fig:mass_2d_3d} to correct the derived stellar masses of the 2D detected clumps if one assumes that the 3D mass is the \emph{true mass}. The figure shows indeed that there is a correlation between the 3D mass and the estimated mass in 2D with secondary dependences on other parameters such as the clump distance. It should be possible therefore to find a function $h_w$ so that $m_{3D}=h_w(m_{2D},\vec{\theta})$ where $m_{3D}$ and $m_{2D}$ are the 3D and 2D clump stellar masses respectively and $\vec{\theta}$ is a vector containing any other secondary parameters. Since we do not know a priori the analytic form of $h_w$, we model it with a mixture density network. We define $\vec{\theta}$ as a 6 dimensional vector containing the clump distance to the center which we have seen correlates with the stellar mass difference, the inclination of the galaxy ($b/a$), the effective radius of the galaxy ($R_e$) and the Sersic index ($n$) derived from the best Sersic fit, the stellar mass of the galaxy and the redshift. We model the posterior distribution with a Gaussian probability density function ($q(m_{3D}|(m_{2D},\vec{\theta}))\sim\mathcal{N}(\mu,\,\sigma^{2})$) and train the neural network to maximize the log likelihood of the \emph{true} clump mass value. We then use the mean of the posterior distribution to compute the estimated $\tilde{m}_{3D}$. We train the network using the clump measurements in the simulated images obtained with all random camera projections except one, which is used to test. Since the different projections are fully random, the clump 2D masses and the best fit Sersic parameters of the galaxies can be considered as independent across projections. Once trained, we use the learned model, $h_w$, to test the correction on the clumps detected on the camera not used for training. The results are shown in the middle panel of figure~\ref{fig:mass_2d_3d} for camera 15, but similar results are obtained for other camera orientations. 

We can easily see that the corrected 2D masses are now in much better agreement with the 3D measurements. This proves that it is possible to recover the intrinsic 3D clump mass (in the $10\times10$ box) given the 2D measurement. We note that the correction does not significantly depend on the exact architecture used for the neural network nor on the initialization of it. We tried 50 different random models, including more complex posteriors modeled with a mixture of Gaussians and changing the number or layers and number of units in each layer. This model uncertainty will be incorporated into the error budget in the following sections. We have also tried adding additional parameters to the network such as clumps colors and luminosities without any significant impact.

Although the correlation shown in the middle panel of figure~\ref{fig:mass_2d_3d} represents a significant improvement compared with the uncorrected values, we notice that the estimator is still biased at the low and high mass ends. As a matter of fact, at the high mass end, the predictions are under estimated and the opposite happens at the low mass end. This is important because it means that the current estimator will tend not to estimate clump masses larger than $10^9$ solar masses which could bias our results when applied to real observations. The reason for this bias is that the posterior distribution $q(m_{3D}|(m_{2D},\vec{\theta}))$ estimated by the mixture density network represents in fact the posterior under the training prior $\tilde{p}(m_{3D})$. Since the VELA sample used for training contains very few clumps with masses at the edges of the distribution, i.e. larger than $10^9$ and smaller than $10^6$ solar masses,  the estimator tends to shift the posterior towards the prior to minimize the risk of failure. 

In order to correct for that effect, we attempt to estimate a posterior $p(m_{3D}|(m_{2D},\vec{\theta}))$ under a flat prior $p(m_{3D})$ by renormalizing the posterior distribution under the VELA prior ($\tilde{p}(m_{3D})$) using the following relation:
$$ p(m_{3D}|(m_{2D},\vec{\theta})) \propto \frac{p(m_{3D})}{\tilde{p}(m_{3D})}q(m_{3D}|(m_{2D},\vec{\theta}))$$ 

Since $p(m_{3D})$ is a constant, it simply implies dividing the posterior estimated by the density network by the VELA prior. The result of applying the posterior under a flat prior to estimate the clump mass is shown in the right panel of figure~\ref{fig:mass_2d_3d}. We can see that the estimator is now less biased especially at the edges of the distribution. The scatter also tends to increase. This is because the estimated values are no longer pushed towards the average value of the prior distribution which is around $10^7$ solar masses.

 In the following sections we will analyze the impact of these different corrections on the derived clump properties in the observations and in the forward modeled simulations.

\subsection{Clump Stellar Mass Function}
\label{sec:lum_funct}

We start our analysis by focusing on the clump stellar mass function (cSMF) which can provide interesting clues about the physical processes governing clump formation. For example, several works have pointed out that the slope of the cSMF might be indicative of the formation mechanisms (e.g. ~\citealp{2006ApJ...644..879E, dz18, 2018ApJ...869..119E}).  If clumps are formed by turbulence driven fragmentation of gas clouds, the resulting slope $\alpha$ of the cSMF is expected to be around $-2$~\citep{2018ApJ...869..119E}. Also, clumps formed ex-situ are expected to be on average more massive than in-situ formed clumps ~\citep{mandelker17} so the distribution of clump stellar mass can also provide clues about the origin of the clumps.
 
 \subsubsection{Impact of observational effects on the clump stellar mass function}
 \label{sec:CANDELS-VELA}
 
\begin{figure*}
	\centering
	 \subcaptionbox{Uncorrected}{\includegraphics[width=0.30\textwidth]{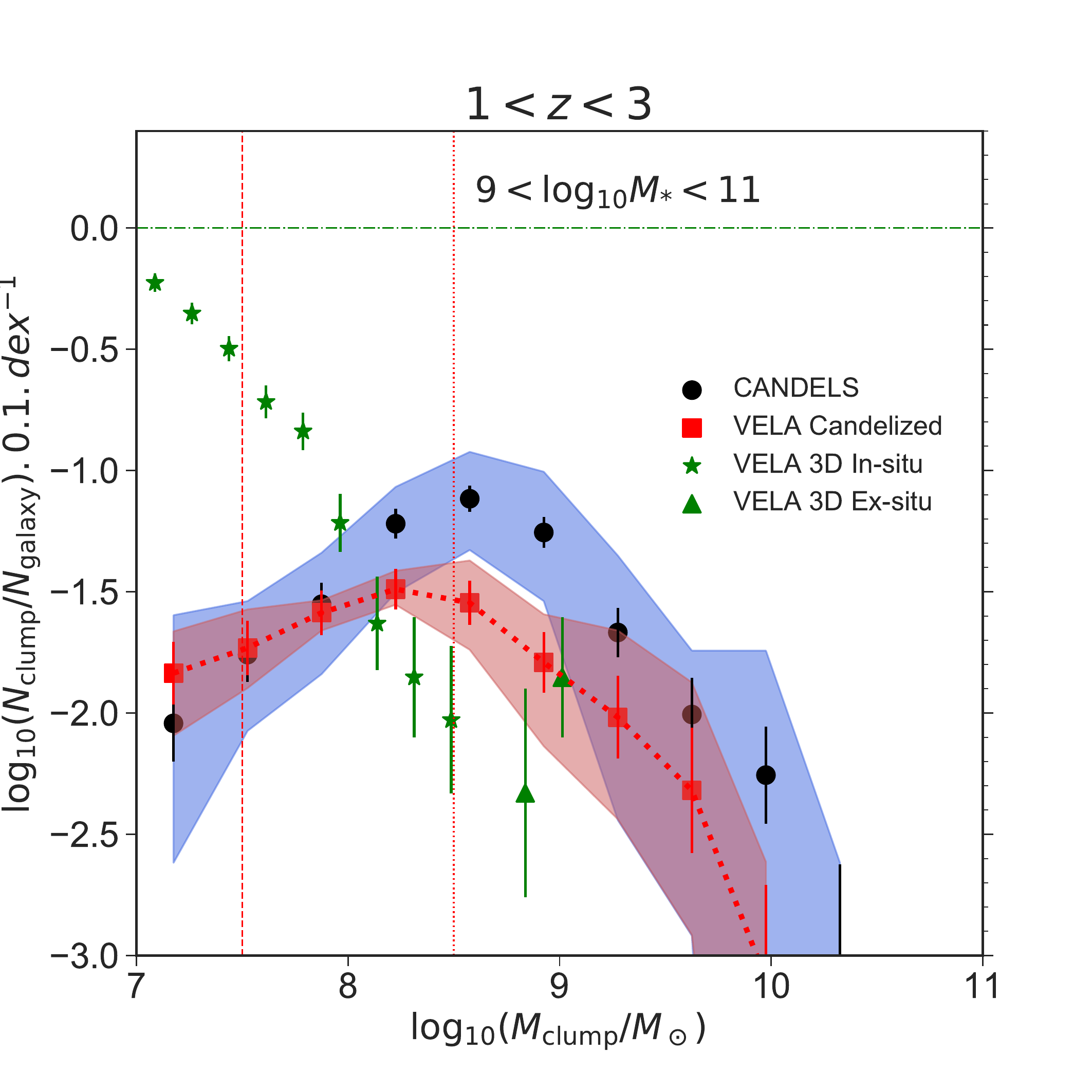}}%
         \qquad
          \subcaptionbox{Corrected with VELA prior}{\includegraphics[width=0.30\textwidth]{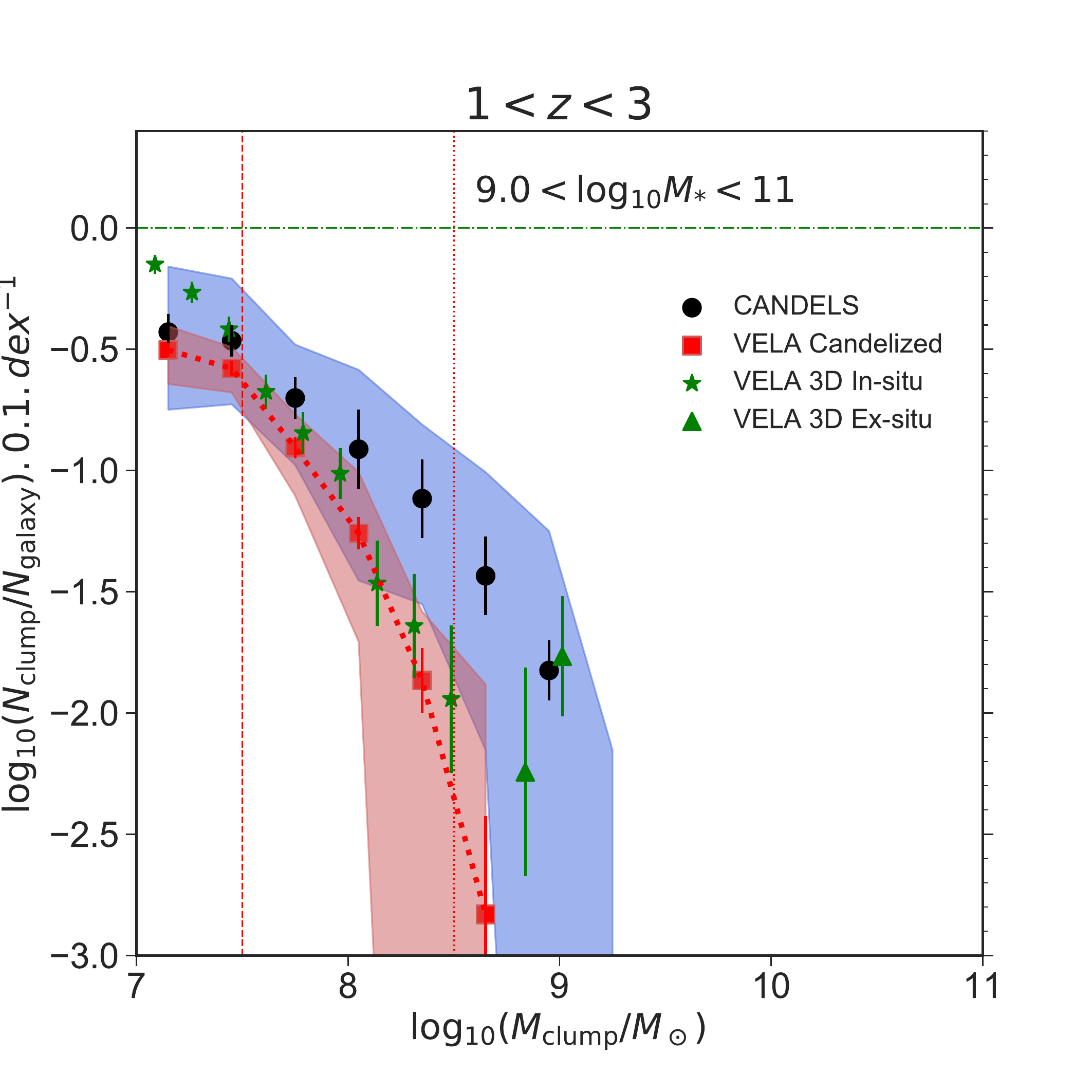}}%
          \qquad
          \subcaptionbox{Corrected with flat prior}{\includegraphics[width=0.30\textwidth]{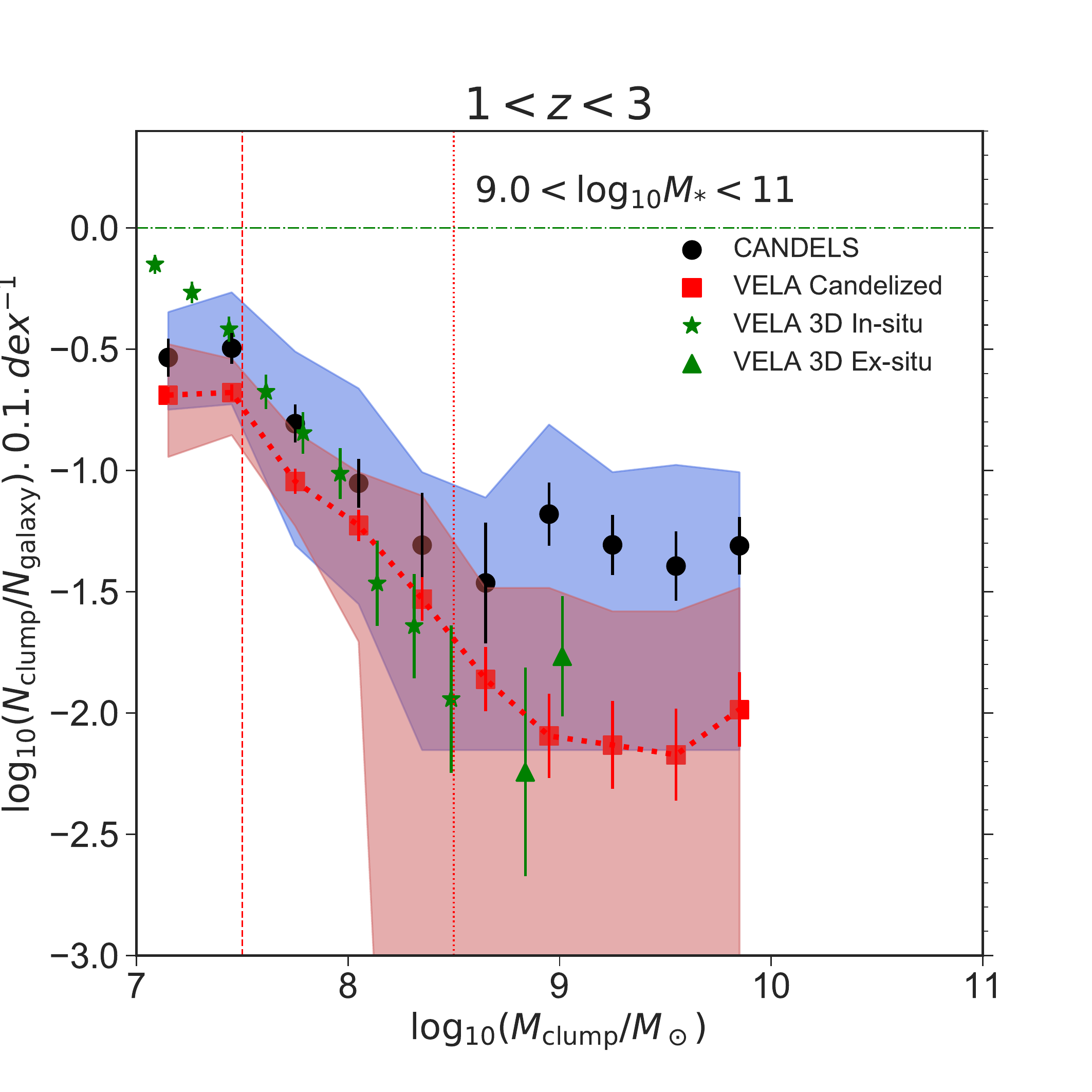}}%
            \qquad

   \caption{Clump stellar mass functions of simulated galaxies (red squares) and observations (black filled circles). (a)  Uncorrected stellar masses. (b) Corrected stellar masses with a mixture density network using the VELA sample as a training set (see text for details). (c) Corrected stellar masses with a flat prior. We include galaxies in the stellar mass range $9<\log_{10}M_*/M_\odot<11$ and with redshifts between 1 and 3. The black filled circles are the average measurement in CANDELS over a set of random samples with the same size and stellar mass and redshift distribution as the simulated sample. The shaded blue region indicates the confidence interval of the observations inferred through MonteCarlo sampling of the main observational sample (see text for details). The filled red squares are the measurements performed on the VELA Candelized images and the shaded red regions indicate confidence intervals from using different camera orientations and different neural network models for correction. The green stars (triangles) are the measurements in the VELA simulation using the simulation metadata in 3D (Mandelker et al. 2017) for in-situ (ex-situ) clumps. The red dashed and dotted vertical lines indicate clump masses of $10^{7.5}$ and $10^{8.5}$ solar masses respectively. The green dashed horizontal line indicates a reference value of one clump / galaxy. }
    \label{fig:cSMFs_VELA}
\end{figure*}

Before analyzing the clump stellar mass function, it is important to first calibrate the amount of information that can be recovered from the CANDELS-like images given the large uncertainties in the clumps stellar masses reported in the previous subsection. To that purpose, we first compare the cSMF derived using the detections performed on the Candelized VELA simulation with the clump stellar mass function obtained using the raw 3D output from the VELA simulations analyzed by~\cite{mandelker17}. This is shown in the left panel of figure~\ref{fig:cSMFs_VELA}. The values are normalized by the total amount of galaxies in a given bin. The sample only contains 35 galaxies.  However, since images are produced in every time step, using only one camera results in a sample of few hundred images in the redshift range of interest ($1<z<3$). For this exercise of building the mass functions, we consider each image as if it was an independent galaxy with a given stellar mass and redshift. Since the galaxy sample is still small, we compute the cSMF for all galaxies together ($9<\log_{10}(M_*/M_\odot)<11$). The shaded region indicates how the results change depending on the camera orientation used. \\

First of all, we see a dramatic difference between the cSMF derived with the Candelized images and the \emph{true} 3D based cSMF. The Candelized cSMF peaks at a clump mass of $\sim10^{8.5}$ while the 3D mass function shows a power-law increase at low masses without any flattening as opposed to the observations.  This big difference between 3D and 2D cSMFs is expected, given the several systematics that affect the measurements in the Candelized images reported in the previous subsections. One is obviously incompleteness. Because Candelized images are affected by noise, low mass clumps are not detected by our clump detector as already discussed in subsection~\ref{sec:compl}.  Additionally, clump blending can also contribute to lower the normalization of the cSMF.  The HST PSF is $\sim 1$ kpc in the redshift range of interest which means that on average one HST clump corresponds to $\sim$ three 3D clumps. Although completeness and blending can contribute to change the slope, they are not enough to change the shape of the cSMF so dramatically. A key and dominant effect comes from the systematic errors in the SED based estimations of clump stellar masses shown in figure~\ref{fig:mass_2d_3d}. Both the systematic clump mass overestimation and Eddington bias~\citep{1913MNRAS..73..359E} will tend to change the shape of the cSMF by over populating the high mass end and decreasing the slope as we seem to see. In appendix~\ref{sec:app} we show indeed that a combination of these effects can reconcile the intrinsic 3D cSMF with the observed one. This suggests that, with current data and no correction, it is difficult to use the shape of the clump stellar mass function as a constraining parameter for clump formation. 

The left panel of figure~\ref{fig:cSMFs_VELA} also includes the measurements obtained on observed CANDELS galaxies. Due to the small sample size and the VELA halo masses being roughly uniformly distributed in log mass, the VELA galaxies’ stellar mass distribution is different from that of our CANDELS mass-complete sample. To take that into account for a fair comparison, we create $100$ samples from the parent CANDELS sample of equal size as the VELA sample to reproduce the number, stellar mass and redshift distributions of the VELA sample. We call these samples the CANDELS-VELA samples. The cSMF for the CANDELS-VELA samples are plotted in the shaded red regions of figure~\ref{fig:cSMFs_VELA} which provide an estimate of the confidence interval. Although the observational points tend to lie above the simulated ones, an important result is that, overall, the simulated points lie within the confidence interval of the observations. Additionally, the shapes of the cSMFs are very similar. This means that the VELA simulations predict globally the correct number of clumps when compared with observations under comparable conditions. It also means that even if the observed cSMF 
differs from the true one, the same systematics seem to apply to both observations and simulations when the latter are forward modeled to the observational plane. This clearly demonstrates that it is of capital importance to compare observations and simulations under the same conditions, especially in the low S/N and low resolution regimes. It also highlights the limitations of current HST data to study the resolved structure of high redshift galaxies and the importance of future JWST observations.

\subsubsection{Corrected clump stellar mass function}

In this work, we go a step further and try to correct the estimated clump stellar masses to recover the intrinsic clump stellar mass function. The middle and right panels of figure~\ref{fig:cSMFs_VELA} show the corrected stellar mass functions using the neural network trained as detailed in section~\ref{sec:compl}. We show the results of the two different priors. We clearly see that the corrected observed cSMFs do agree significantly better with the intrinsic 3D mass function. This confirms that the correction applied is effective. In particular, the peak of the cSMF is now shifted to lower masses and the mass function presents a steep decrease between $10^7$ and $10^9$ solar masses. Between $\sim10^7$ and $\sim10^9$, the impact of the prior applied in the correction is rather small. Most of the differences are seen in the high mass end, where the flat prior correction tends to predict more clumps. The uncertainties are also large. As we have seen, the VELA prior prediction tends to underestimate the number of clumps above $10^{8.5-9}$ solar masses. The flat prior prediction will tend to compensate for the lack of training points on the massive end. By doing so, it will also tend to over estimate the number of massive clumps as the correction boosts the posterior distribution outside the training values. Overall, the exact abundances remain quite uncertain given the lack of training points. However, both corrections should bracket the range of possible values. 

As a way of quantifying how well the intrinsic cSMF can be recovered at least in the range of masses in which the correction is more stable, we also report in table~\ref{tbl:slopes} the slopes of the best power law model fits ($\log (dN/dM)=-\alpha\times \log M + \log const$) to the VELA 3D, VELA Candelized and VELA-CANDELS samples after the correction of the stellar masses is applied. We restrict to the mass range $10^{7.5}-10^{8.5}$ where the mass function is better constrained. Although the \emph{true} 3D clump mass function has a steeper slope than the measured with the corrected values, it is within the confidence intervals of the derived slopes with both the VELA and flat prior. This suggests that the correction enables us to recover to some extent the intrinsic slope and shape of the 3D cSMF. The reason why the recovered slope is still shallower than the intrinsic one might be related to both errors in the correction and also completeness and blending effects. We note however that the relative agreement between CANDELS and VELA is independent of the correction applied as both cSMFs agree even when uncorrected values are used. It is also worth noticing that the slope measured for the 3D cSMF is shallower than the value of $-2$ (see table~\ref{tbl:slopes}) even if the clumps are known to be formed in-situ. This is not entirely surprising. The measured slope is expected to be shallower than the \emph{birth} slope given that clumps of different masses have different lifetimes~\citep{2020MNRAS.497..698T}.

\begin{figure}
	\centering
	\includegraphics[width=0.45\textwidth]{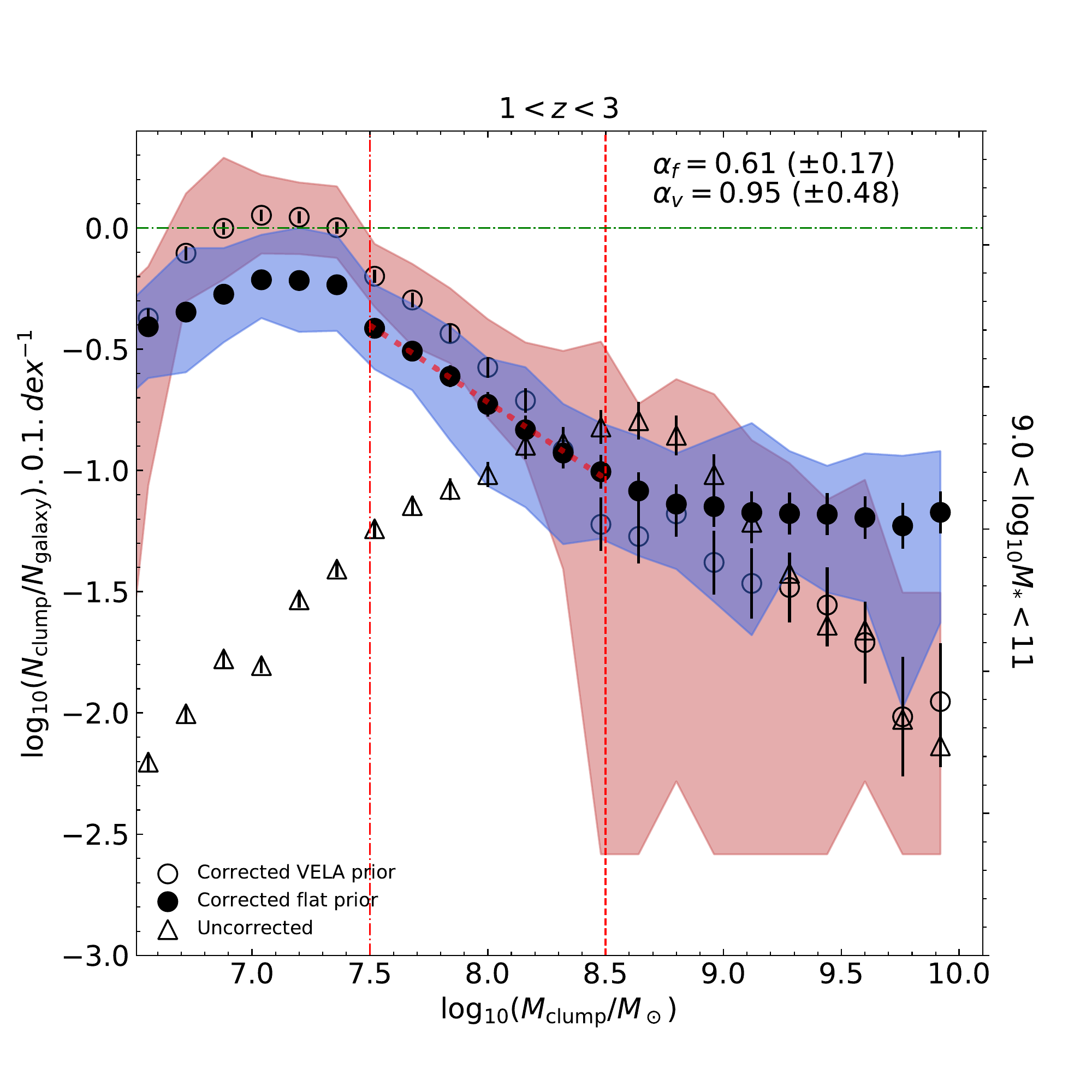}
    \caption{Clump stellar mass function in CANDELS. Empty triangles indicate uncorrected values. Filled circles indicate corrected values with a flat prior and empty circles show corrected clump masses using a VELA prior (see text for details). The y-axis is normalized with the total number of galaxies in that bin so that it provides an indication of the average number of clumps per galaxy. Error bars indicate Poisson uncertainties. The blue and red shaded regions indicate the range of solutions depending on the neural network model applied for correction. The dashed red line indicates the best power-law fit: $\log (dN/dM)=-\alpha\times \log M + \log const$ for clumps more massive than $10^{7}$ solar masses. The best fit value of the slopes $\alpha$ for  both the flat ($\alpha_f$) and VELA ($\alpha_v$) priors based corrections are given in table~\ref{tbl:slopes}. The dashed and dotted vertical red lines indicate clump stellar masses of $10^{7.5}$ and $10^{8.5}$ respectively. The dash-dotted green horizontal line is shown for reference and indicates a value of one clump / galaxy.  }
    \label{fig:clump_LF_mstar}
\end{figure}

Given that the correction applied seems to allow the recovery of the intrinsic cSMF at least in broad terms, we now move to explore the corrected cSMF using the whole CANDELS dataset. This is of course more risky, as the CANDELS sample necessarily contains more galaxies not well covered by the VELA training set,
so the results need to be taken with caution. This is shown in figure~\ref{fig:clump_LF_mstar}. We show both the uncorrected and corrected cSMFs with the two different priors. The shaded region in the figure shows the range of solutions obtained using $50$ different neural network models and the black points are the average values. 

First of all we see that, although there is some range of possible solutions, the global shape of the mass function is consistent across the different neural network models. Overall, the correction has a similar effect on the cSFMs as the one reported for VELA which is reassuring. Namely, the peak of the cSMFs is displaced towards lower masses from $\sim10^{8.5}$ to $\sim10^{7.5}$. Overall, the cSMF presents a steep increase with decreasing clump mass until  reaching a peak close to the $10^{7}$ solar masses for corrected masses and starts declining again at lower masses with a shallower slope. The peak of the cSMF is most probably due to incompleteness as it matches reasonably well the value estimated in figure~\ref{fig:VELA} and there is no particular physical reason to expect a flattening at the low mass end. 

Another interesting result is that regardless of the correction applied, the majority of the clumps have stellar masses lower than $10^9$ solar masses. The abundances of massive clumps are poorly constrained though, as they depend on the prior used to correct masses. The correction using flat prior tends to estimate large abundances of massive clumps. As previously explained, this is an upper limit, as the correction applied boosts the posterior distribution beyond the training regime. However, the fact that the cSMF tends to peak at masses close to $10^{7}$ is a robust result that is not significantly affected by the correction applied. This order of magnitude for the typical clump masses is also in better agreement with recent independent measurements of clump masses using gravitational lensing techniques. 

 \subsubsection{Implications for the formation and origin of clumps}

Although significantly affected by uncertainties, the derived clump stellar mass function and the comparison with simulations allow us to speculate about the nature of giant clumps in high redshift galaxies. 

 First of all, our results suggest that the majority of the clumps detected in high redshift galaxies are less massive than $10^9$ solar masses. Although our results do not allow us to put strong constraints on the abundances of very massive clumps, it seems clear that the bulk of the population has lower masses. This is an important assessment, as some previous works have estimated that high redshift clumps are frequently more massive than $10^{9}$ solar masses (e.g.~\citealp{fs11b, 2019MNRAS.489.2792Z}). Our results suggest that these measurements might have been affected by similar systematic biases raised in this work, as pointed out for example in~\cite{2018NatAs...2...76C} using gravitationally lensed data.  The different selections can perhaps also explain some of the differences, as the aforementioned works do not analyze complete samples and might have biased their selection towards galaxies with prominent clumps. 

Secondly, our results also show that the simulations globally reproduce the shape of the observed cSMF irrespective of any correction applied. This agreement between simulated and observed cSMFs potentially puts some constraints on the origin of clumps detected in the observations. In the VELA simulations, the majority of the clumps are formed in-situ and therefore ex-situ clumps alone could not account for the agreement observed in figure~\ref{fig:cSMFs_VELA}. Furthermore, ex-situ clumps are predicted to be significantly more massive than in-situ clumps. This implies that the majority of observed clumps in CANDELS must have an in-situ origin. In order to further assess this, we perform a match of the 2D Candelized clumps with the 3D clumps as done in subsection~\ref{sec:compl} and flag all 2D clumps which correspond to an ex-situ clump in the simulation. We find that only 5 clumps (4 at $z<1$ and 1 at $z>2$) among the 186 detected clumps are flagged as ex-situ. This represents $\sim2\%$ of the clumps and has basically no effect in the derived cSMF. Therefore the observed agreement between observations and simulations must be driven by in-situ clumps. This is the first time that evidence of an in-situ origin for most clumps is shown through direct comparison of observations with numerical simulations. 

\begin{figure*}     
    	 \subcaptionbox{Uncorrected}{\includegraphics[width=0.30\textwidth]{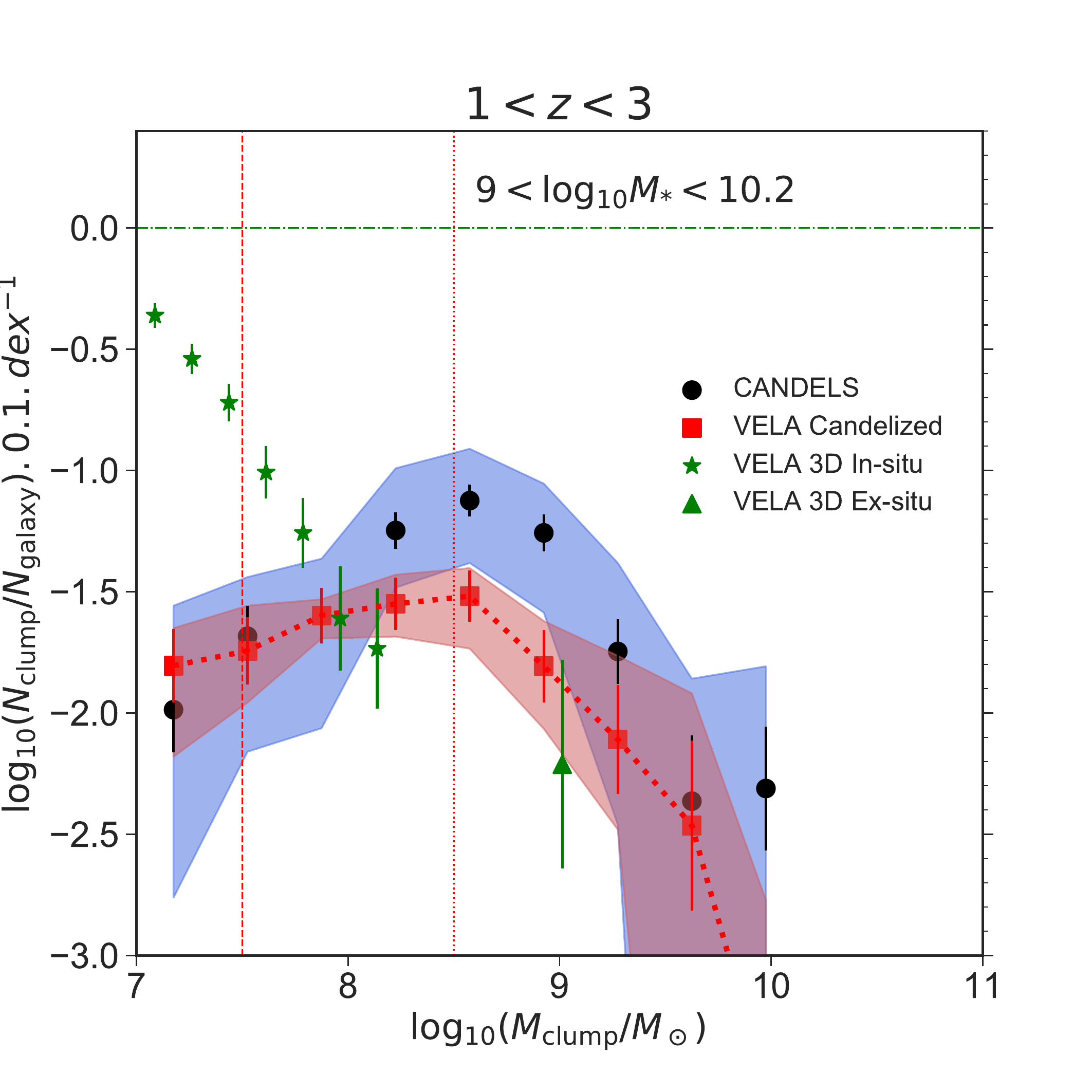}}%
         \qquad  
          \subcaptionbox{Corrected with VELA prior}{\includegraphics[width=0.30\textwidth]{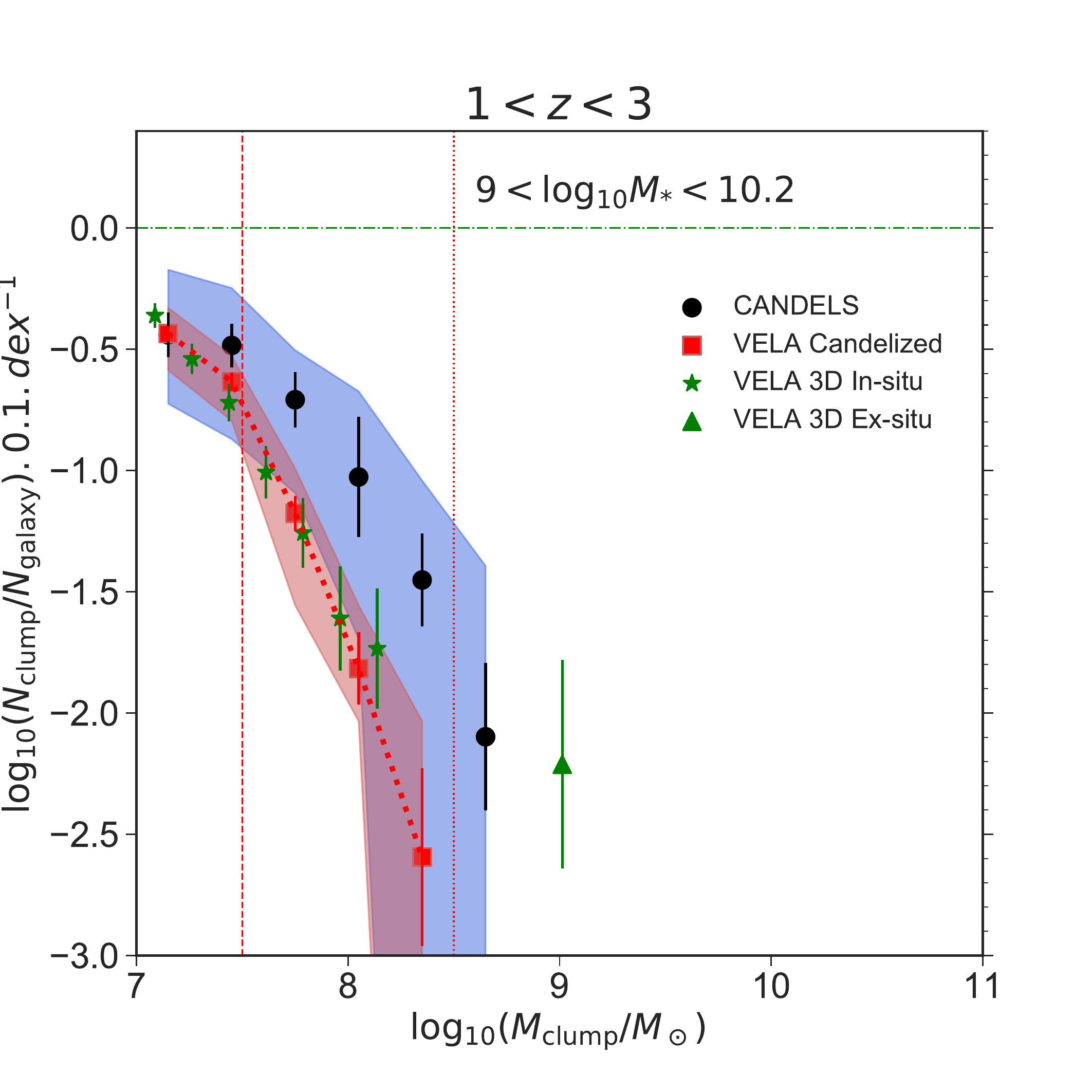}}%
          \qquad
          \subcaptionbox{Corrected with flat prior}{\includegraphics[width=0.30\textwidth]{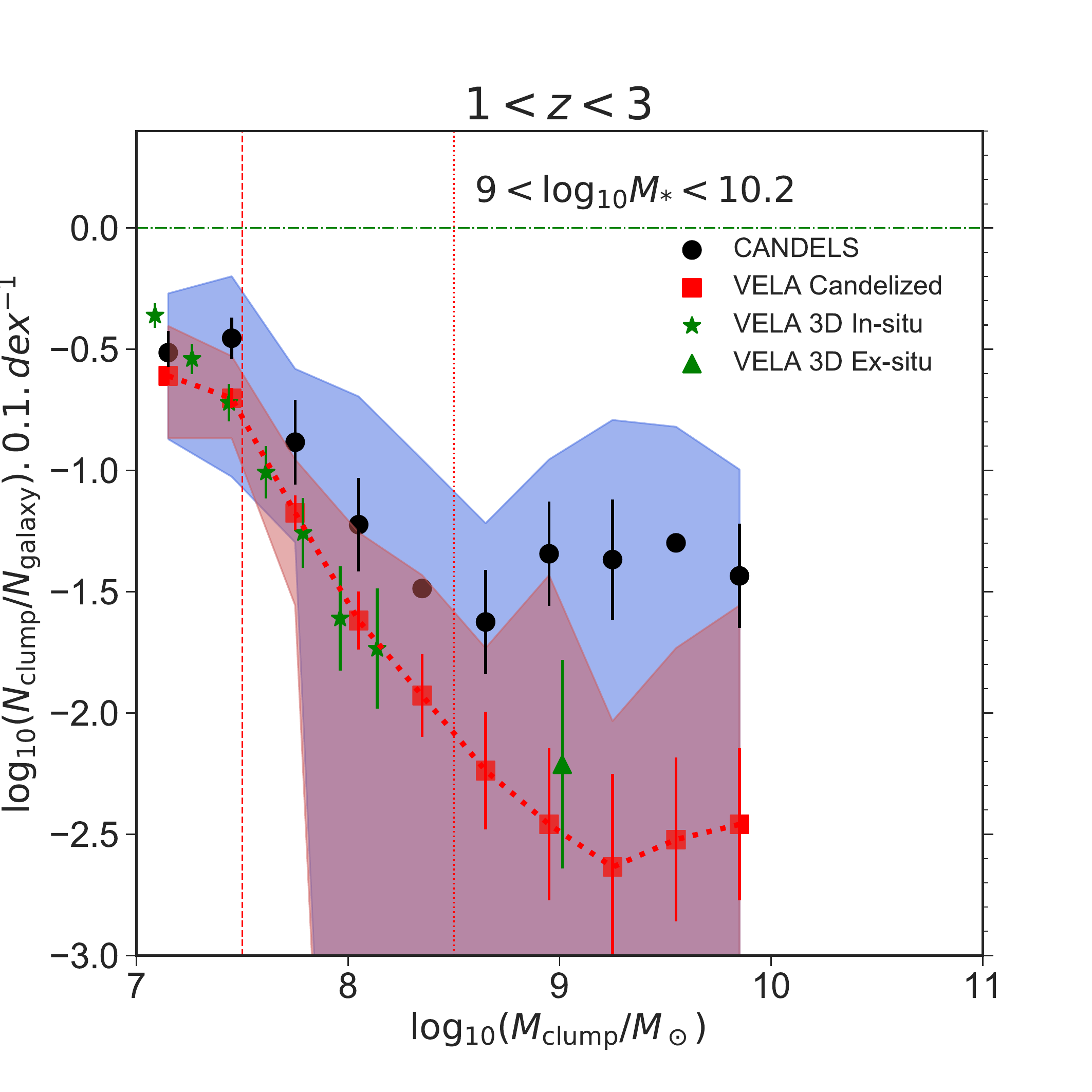}}%
            \qquad
                    	 \subcaptionbox{Uncorrected}{\includegraphics[width=0.30\textwidth]{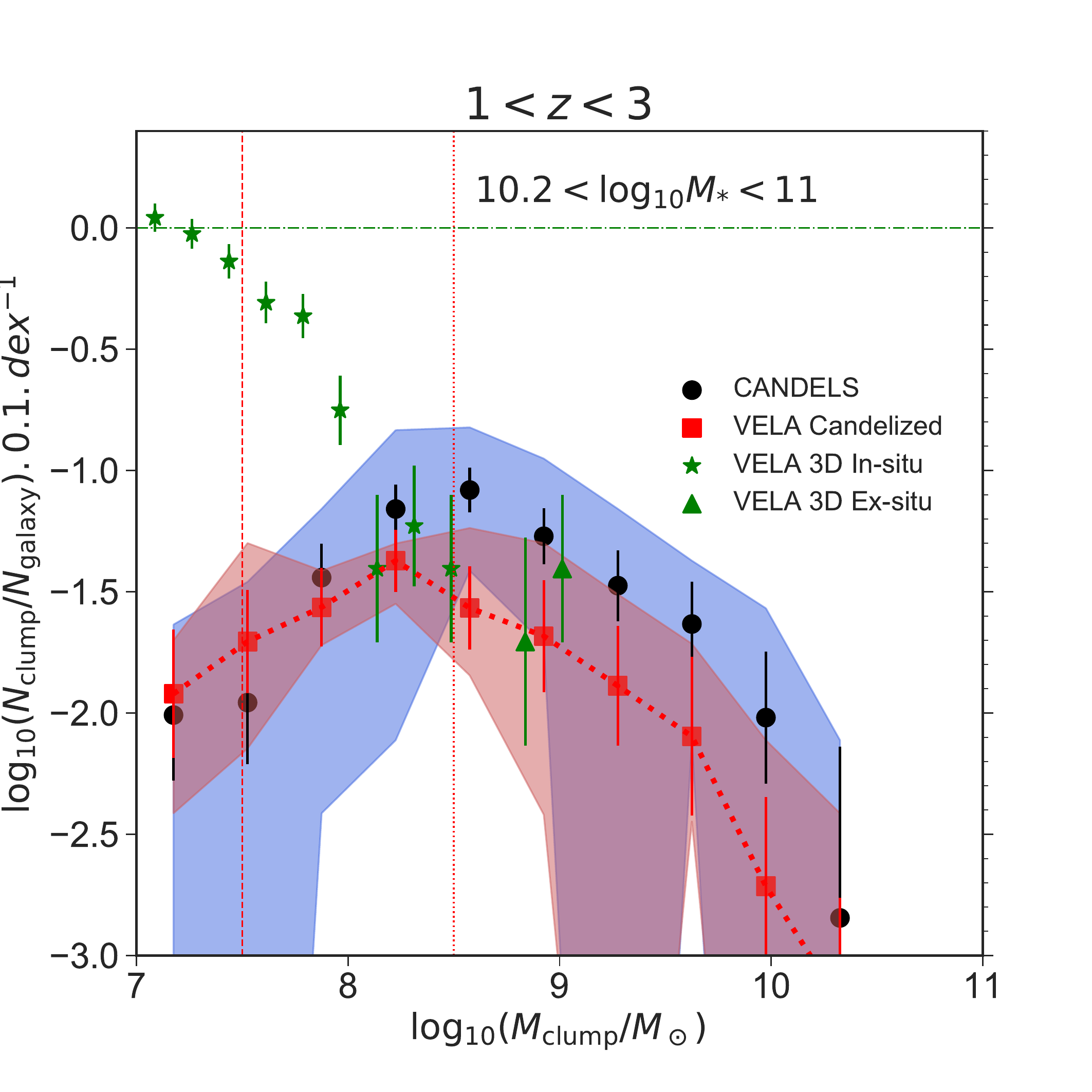}}%
         \qquad
          \subcaptionbox{Corrected with VELA prior}{\includegraphics[width=0.30\textwidth]{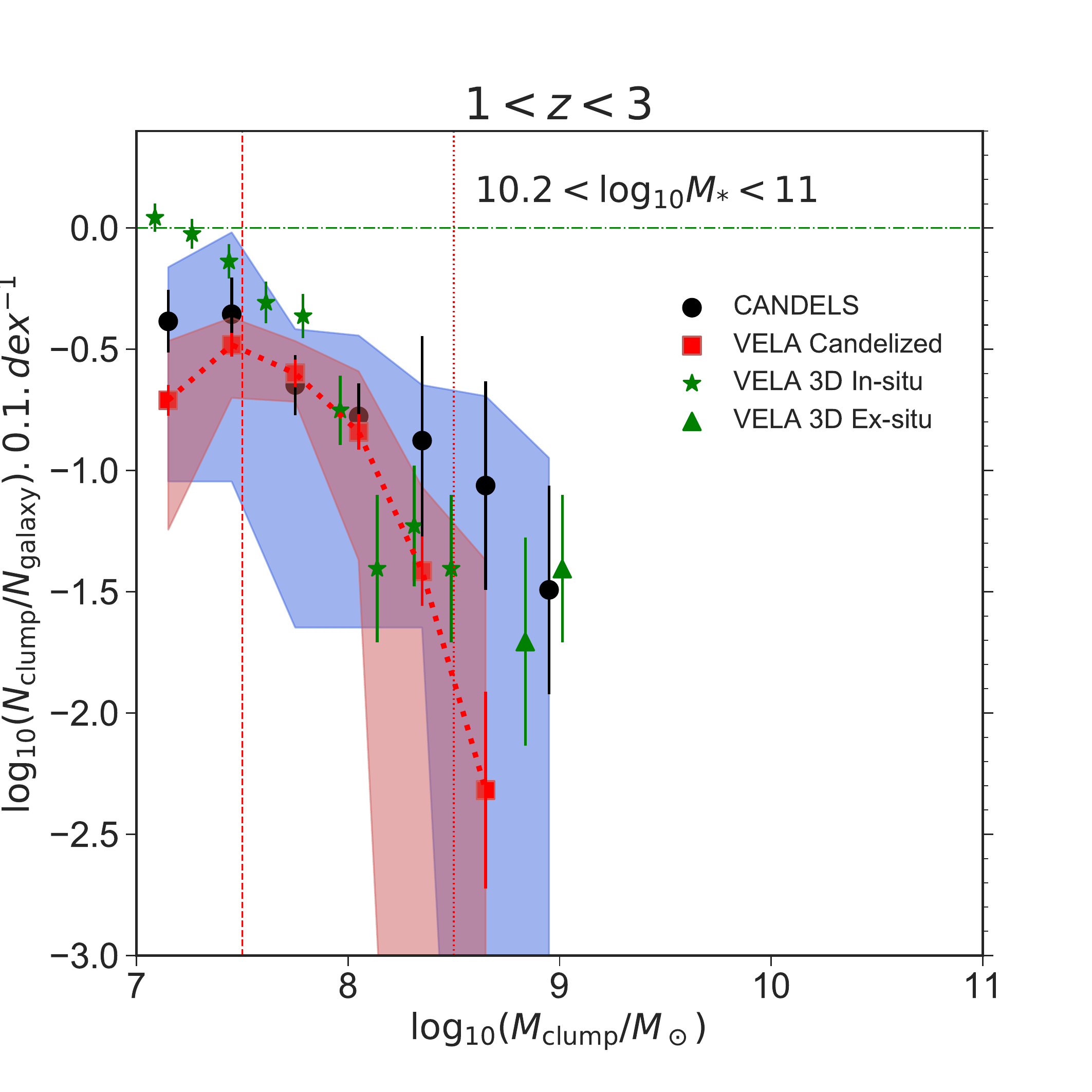}}%
          \qquad
          \subcaptionbox{Corrected with flat prior}{\includegraphics[width=0.30\textwidth]{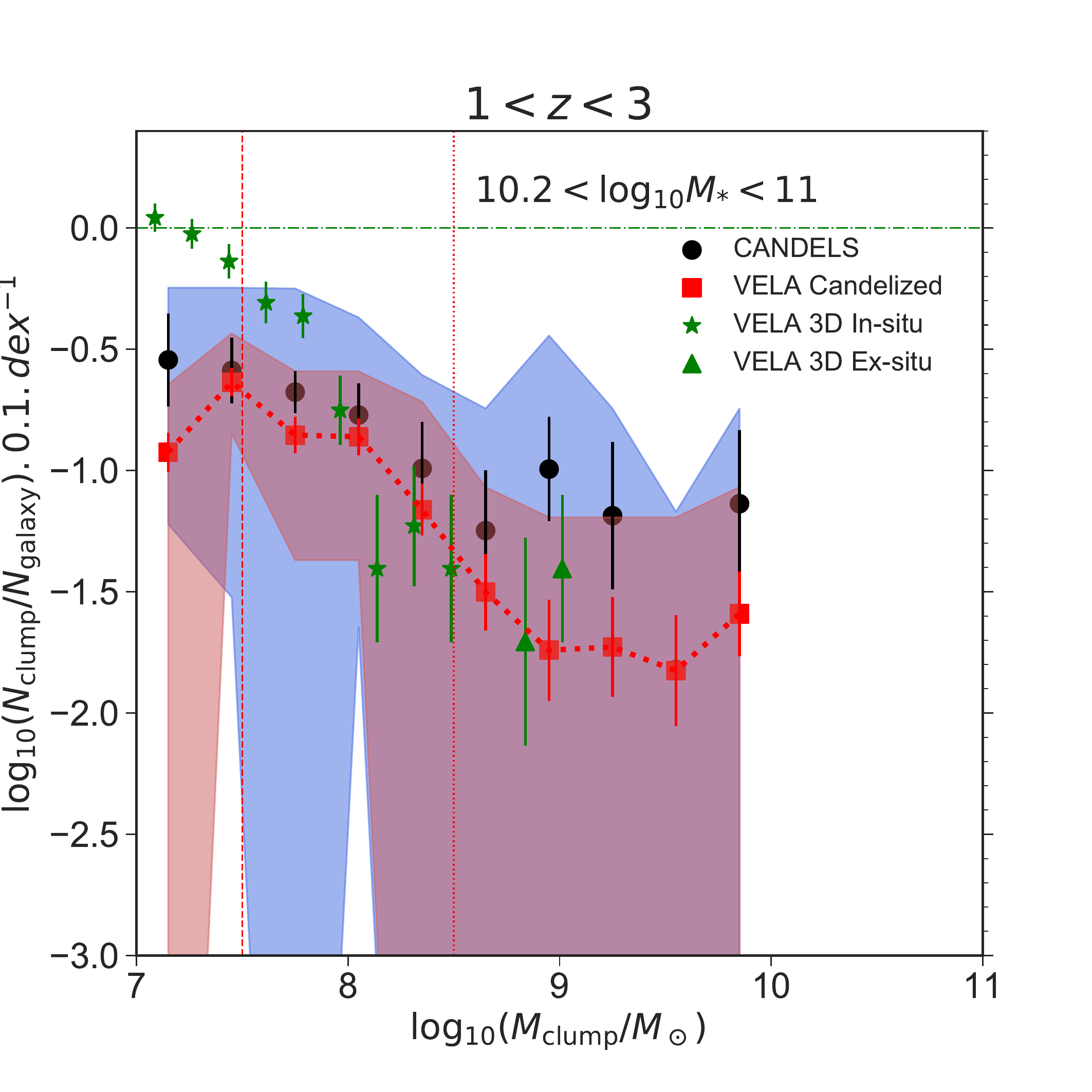}}%
           \qquad
   \caption{Same as figure~\ref{fig:cSMFs_VELA} but divided in bins of galaxy stellar mass. The top row shows clumps in low mass galaxies ($9< \log(M_*/M_\odot<10.2$). The bottom row shows massive galaxies ($10.2< \log(M_*/M_\odot<11$). ~Since the detection of clumps is less complete for massive galaxies, we measure a shallower slope of the cSMF in massive galaxies (see text for details). }
    \label{fig:cSMFs_VELA_massbins}
\end{figure*}

 An interesting question related to the clump origin is whether it depends on the galaxy properties such as stellar mass. Given that our results seem to point towards an in-situ origin of clumps, one might wonder if the galaxy properties affect the shape of the cSMF. Although our sample is small, we address this issue in figure~\ref{fig:cSMFs_VELA_massbins} where we show the cSMFs for simulated and observed galaxies in two broad galaxy stellar mass bins. The different slope measurements are also reported in table~\ref{tbl:slopes}. The effect of galaxy stellar mass on the intrinsic 3D cSMF is very mild, suggesting that galaxy stellar mass has little effect on the properties of formed clumps. This is not entirely surprising. The work by~\cite{inoue16} shows that external events such as minor mergers often trigger the formation of clumps. However, we see that the slopes derived from the 2D Candelized data tend to be shallower in massive galaxies. We argue that this is most likely related to incompleteness. As discussed in section~\ref{sec:compl}, lower mass clumps detected in massive galaxies are less complete. This implies a decrease of the cSMF at the low mass end as seen in figure~\ref{fig:cSMFs_VELA_massbins}, which in turn results in a shallower slope. We do not think it has a physical origin as the effect is not seen in 3D. It illustrates how completeness can affect the measurement of the slope.

 \begin{table}
\begin{center}
\begin{tabular}{ccc} 
\hline
 & $-\alpha$ w/ VELA prior & $-\alpha$ w/ flat prior  \\
 \hline
$9< \log(M_*/M_\odot)<11$ && \\
\hline 
VELA in-situ  & $-1.35\pm0.15$ &$-1.35\pm0.15$  \\ 
VELA Candelized & $-1.55\pm0.34$ & $-0.79\pm0.54$  \\ 
CANDELS (VELA sample) & $-0.71\pm.0.14$ &$-0.72\pm0.13$  \\ 
CANDELS (All sample) & $-0.95\pm0.48$ & $-0.61\pm0.17$  \\ 
\hline
$9< \log(M_*/M_\odot)<10.2$ && \\
\hline 
VELA in-situ & $-1.49\pm0.02$ & $-1.49\pm0.02$ \\
VELA Candelized  & $-1.06\pm0.76$ & $-1.25\pm0.52$  \\ 
CANDELS (VELA sample) & $-0.45\pm.0.21$ & $-1.00\pm0.23$  \\ 
\hline
$10.2< \log(M_*/M_\odot)<11$ && \\
\hline 
VELA in-situ & $-1.38\pm0.07$ & $-1.38\pm0.07$ \\
VELA Candelized  & $-0.80\pm0.38$ & $-0.51\pm0.59$  \\ 
CANDELS (VELA sample) & $-0.38\pm0.19$ & $-0.52\pm.0.37$  \\ 
\end{tabular}
\end{center}
\caption{Measured slopes of the best power-law fit ($\log (dN/dM)=-\alpha\times \log M + \log const$) to the clump stellar mass function for different samples. The columns indicate two different corrections, using a flat or VELA prior (see text for details).  The first row reports the slope measured using the 3D identified clumps in the VELA simulation. The second row shows the slope measured when the corrected Candelized measurements are used on VELA. The third and forth rows indicate the slopes for observed galaxies sampled according to the VELA distribution and for all CANDELS data respectively. We show the results for all galaxies and also divided in two broad stellar mass bins as labeled. }
\label{tbl:slopes}
\end{table}

\subsection{Contribution of clumps to stellar mass}
\label{sec:clump_contrib}

In order to better investigate how the clump abundances depend on different galaxy properties without being affected by small numbers, we adopt a summary statistic which measures the total fraction of stellar mass ($C_*$) contained in clumps. $C_*$ is directly related to the integral of the cSMF and can be estimated by dividing the total clump stellar mass by the total galaxy mass in a given sample. For the sake of clarity we only report results using corrected masses with a flat prior, although the main trends do not change if a VELA prior is used. We therefore set here a lower limit of $10^7$ solar masses for the clumps to avoid incompleteness issues. 

\subsubsection{Observations}

\begin{figure*}
	\centering
	 \subcaptionbox{}{\includegraphics[width=0.31\textwidth]{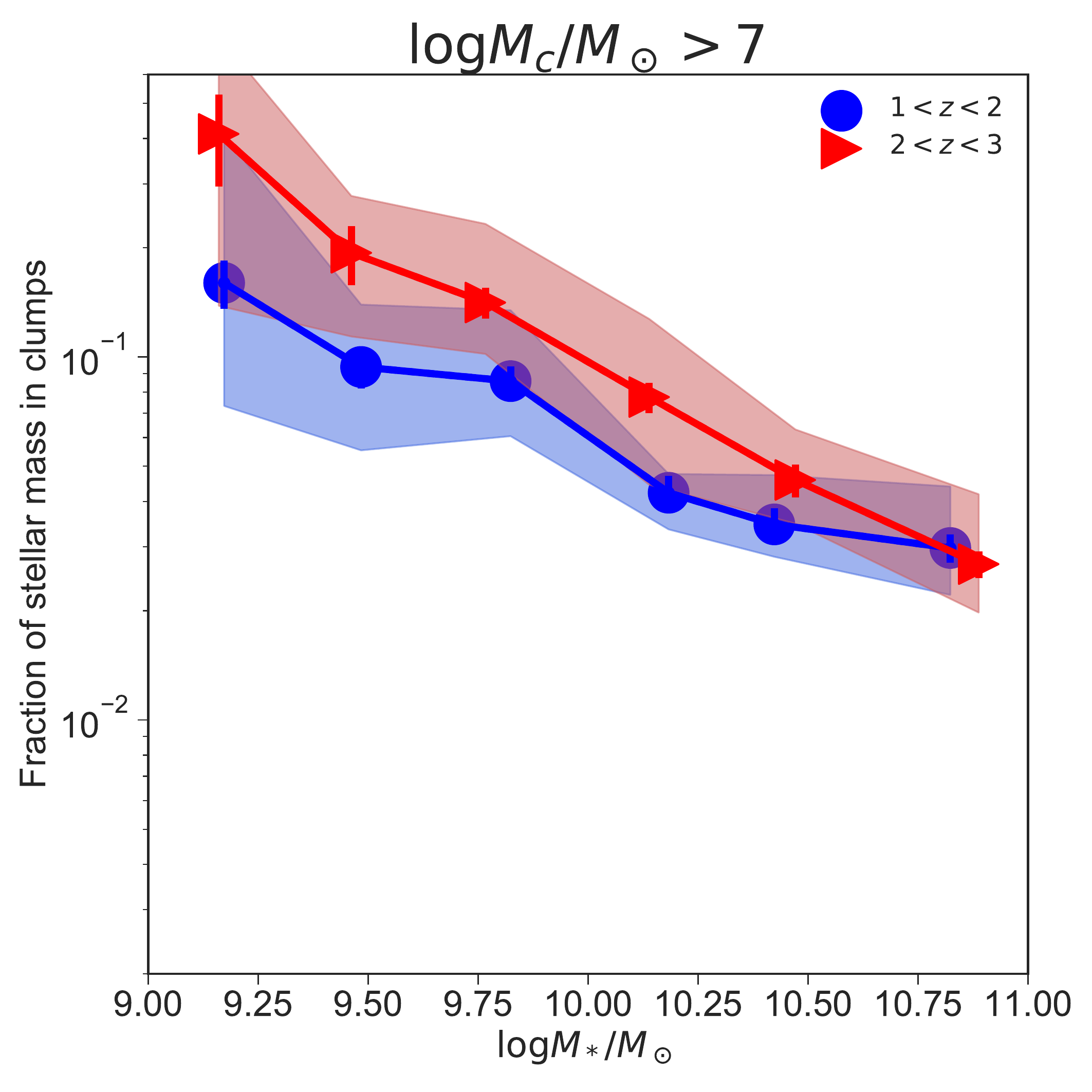}}%
         \qquad
          \subcaptionbox{}{\includegraphics[width=0.31\textwidth]{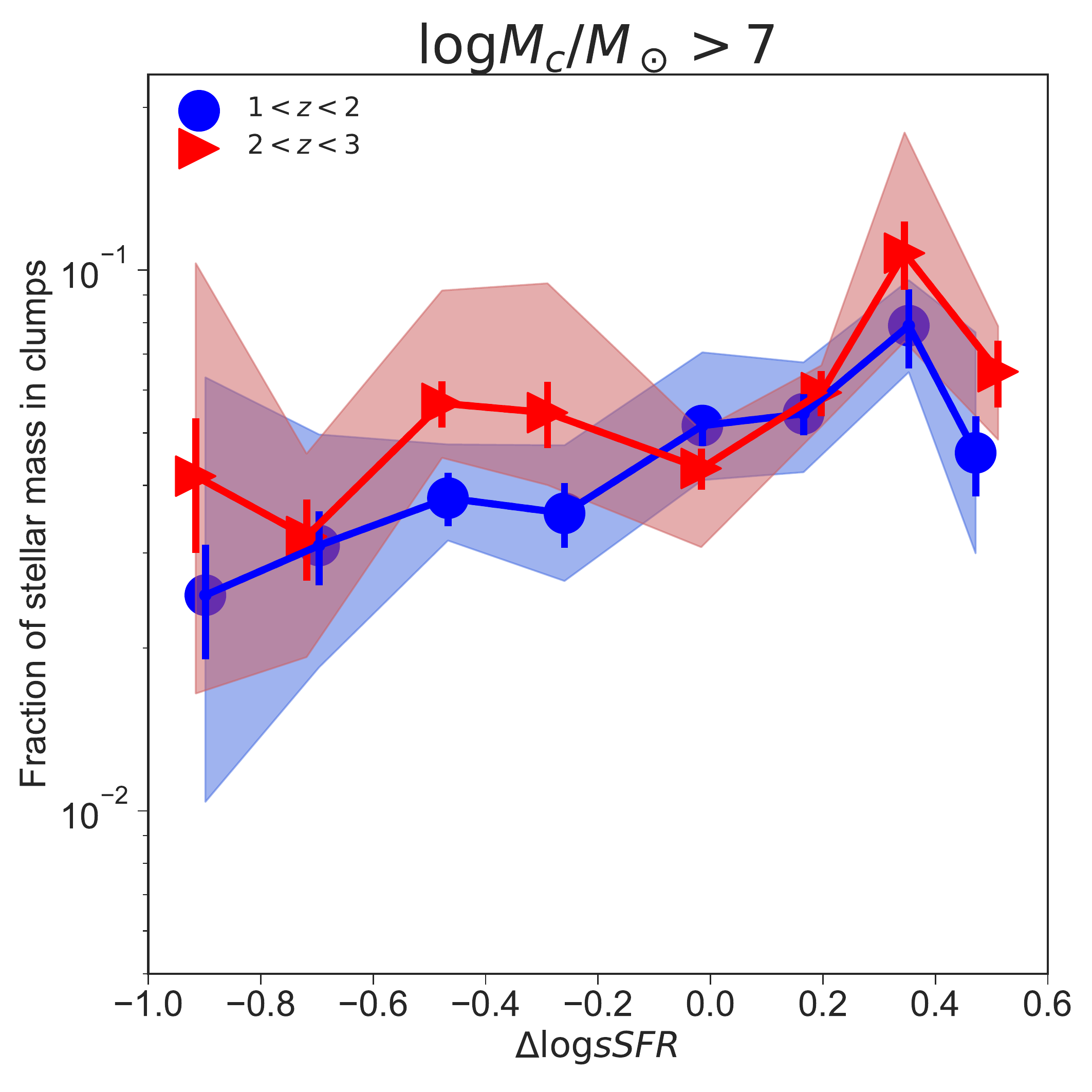}}%
         \qquad
 \subcaptionbox{}{\includegraphics[width=0.31\textwidth]{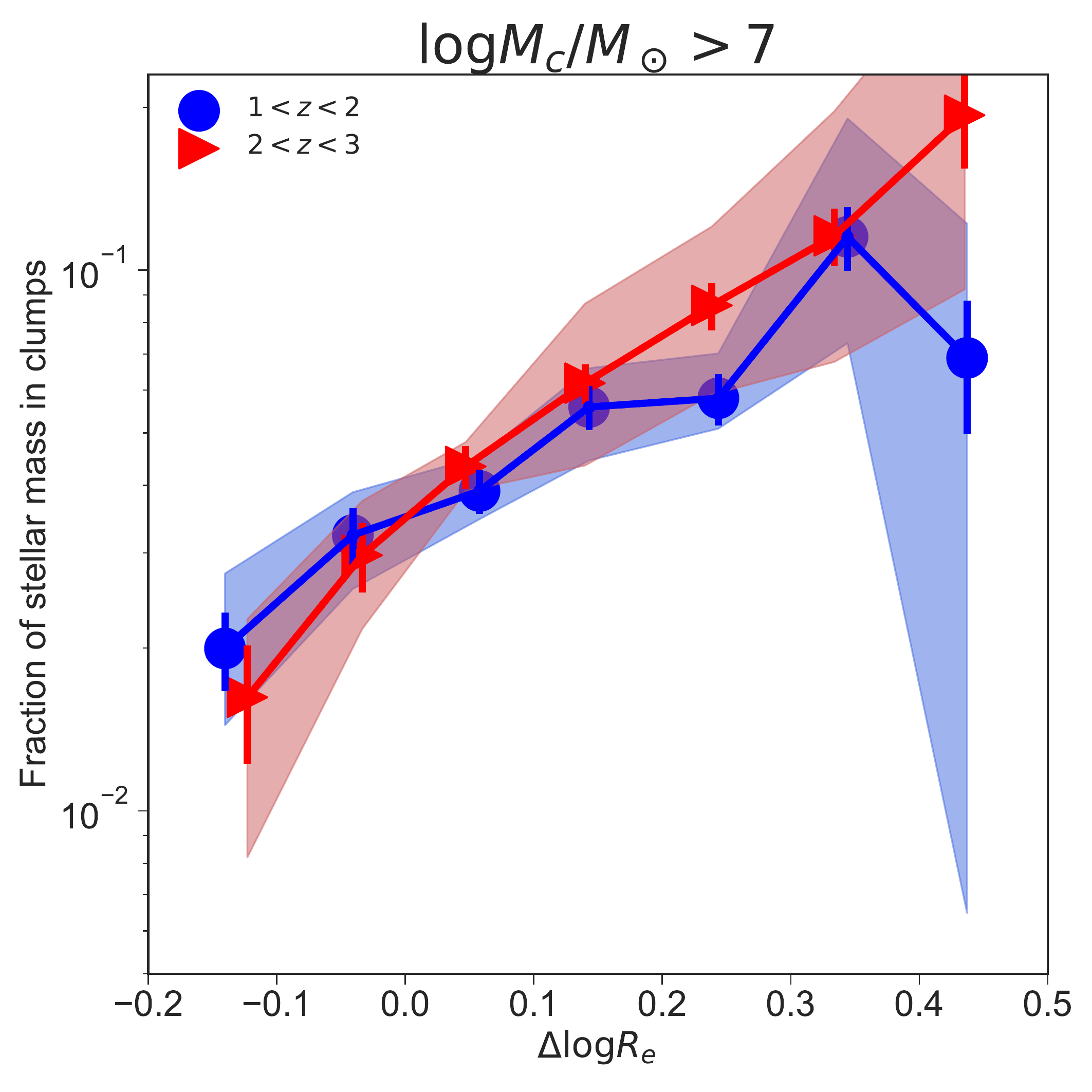}}%
    \qquad
    \caption{Total fraction of stellar mass in clumps more massive than $10^{7}$ solar masses in CANDELS star-forming galaxies as a function of (a) stellar mass, (b) $\Delta\log sSFR$,  and (c) $\Delta \log R_e$. Each color/symbol shows a different redshift bin as labelled. Error bars are computed through multiple samples of the stellar mass posterior distributions. Shaded regions are model uncertainties due to the correction applied to the clump stellar masses. On average clumps account for $\sim 2\%-5\%$ of the total stellar mass. Clumps tend to represent a larger fraction of the stellar mass in low mass, high specific star formation rate, and large radius galaxies. }
    \label{fig:clump_mass_contrib}
\end{figure*}

The left panel of figure~\ref{fig:clump_mass_contrib} shows $C_*$ as a function of galaxy stellar mass and redshift for all our CANDELS sample.  Error bars are computed by performing $100$ MonteCarlo simulations through the sampling of the posteriors of the clump stellar masses. For every iteration we sample the posterior to associate a stellar mass to every clump and recompute $C_*$. The reported values of $C_*$ are then the median values of the different realizations and the error bars correspond to the standard deviation from the different samples. The shaded regions show the range of values depending on the neural network model used to correct the stellar masses among the 100 realizations. The figure first shows that $C_*$ decreases with increasing galaxy mass. For galaxies more massive than $10^{10}$ solar masses, clumps contain on average between $\sim2\%$ and $\sim5\%$ of the total stellar mass. The fraction increases for low mass galaxies, reaching $10-15\%$ although the uncertainties are large. This large uncertainty at low masses is driven by low statistics and the uncertainties regarding the amount of massive clumps reported in the previous section. Higher redshift galaxies tend also to have larger fractions of mass in clumps since the red points in figure~\ref{fig:clump_mass_contrib} are systematically above the blue ones. 

The average value of $\sim2-5\%$ of the stellar mass seems to be smaller than values  reported in the literature with smaller samples too. \cite{fs11b} quotes for example fractions of $\sim10-20\%$ of the galaxy stellar mass in clumps using a sample of 6 galaxies. In the recent work by~\cite{2019MNRAS.489.2792Z}  the authors analyzed $\sim50$ star-forming galaxies and found that $\sim20\%$ of the stellar mass is in compact clumps which should be comparable to our selection. There are several reasons why our measurements are smaller. First of all, and most importantly, we are correcting the clump stellar masses for the overestimation reported in section~\ref{sec:compl}. This correction primarily reduces the clump stellar masses by an order of magnitude, therefore reducing their contribution to the total galaxy mass. We assume that previous measurements could have suffered from similar overestimations and thus reported larger values. However, even if we use uncorrected clump mass measurements, $C_*$ typically reaches values of $\sim5\%-7\%$ which is still smaller than other reported values. Another factor is that our analysis is made on a complete sample of galaxies ($1,500$ as opposed to a few tens). Our sample thus contains clumpy galaxies but also galaxies which do not host any clump, so the overall stellar mass fraction decreases. \cite{wuyts12} measured indeed a value of $C_*\sim7.5\%$ using both clumpy and non-clumpy galaxies which is in better agreement with our uncorrected measurements. Additionally the ~\cite{2019MNRAS.489.2792Z} sample for example  is dominated by galaxies with stellar masses lower than $10^{10}$ solar masses for which we also measure a larger contribution of clumps (figure~\ref{fig:clump_mass_contrib}(a)). Approximately $10\%$ of their sample is made of starbursts which can also boost the obtained stellar mass fractions. 

Our mass complete sample allows us to also investigate how the abundance of clumps depends on  galaxy properties, which is more difficult with incomplete samples such as the ones usually explored in the previous works. We focus here on effective radius ($R_e$) and specific-star formation rate (sSFR). There are indeed well established scaling relations between stellar mass and size and stellar mass and star-formation rate (the star-formation main sequence) which have been extensively studied \citep[e.g.][]{2014ApJ...788...28V, 2012ApJ...754L..29W} and are thought to be central for describing major evolutionary tracks of galaxies (e.g.~\citealp{2014ApJ...791...52B, 2017MNRAS.470..651R, 2019arXiv190910817C,2019arXiv191010947L}). Our goal is to investigate whether clumps can provide clues about the physics of galaxies along these tracks. 

We therefore start by analyzing whether clumps are more likely to be formed in galaxies which deviate from these median scaling relations. We adopt here $\Delta\log sSFR$ and $\Delta\log R_e$ as main proxies, which precisely measure how far the specific star formation rate and the effective radius are from the median $\log_{10} (M_*/M_\odot)$-$\log_{10} sSFR$ and $\log_{10} (M_*/M_\odot)$-$\log_{10} R_e$ relations. $\Delta$ values have the advantage of taking into account the mass dependence and allow us to explore the average relation without further binning in galaxy stellar mass. We use the best fit relations from \cite{2018ApJ...858..100F} as reference values for the CANDELS galaxy scaling relations. Positive / negative values of  $\Delta$ values therefore indicate galaxies above / below the median scaling relation at fixed stellar mass. 

The middle and right panels of figure~\ref{fig:clump_mass_contrib} show the dependence of $C_*$ with $\Delta\log sSFR$ and $\Delta\log R_e$ respectively. First we see that there is very little dependence of the clump contribution with the relative position of galaxies in the main sequence. The value of $C_*$ is almost constant, although there is a hint of a slight increasing trend in galaxies with a larger sSFR than the average of the same mass. Since the sSFR is expected to correlate with the gas mass density, one would naturally expect a larger contribution of clumps in galaxies above the main sequence. Several reasons can account for this weak correlation. Errors in both quantities (clump mass fractions and sSFR)  can wash out the trend. The difference in the star-formation rates in our sample might not be large enough to notice a strong effect on the clump properties (recall we select only galaxies with $\log sSFR>-10$). Some theoretical works (e.g.~\citealp{inoue16}) have also shown that clump formation is generally triggered by external perturbations such as minor mergers. This can also contribute to explain the lack of correlation with the main sequence offset. We also note that this weak correlation is independent of the galaxy stellar mass. If the sample is restricted to massive galaxies only we see a similar trend. 
 
In the right panel of figure~\ref{fig:clump_mass_contrib} we plot the dependence of  $C_*$ with  $\Delta\log R_e$. We observe here a clear dependence. Galaxies with larger effective radii than the average in galaxies of the same stellar mass have a higher fraction of their mass in clumps reaching up to $\sim8\%$ at $z>2$ which is three times the value for smaller galaxies. We recall that the stellar mass dependence has been removed. If this increasing trend was driven by different mass distributions, the opposite trend would have been expected, since large galaxies are also more massive and thus $C_*$ is smaller. This result is therefore suggesting that clumps are perhaps more efficiently detected in large galaxies. This is interesting, because large and small galaxies of the same mass are expected to have similar gas mass densities given the lack of size gradients in the star-formation main sequence~\citep{2019arXiv191010947L} so, if anything, large galaxies should have lower densities. The fact that clumps seem to represent a higher fraction of the stellar mass in large galaxies cannot be explained only by global differences in the gas densities and must be a consequence of other physical processes or some sort of observational bias. Limited spatial resolution could for example contribute to such a trend. Small galaxies are indeed less well resolved and it is therefore harder to detect clumps in those systems. We recall however that we have already selected galaxies with effective radii larger than $0.2$ arcsec (4 pixels). We also exclude all clumps in the central regions ($< 0.5R_e$) which should be more affected by limited spatial resolution in small galaxies.  A visual inspection shows that even the smallest galaxies in our selection appear to be well resolved. We hence think that this increasing trend is not driven by central clumps not being detected in small galaxies.  We investigate this further in the following sections.

\subsubsection{Simulations}

\begin{figure*}
    \centering
\centering
	 \subcaptionbox{}{\includegraphics[width=0.31\textwidth]{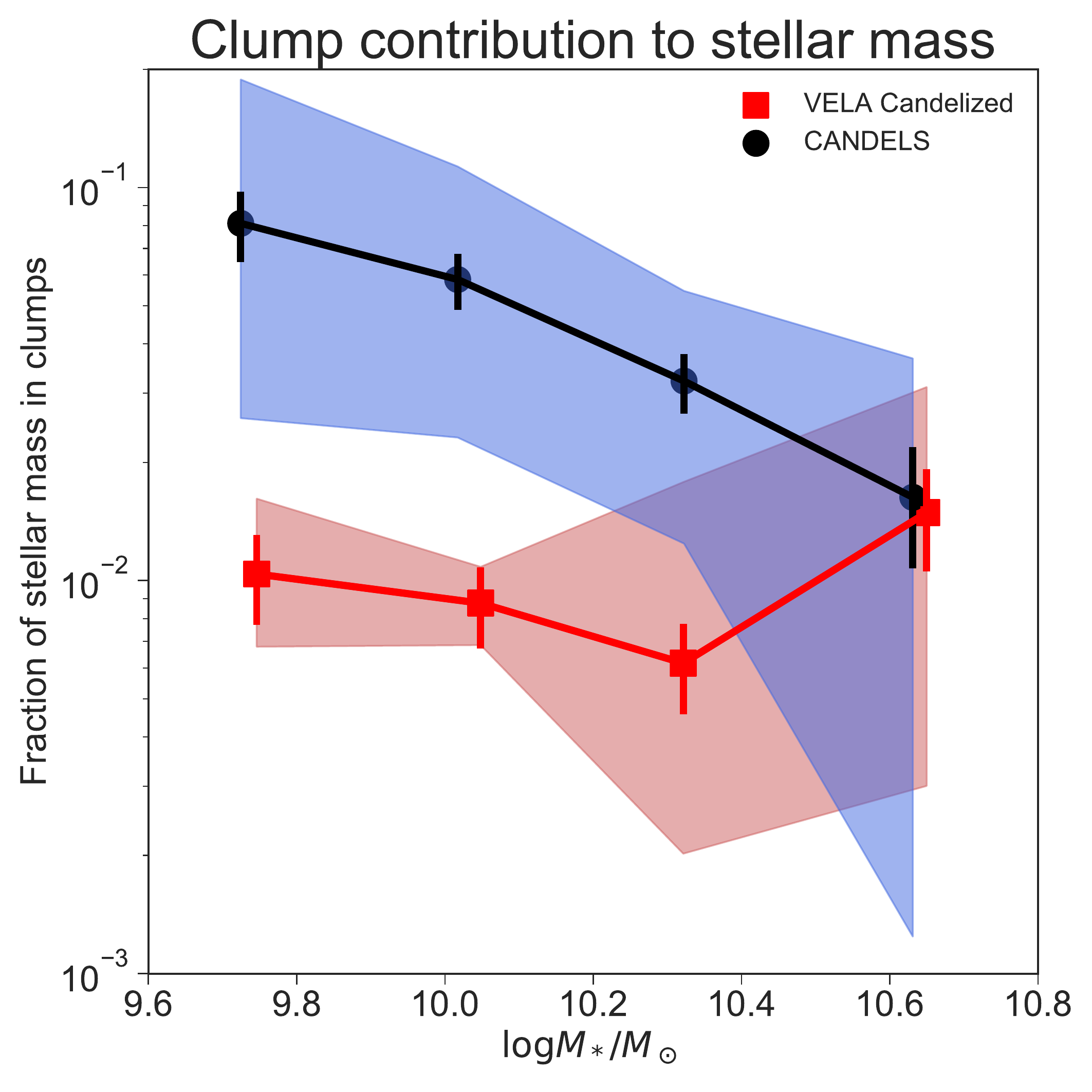}}%
         \qquad
     \subcaptionbox{}{\includegraphics[width=0.31\textwidth]{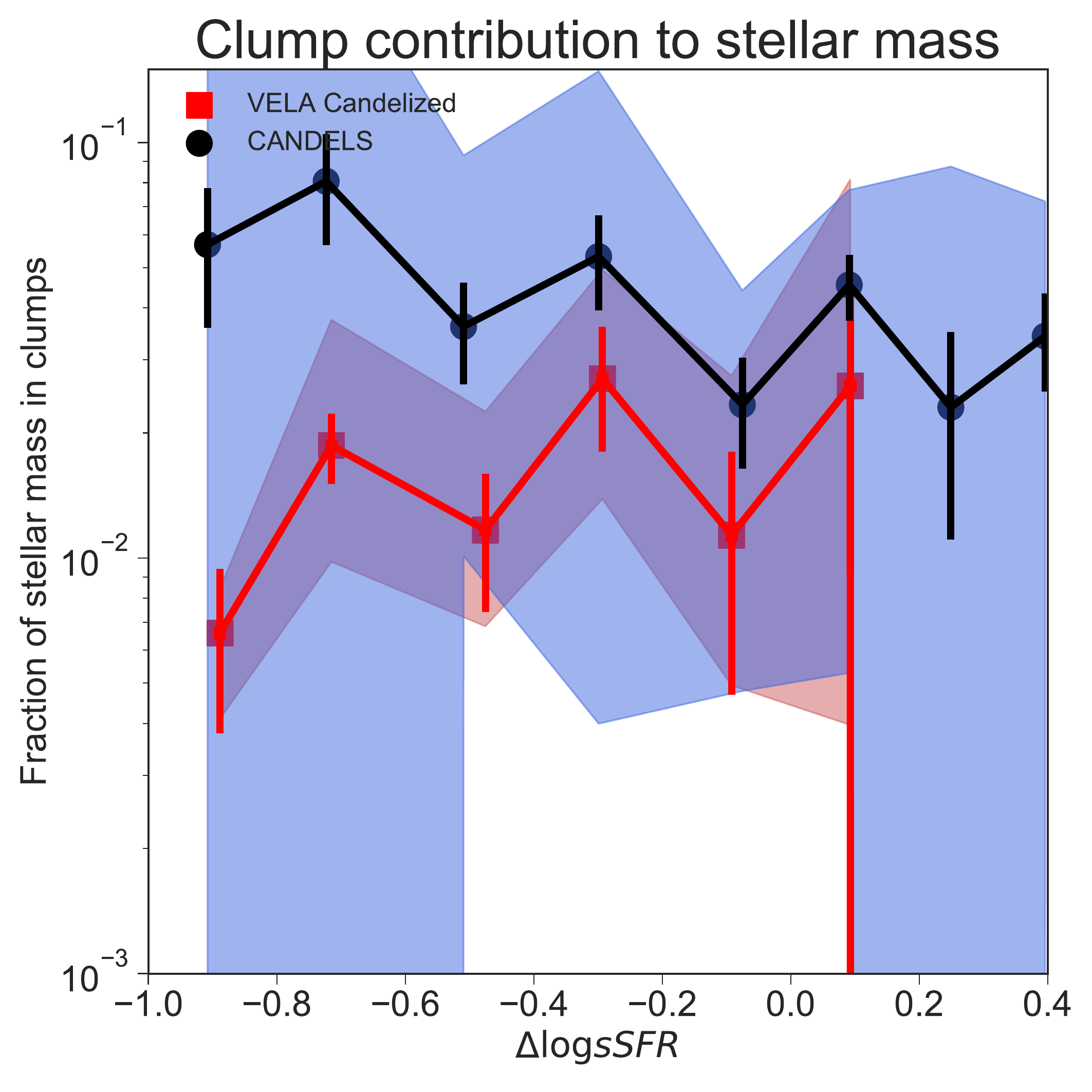}}%
     \qquad
\subcaptionbox{}{\includegraphics[width=0.31\textwidth]{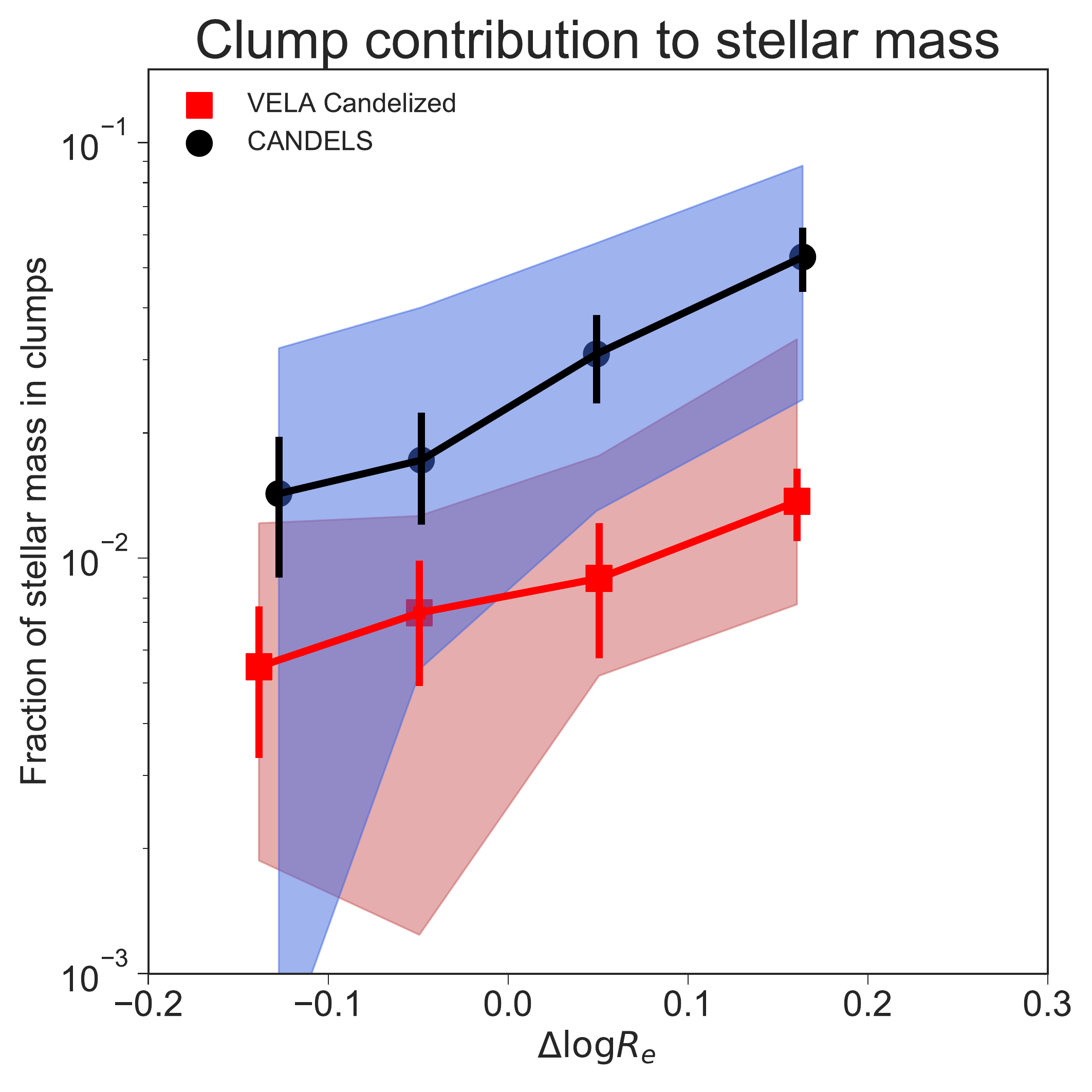}}%
    	\qquad
    \caption{Fraction of stellar mass in clumps more massive than $10^{7}$ solar masses in CANDELS and VELA star-forming galaxies as a function of (a) stellar mass, (b) $\Delta\log sSFR$,  and (c) $\Delta \log R_e$. All galaxies are in the redshift range $1<z<3$. The red filled squares are the measurements for simulated galaxies and the light red shaded region represents the uncertainty due to the different camera projections and corrections applied. The light blue shaded region indicates the range populated by observations sampled with the same stellar mass and redshift distributions as the simulated galaxies (see text for details). The black filled circles show the average values for the observations. Overall simulations tend to lie in the lower limits of the observations but are still compatible, although the dependence with galaxy stellar mass seem to differ.  }%
    \label{fig:clump_mass_VELA_CANDELS}%
\end{figure*}

With the purpose of better interpreting the observed trends, we investigate now the behavior of $C_*$ in the VELA Candelized simulations.  As done in subsection~\ref{sec:CANDELS-VELA}, in order to perform a fair comparison, we sample the observational dataset 100 times to build CANDELS-VELA subsamples of comparable size to the simulated one and with the same stellar mass and redshift distributions.  Figure~\ref{fig:clump_mass_VELA_CANDELS} shows the fraction of mass in clumps for VELA and CANDELS as a function of stellar mass, $\Delta\log sSFR$, and $\Delta\log R_e$ bins. $\Delta\log sSFR$ and $\Delta\log R_e$ are computed in VELA using the same reference fit used for the observations from~\cite{2018ApJ...858..100F}. We also checked that the distributions in both datasets are comparable. We do not separate galaxies in redshift bins because of the small statistics. The width of the shaded regions in the figure indicate the range of values obtained in the observations within the 100 samples and in the different camera orientations for the simulations. The uncertainty can be quite large given the small size of the considered samples. 

We see that the simulations overall predict a fraction of $\sim2-3\%$ of mass in clumps. This is in the lower limit of the confidence values from the observations. This is expected given the agreement of the cSMFs shown in the previous subsection. Despite the small numbers, there are two main trends that are still visible in the VELA-CANDELS sample. $C_*$ decreases with increasing galaxy stellar mass. This trend does not seem to be seen in VELA though, where the relation with stellar mass is flat and even slightly increases at the high mass end. This is a potential element which might be worth investigating further with increased statistics as it seems to point out that low mass galaxies do not form clumps as efficiently as in the observations. However, one should keep in mind the large uncertainties at low masses. Another trend which remains in CANDELS is the tendency for a larger fraction of mass in clumps for large galaxies at fixed mass, as can be seen in the rightmost panel of figure~\ref{fig:clump_mass_VELA_CANDELS}. A similar trend appears in VELA although with a shallower slope. Notice that if the trend with galaxy size was uniquely due to an observational bias, one would have expected to see a similar trend in the VELA Candelized data. The fact that the trends are different suggests that there is indeed a physical origin for the correlation. Finally, both observations and simulations show no dependence of $C_*$ with $\Delta\log sSFR$ which confirms that even if clumps are formed in-situ, the dependence with sSFR is not necessarily strong.

\subsection{Clumpy fractions}

In order to gain insights into how clumps are formed, it is also useful to quantify the abundance of clumps in galaxies of different properties. This is typically done with the so-called \emph{clumpy fraction} which measures the fraction of galaxies considered as \emph{clumpy}, i.e. which host at least one off-centered clump. This quantity is of course related to the fraction of stellar mass in clumps discussed in the previous subsection, but it is not exactly the same. The clumpy fraction is an indicator of the frequency at which galaxies form clumps. However, this quantity strongly depends on the stellar mass threshold used to select clumps and define a galaxy as clumpy. With low enough threshold, a normal, nearby star forming region could be called \emph{clumpy}.  In Guo15, the authors emphasize thus the importance of setting a threshold to distinguish actual giant clumps from regular star forming regions observed in nearby spirals. By using simulations of redshifted local galaxies, they find that an $\sim8\%$ threshold of the UV luminosity is a good choice to select giant clumps. 

In this work, we have access to the stellar masses of clumps in addition to luminosities and therefore we can establish clumpy fractions defined by stellar mass instead of luminosity, which should be more directly comparable to simulations of galaxy formation. Given our completeness limits, we define a galaxy as clumpy if it contains at least one clump more massive than $10^{7}$ solar masses. This mass threshold removes \emph{normal} local-like star forming regions  which are rarely more massive than $10^6$ solar masses. The definition is of course arbitrary, but it is physically motivated and provides a benchmark for comparison with simulations.

\subsubsection{Observations}

\begin{figure*}
	\centering
	 \subcaptionbox{}{\includegraphics[width=0.31\textwidth]{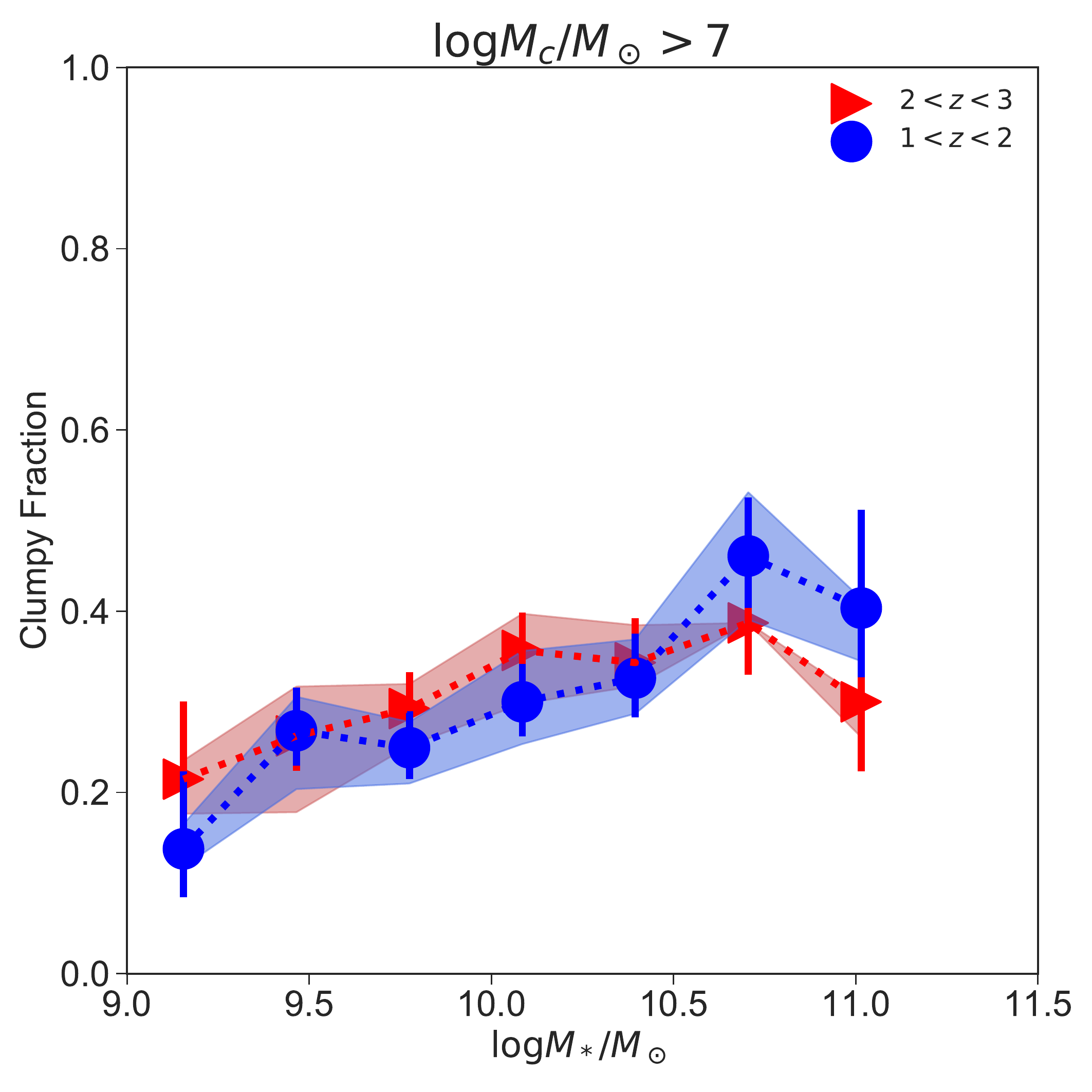}}%
         \qquad
     \subcaptionbox{}{\includegraphics[width=0.31\textwidth]{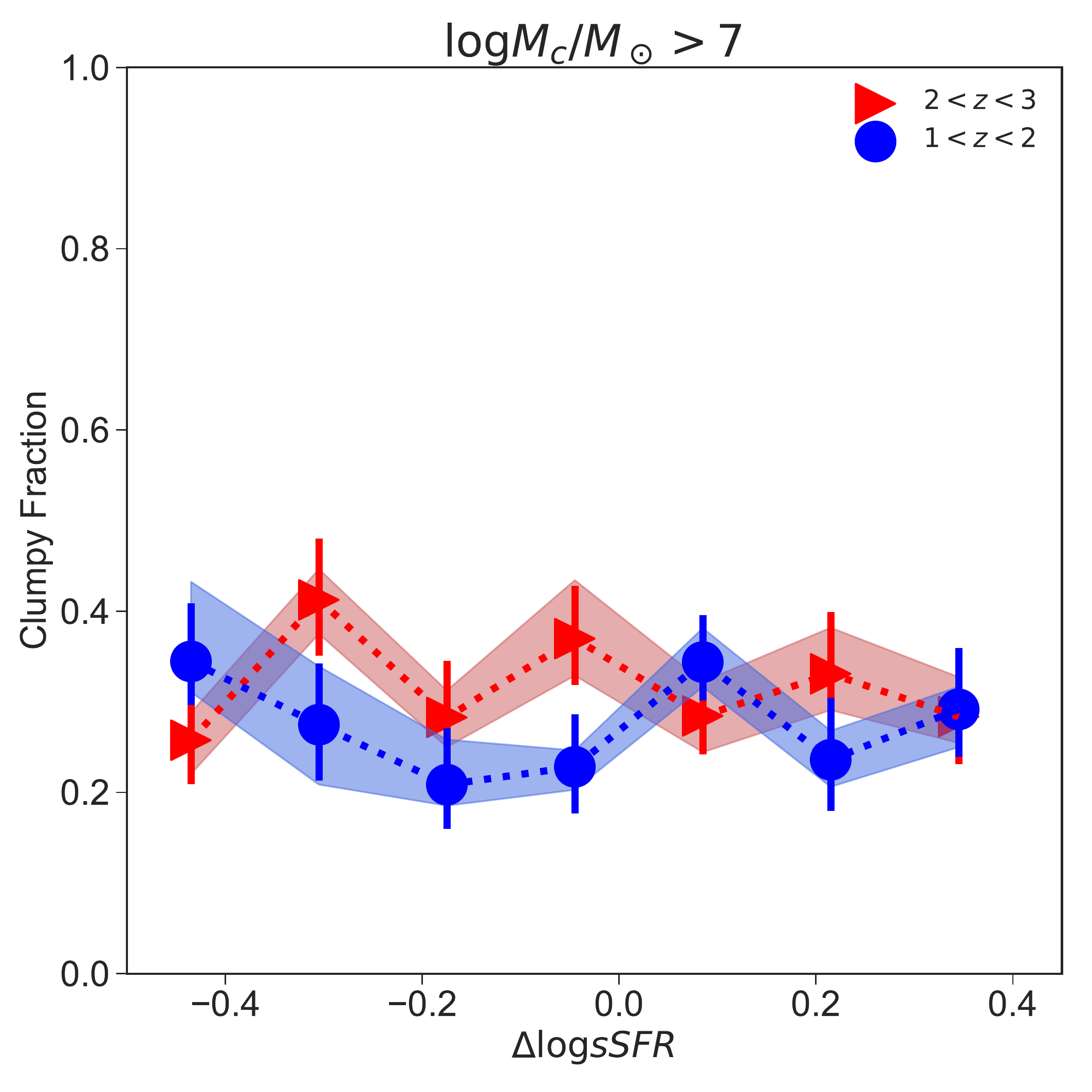}}%
     \qquad
\subcaptionbox{}{\includegraphics[width=0.31\textwidth]{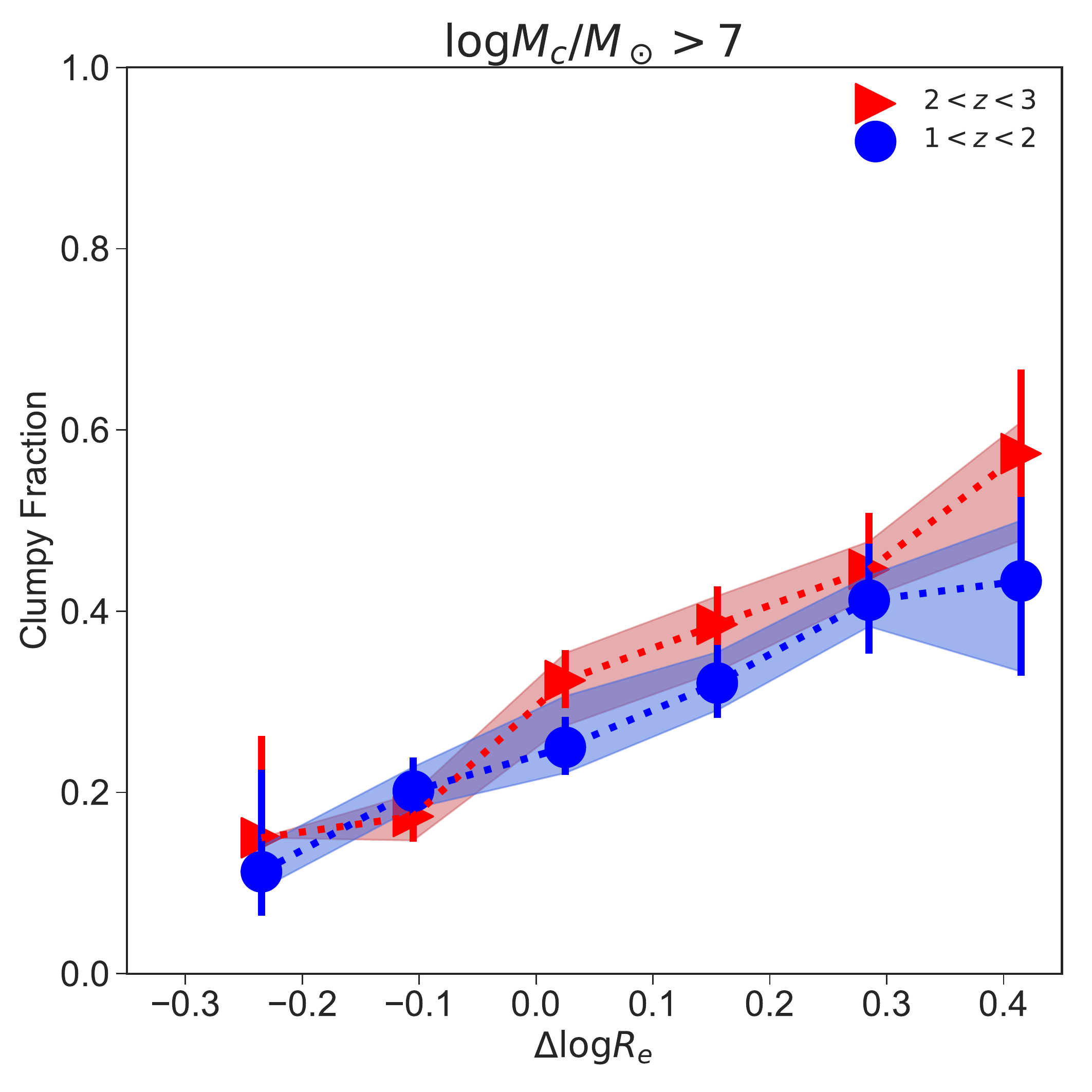}}%
    	\qquad

    \caption{Clumpy fraction in star-forming CANDELS galaxies as a function of (a) stellar mass, (b) $\Delta\log sSFR$,  and (c) $\Delta \log R_e$. The clumpy fraction measures the fraction of galaxies with at least one massive ($log_{10}M_c/M_\odot>7$) off centered clump. Different symbols show different redshift bins as labeled. Error bars are obtained by performing multiple samples of the posterior distributions  and the shaded red regions indicate uncertainties arising from the correction applied to clump masses (see text for details). On average $\sim20\%$ of the galaxies, in the redshift range considered, host massive  off-centered clumps with a trend of larger fractions in massive and large galaxies. The clumpy fraction does not depend on the relative position of galaxies in the star-forming main sequence. }
    \label{fig:clumpy_fractions}
\end{figure*}

Figure~\ref{fig:clumpy_fractions} shows the clumpy fractions as a function of stellar mass and redshift. As for the total mass contribution, the reported values are obtained through MonteCarlo sampling of the posterior distribution of the stellar masses derived through SED fitting. We first see that massive galaxies ($\log(M_*/M_\odot)>10$) have larger clumpy fractions than lower mass galaxies ($\log(M_*/M_\odot)<10$). Around $\sim40\%$ of galaxies more massive than $10^{10}$ solar masses present massive clumps. This fraction drops to $10-20\%$ for $\sim10^9$ solar mass galaxies. This is a direct consequence of the clump mass threshold used to define clumpy galaxies. It is less frequent to have such massive clumps in low mass galaxies since they represent a significant fraction of the total galaxy mass. However, figure~\ref{fig:clump_mass_contrib} shows that even if less frequent, the integrated contribution of clumps to the stellar mass budget is larger in these low mass systems. Another interesting result of figure~\ref{fig:clumpy_fractions} is that there is a very mild evolution with redshift especially for massive galaxies. The two redshift bins considered present very similar values.This behavior is in contrast with the clumpy fractions reported using UV luminosities (Guo15, ~\citealp{shibuya16}, among others) which can reach up to $65\%$ values at $z\sim2$ and significantly decrease towards low redshifts, down to $<20\%$ especially for massive galaxies. Our results imply that, even if UV clumps are very frequent at high redshift, the clump  contribution to the stellar mass does not evolve much.  



We additionally explore in figure~\ref{fig:clumpy_fractions} the dependence of the clumpy fractions with $\Delta\log R_e$ and $\Delta\log sSFR$. We observe that the clumpy fraction does not significantly depend on the specific star formation rate. Galaxies above and below the main sequence of star formation have very similar clump frequencies. Even if this might appear surprising considering that in-situ clump formation is very dependent on the gas mass density which is expected to be higher in galaxies with large sSFRs, we have already seen that the contribution of clumps to the stellar mass also shows very little dependence on the sSFR. 


In the right panel of figure~\ref{fig:clumpy_fractions} we plot the clumpy fraction as a function of $\Delta\log R_e$. We report an increasing clumpy fraction in large galaxies as compared with small galaxies of the same mass. Around forty percent of galaxies above the median mass-size relation have at least one massive clump. The fraction decreases to $\sim20\%$ for galaxies below the mass-size relation. This result, combined with the plot of figure~\ref{fig:clump_mass_contrib} implies that large galaxies not only have more massive clumps but also are more likely to host at least one such clump. We recall that the stellar mass dependence has been removed, so the dependence with size is at fixed mass and it is not a mere consequence of the mass dependence reported.  

This can again be a consequence of spatial resolution. Clumps too close to the galaxy centers in physical distance are difficult to detect because of the central regions being too bright. Alternatively, a more physical explanation implies that the outskirts of big disks seem to have physical conditions which favor the formation of clumps. This might appear counter intuitive in the first place. At fixed galaxy properties (e.g. stellar mass, velocity dispersion), increasing the size  will result in an increase of the Toomre parameter $Q$ (i.e. increased stability), which should result in turn in fewer clumps.  However, \cite{inoue16} have shown that $Q$ is not necessarily a good predictor for clumps as in many cases their formation is triggered by minor mergers. In some cases of protoclumps Q is significantly above unity.  Additionally, in a recent paper,~\cite{2020arXiv200308984D} suggest that clumpy star forming rings formed by high angular momentum streams can survive in large disks after the main compaction event of galaxies. This mechanism could also explain the larger number of clumps in large galaxies that we observe.

\subsubsection{Simulations}

\begin{figure*}
    \centering
\centering
	 \subcaptionbox{}{\includegraphics[width=0.31\textwidth]{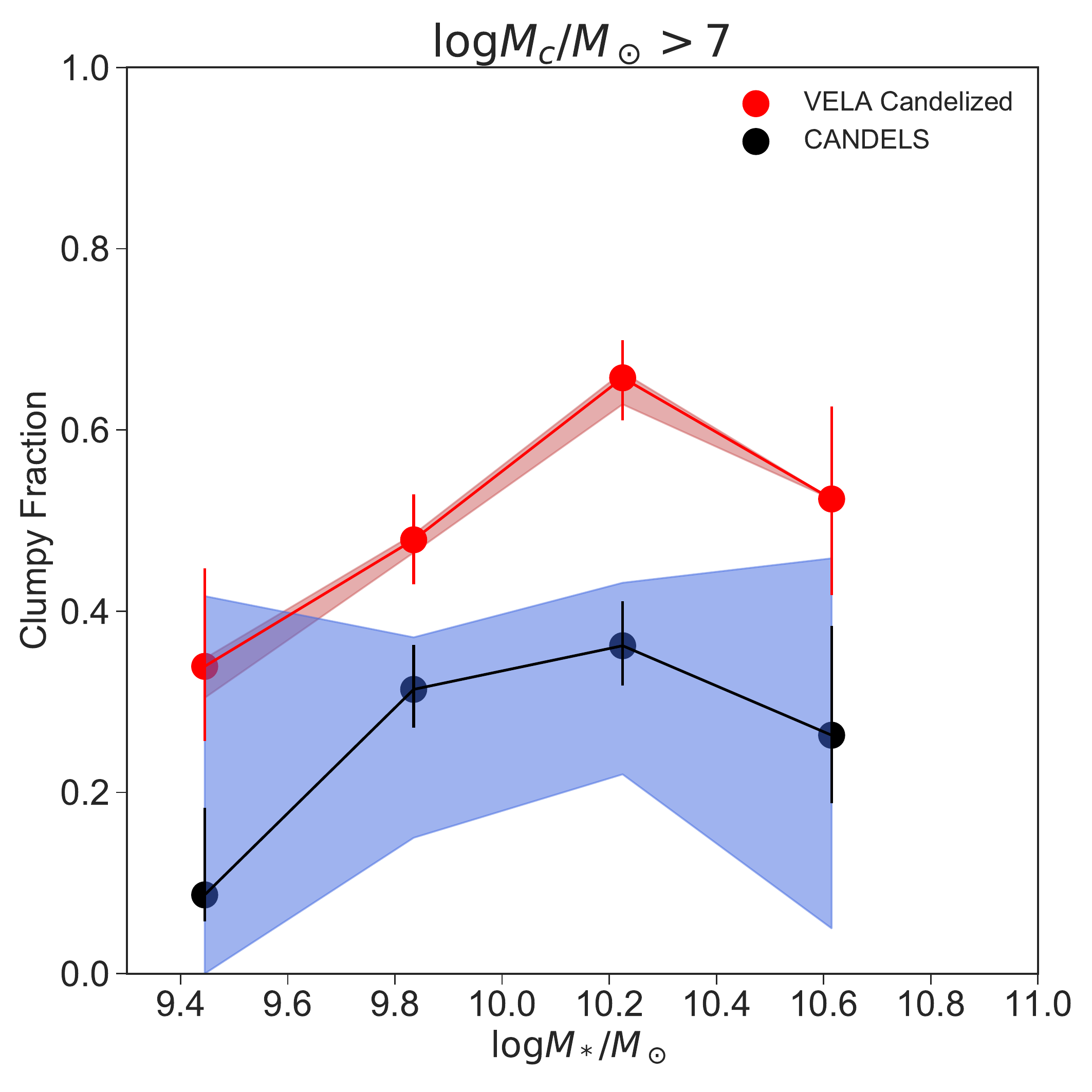}}%
         \qquad
     \subcaptionbox{}{\includegraphics[width=0.31\textwidth]{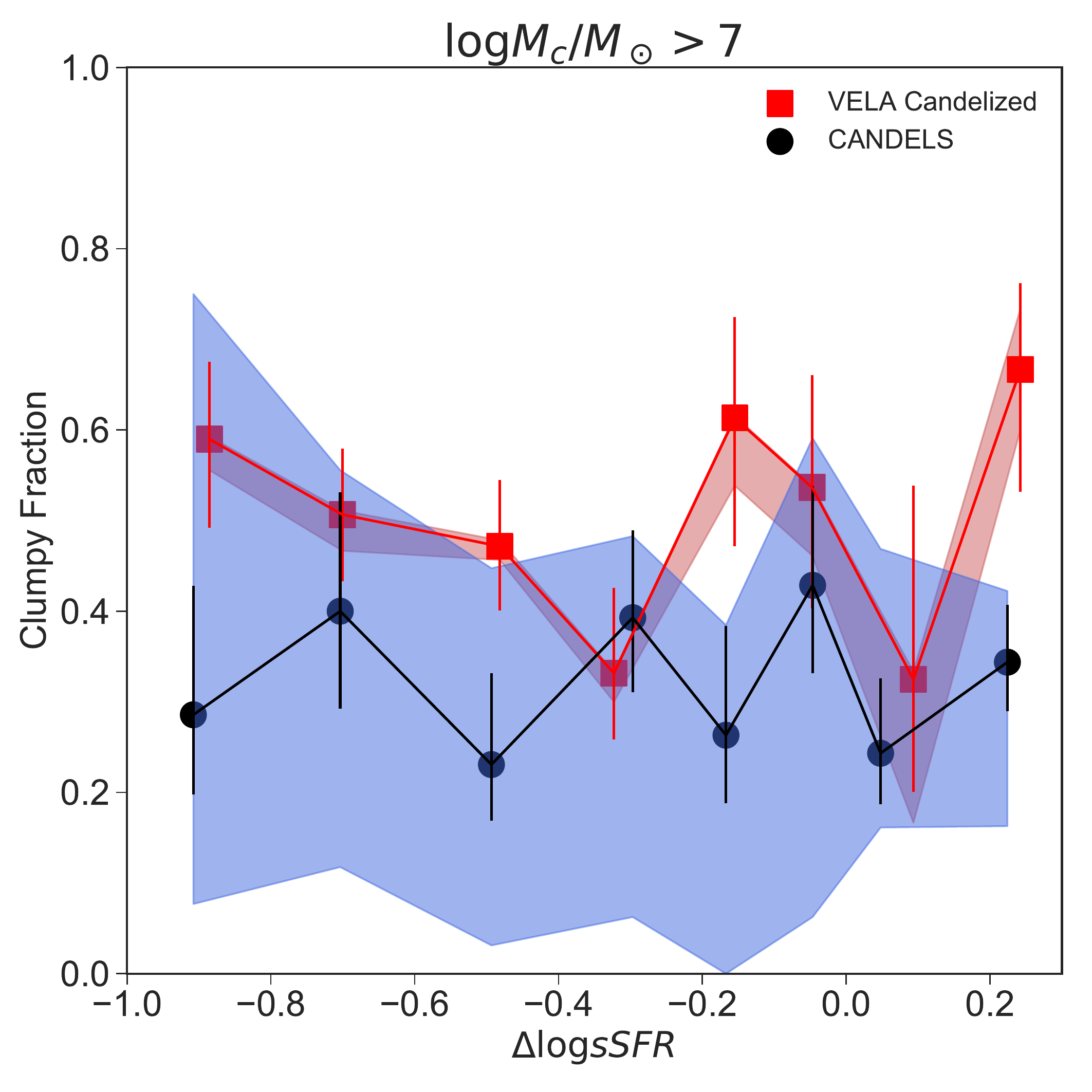}}%
     \qquad
\subcaptionbox{}{\includegraphics[width=0.31\textwidth]{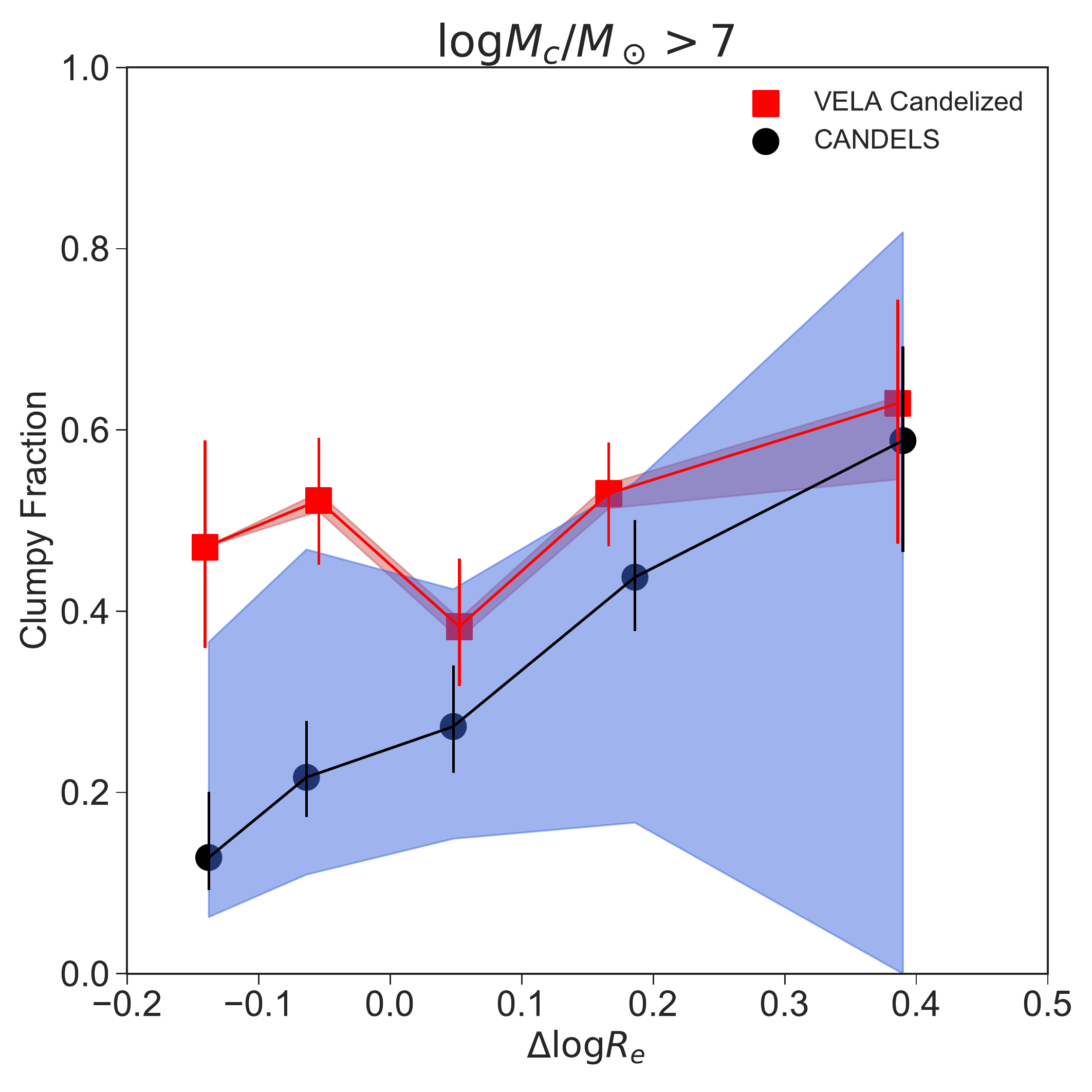}}%
    	\qquad
    \caption{Clumpy fraction in CANDELS and VELA star-forming galaxies as a function of (a) stellar mass (b) $\Delta\log sSFR$  and (c) $\Delta \log R_e$. The clumpy fraction measures the fraction of galaxies with at least one massive ($\log_{10}M_c/M_\odot>7$) off centered clump. The shaded blue regions are the confidence intervals derived for the observations by randomly creating CANDELS-VELA samples with the same size, mass and redshift distribution as the simulated galaxies (see text for details). The black filled circles show the average values. The red filled squares are the values measured in the VELA simulations. Overall simulations tend to be in the confidence interval of the observations.  }%
    \label{fig:clump_fraction_VELA_CANDELS}%
\end{figure*}

 We now compare the clumpy fractions, defined in both VELA and CANDELS as the fraction of galaxies hosting at least one massive clump. This is shown in figure~\ref{fig:clump_fraction_VELA_CANDELS}. We see a general good agreement between observed and simulated datasets although simulations tend to predict clumpy fractions in the higher region of the confidence interval. Both samples indeed present clumpy fractions around $20-40\%$ in the considered redshift range. The VELA simulation presents a clear dependence of the clumpy fraction with galaxy stellar mass. More massive galaxies are almost twice as likely to host a massive clump than lower mass ones, which is also seen in CANDELS. The left panel of figure~\ref{fig:clumpy_fractions} in which the complete CANDELS sample is included shows this trend more clearly. Massive galaxies have larger clumpy fractions because we set a mass limit for the clumps and it is obviously less frequent to have such a massive clump when the galaxy mass is smaller. However, it seems that the VELA simulation over predicts the fraction of massive clumpy galaxies.
 
 Interestingly, the clumpy fraction does not seem to depend on sSFR either in the simulations even if most of the clumps are formed in-situ. Regarding size dependence, we observe again that observations present a clear trend. The trend is again less pronounced in the VELA simulations.

\section{Summary and conclusions}
\label{sec:summary}

In the first part of this work, we have presented a neural network based clump detector. The neural network takes a stamp of a galaxy and produces a binary image with the pixels belonging to a clump set to one. The model is trained on simple analytic simulations made of Sersic profiles with added clumps, and it reaches a purity and completeness around $\sim90\%$ based on an independent test set. We have shown that the neural network model generalizes well to real data, even if trained on simulations, reaching comparable and even higher sensitivity on real observations than previously published methods and requiring less computational time. \\

Based on these results, we have applied the clump detector to a sample of $\sim9,000$ star-forming galaxies in CANDELS in up to seven different photometric bands. The catalog of detections is released with the present work. We then derive the stellar masses  of optically selected clumps through a Bayesian fit to the SEDs in two CANDELS fields (GOOD-S and GOODS-N) for which 7 photometric bands are available in the redshift range $1<z<3$. This corresponds to $\sim 1,500$ galaxies and $\sim3,000$ clumps. The same procedure is applied to forward modeled zoom-in cosmological simulations of 35 galaxies as they evolve in the redshift range $z=3$ to $z=1$, including all observational effects. Properties of the clumps in the VELA simulations are 
reported in~\cite{mandelker14,mandelker17}, which found that the vast majority of clumps form in-situ. We analyze simulated galaxies with exactly the same methods as for the observations and compare with  an observational dataset of the same size and similar redshift and stellar mass distributions.\\


Our main results are:

\begin{itemize}

\item Limits in angular resolution and signal to noise ratio in the photometry have a dramatic impact on the derived clumps stellar masses and therefore on the measured clump stellar mass function (cSMF) by flattening the low mass end and moving clumps towards larger masses. We have shown that this big difference between intrinsic and measured clump stellar mass functions is essentially due to a combination of incompleteness, clump blending, background light contamination, and overestimation of the clump stellar masses. This highlights the importance of comparing observations and simulations of galaxy formation under comparable conditions. It also suggests that it is difficult to extract accurate properties of individual high redshift clumps with the currently available data.
\item By calibrating with numerical simulations, we have shown that these effects can be partially corrected using a simple multi-layer perceptron to recover the intrinsic clump stellar mass function. We essentially perform a simple regression between the estimated SED clump mass and the 3D clump masses including some additional galaxy parameters. Once corrected, we find that the clump stellar mass function follows a power law with a slope of $-0.61\pm0.17$ above the completeness limit, which we estimate at $\sim10^7$ solar masses, with the majority of the clumps less massive than $10^9$ solar masses. Although the exact abundance of very massive clumps remains unconstrained with our data and methodology, this result tends to challenge previous observational works which found that many of the observed clumps at high redshift could be more massive than $10^9$ solar masses. Our clump stellar masses are also in better agreement with recent measurements based gravitational lensing.
\item The cSMF of simulated galaxies overall agrees with the corrected observed one when compared under the same conditions, suggesting an in-situ origin for the majority of the observed clumps. We emphasize that this agreement is independent of the corrections applied to the clump stellar masses. If simulations and observations are compared under the same conditions, without corrections, both stellar mass functions follow very similar trends. 
\item Using corrected clump masses, we estimate that the fraction of galaxy stellar mass in massive clumps ($>10^7$ solar masses) is of the order of $\sim2-5\%$, with a slight increase at $z>2$. This is smaller than the values reported for the stellar mass fractions in previous studies with incomplete datasets which did not correct for clump mass overestimation. The simulations analyzed also predict a similar fraction of mass in clumps.
\item We find indications that low mass galaxies ($<10^{10}$ solar masses) and larger-radius galaxies at fixed mass tend to present a larger clump contribution, which can reach up to $\sim7-15\%$. However, the mass fraction in clumps shows very little dependence on sSFR. The explored simulations also find a weak dependence with sSFR and an increasing contribution of clumps to the galaxy stellar mass in larger galaxies although with a weaker trend. This might be an indication of enhanced clump formation efficiency at large galacto-centric distances. 
\item We measure that $\sim20-40\%$ of star-forming galaxies present an off-centered clump more massive than $10^7$ solar masses, which also agrees reasonably well with the predictions of numerical simulations explored in this work. 
\end{itemize}

The present work highlights the importance of comparing simulations and observations in a consistent way. We also have shown that with the currently available data it is very difficult to establish accurate properties of clumps, and hence the need for new observing facilities such as JWST. 

 In future work, we will discuss clump life times using a similar approach as the one presented here (Ginzburg et al. in prep.). We also plan to extend the comparison of clump properties to other recent simulated datasets such as the new generation of VELA simulations which uses a stronger feedback and hence has a potential impact on clumps, in order to establish statistical constraints on the feedback mechanisms. A more refined SED fitting method involving more complex star-formation histories is another potential follow-up of this work.

\section*{Acknowledgements} This material is based upon C.T. Lee's research supported by the Chateaubriand Fellowship of the Office for Science \&
Technology of the Embassy of France in the United States.  The research is also supported by a Google Faculty Grant awarded to Prof. Joel Primack.  MHC is grateful to F. Lanusse for enlightening discussions about bayesian statistics with neural networks. NM acknowledges support from the Klauss Tschira Foundation through the HITS Yale Program in Astrophysics (HYPA). We acknowledge use of observations with the NASA/ESA Hubble Space Telescope obtained from the MAST Data Archive at the Space Telescope Science Institute, which is operated by the Association of Universities for Research in Astronomy, Incorporated, under NASA contract NAS5-26555. Support for Program number HST-AR-15798 was provided through a grant from the STScI under NASA contract NAS5-26555. The work of AD and OG was partly supported by the grants  BSF 2014-273, NSF AST-1405962, GIF I-1341-303.7/2016,  and DIP STE1869/2-1 GE625/17-1

\section*{Data Availability}

The observational data underlying this article are part of the CANDELS survey and can be accessed here: \url{http://arcoiris.ucolick.org/candels/}. The simulations are part of the VELA simulations suite. Catalogs of metadata are available upon request. Surnise images of the VELA simulated galaxies used in this work are available online at MAST -- see \url{https://archive.stsci.edu/prepds/vela/}. The derived data generated in this research, including deep learning models together with training sets and clump catalogs will be shared on reasonable request to the corresponding author.






\appendix

\section{Effects on the clump stellar mass function of measurement errors and clump blending}
\label{sec:app}
\begin{figure}
	\centering
	\includegraphics[width=0.45\textwidth]{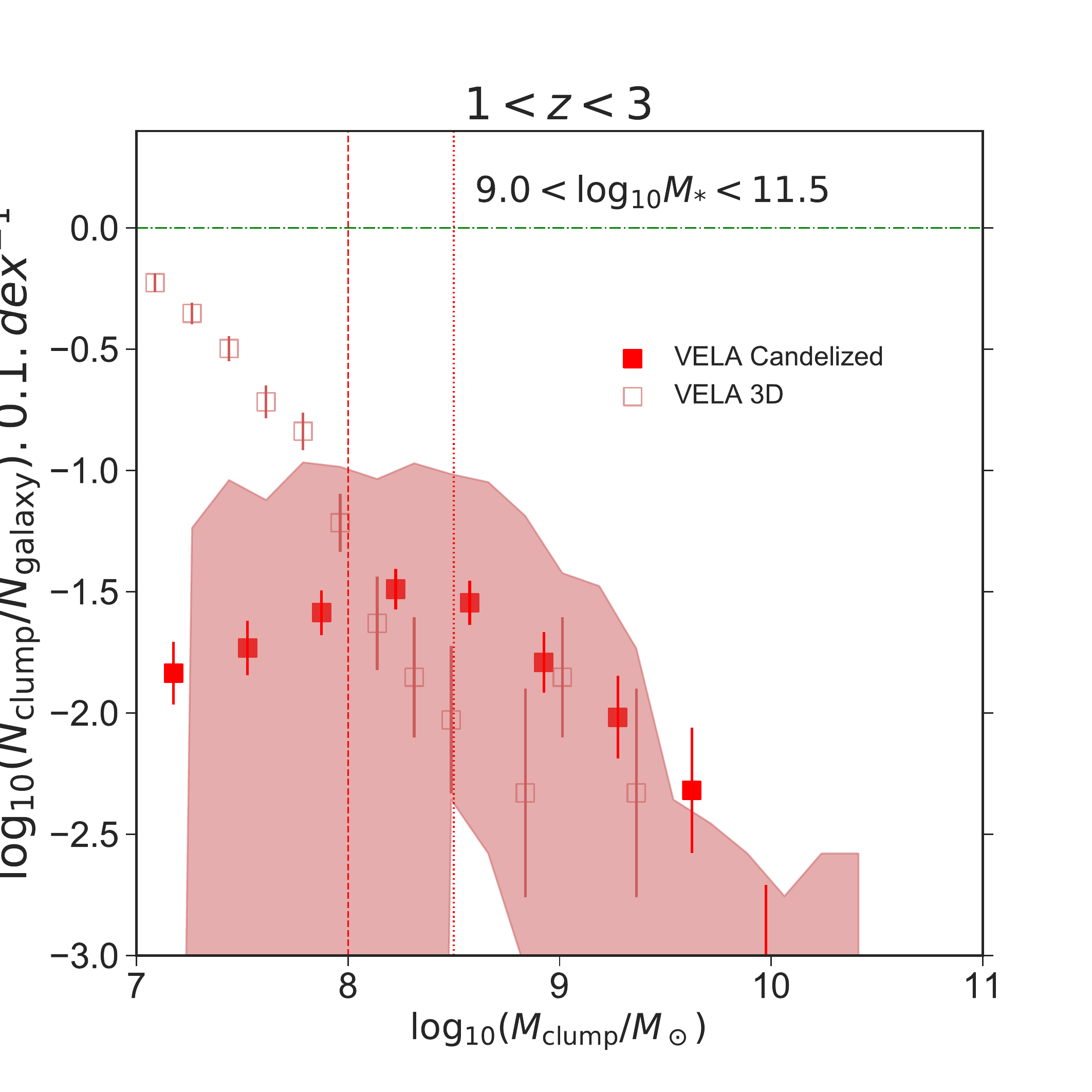}%
    \caption{Effect of mass measurement errors and clump blending on the derived cSMF. Empty squares show the cSMF derived from the simulation metadata in 3D~\citep{mandelker17}. The red filled squares show the cSMF measured from the Candelized images. The shaded region indicates the range of obtained cSMFs when mass errors and blending effects are added to the intrinsic 3D VELA cSMF (see text for details). A combination of severe measurement errors and clump blending can globally explain the differences between the Candelized and 3D clump stellar mass functions. }
    \label{fig:app_1}
\end{figure}

One of the important results we have highlighted in this work is that the cSMFs estimated from the Candelized images differ significantly from the ones derived directly from the simulation output. In section~\ref{sec:CANDELS}, we have shown that this large difference might be due to a combination of mass measurement errors and clump blending. In this appendix, we further explore the impact of these two effects in the cSMFs using simple simulations. We start from the intrinsic 3D VELA cSMF and randomly add to the stellar mass of every clump more massive than $10^7$ solar masses a gaussian error of $1$ dex with a standard deviation of $0.3$ dex based on the results of figure~\ref{fig:mass_2d_3d} and recompute the cSMF with the randomly assigned masses. We repeat this 50 times. We then assume than only $\sim80\%$ of the clumps are detected (figure~\ref{fig:VELA}) and that, on average every 2D detection corresponds to $\sim3$ 3D clumps because of blending (fig~\ref{fig:mass_2d_3d}) and renormalize the cSMF accordingly. The results of this exercise are shown in figure~\ref{fig:app_1}. The shaded region indicates the region occupied by the different realizations. We see that the Candelized cSMF and the 3D cSMF after applying the observational effects just described tend to agree much better. In particular, we see that the high mass end gets populated and that the slope is flattened. This supports our idea that the big difference between 3D and Candelized cSMFs comes from a combination of blending and mass measurement errors. 

\bsp	
\label{lastpage}
\end{document}